%% file: InclJetPaper.tex
\documentclass[a4paper,11pt]{article}

\usepackage{jheppub} 
\RequirePackage{atlasphysics}
\RequirePackage{subfigure}
\RequirePackage{afterpage}
\usepackage[T1]{fontenc} 
\usepackage{lscape}
\usepackage{multirow}
\usepackage{placeins}
\usepackage{graphicx}

\usepackage{preprintcover}
\PreprintCoverPaperTitle{Measurement of the inclusive jet cross-section in proton--proton collisions at $\sqrt{s}=7$~TeV using 4.5~$\mbox{fb}^{-1}$ of data with the ATLAS detector}  

\PreprintIdNumber{CERN-PH-EP-2014-155}  

\PreprintCoverAbstract{
The inclusive jet cross-section is measured in proton--proton collisions at a centre-of-mass energy of 7~TeV using a data set corresponding to an integrated luminosity of 4.5 $\mbox{fb}^{-1}$ collected with the ATLAS detector at the Large Hadron Collider in 2011.
Jets are identified using the anti-$k_t$ algorithm with radius parameter values of 0.4 and 0.6. 
The double-differential cross-sections are presented as a function of the jet transverse momentum and the jet rapidity, covering jet transverse momenta from 100~GeV to 2~TeV. 
Next-to-leading-order QCD calculations corrected for non-perturbative effects and electroweak effects, as well as Monte Carlo simulations with next-to-leading-order matrix elements interfaced to parton showering, are compared to the measured cross-sections. 
A quantitative comparison of the measured cross-sections to the QCD calculations using several sets of parton distribution functions is performed.
}

\PreprintJournalName{J. High Energy Phys.}

\title{\boldmath Measurement of the inclusive jet cross-section in proton--proton collisions at $\sqrt{s}=7$~TeV using 4.5~$\mbox{fb}^{-1}$ of data with the ATLAS detector}

\author{The ATLAS collaboration}

\emailAdd{atlas.publications@cern.ch}

\abstract{
The inclusive jet cross-section is measured in proton--proton collisions at a centre-of-mass energy of 7~TeV using a data set corresponding to an integrated luminosity of 4.5 $\mbox{fb}^{-1}$ collected with the ATLAS detector at the Large Hadron Collider in 2011.
Jets are identified using the anti-$k_t$ algorithm with radius parameter values of 0.4 and 0.6. 
The double-differential cross-sections are presented as a function of the jet transverse momentum and the jet rapidity, covering jet transverse momenta from 100~GeV to 2~TeV. 
Next-to-leading-order QCD calculations corrected for non-perturbative effects and electroweak effects, as well as Monte Carlo simulations with next-to-leading-order matrix elements interfaced to parton showering, are compared to the measured cross-sections. 
A quantitative comparison of the measured cross-sections to the QCD calculations using several sets of parton distribution functions is performed.
}

\begin{document} 
\maketitle
\flushbottom

\input{intro}
\input{definition}

\input{atlas}
\input{montecarlo}
\input{theory}

\input{dataselec}
\input{unfold}

\input{syst}

\input{result}

\input{conclusion}

\input{Acknowledgement}

\appendix
\input{appxsec}
\clearpage
\bibliographystyle{JHEP_mod} 
\bibliography{InclJetPaper}
\clearpage

\input{atlas_authlist}

\end{document}

%% file: intro.tex
\section{Introduction}
\label{sec:intro}
At the Large Hadron Collider (LHC)~\cite{1748-0221-3-08-S08001}, jet production in proton--proton collisions can be explored in the TeV regime. 
In quantum chromodynamics (QCD), jet production can be interpreted as the fragmentation of quarks and gluons produced in the scattering process and its measurement provides information about 
the colour-exchange interaction.
Therefore, the measurement of the inclusive jet cross-section at the LHC provides a test of the validity of perturbative QCD (pQCD) and the results can contribute to the determination of the parton distribution functions (PDFs) in the proton, in the pQCD framework.
The ALICE, ATLAS and CMS Collaborations have measured inclusive jet cross-sections at centre-of-mass energies, $\rts= 2.76 \TeV$~\cite{Abelev:2013fn, Aad:2013lpa} and $\rts=7~\TeV$~\cite{Aad:2010wv, Aad:2011fc, CMS:2011ab, Chatrchyan:2012gwa, Chatrchyan:2012bja}. 
These data are generally well described by next-to-leading-order (NLO) pQCD calculations to which corrections for non-perturbative effects from hadronisation and the underlying event are applied. 

In this paper, the measurement of the double-differential inclusive jet
cross-section is presented as a function of the transverse momentum of the jets, \pt{}, and their rapidity,\footnote{ 
ATLAS uses a right-handed coordinate system with its origin at the nominal interaction point (IP) in the centre of the detector and the $z$-axis along the beam pipe. 
The $x$-axis points from the IP to the centre of the LHC ring, and the $y$-axis points upward. 
Rapidity is defined as $y=0.5\ln\frac{E+p_z}{E-p_z}$ where $E$ denotes the energy and $p_z$ is the component of the momentum along the beam direction.
The pseudorapidity, $\eta$, is defined in terms of the polar angle $\theta$ as $\eta=-\ln\tan(\theta/2)$.
} 
$y$, at $\sqrt{s}=7$~\TeV{} using the data collected by the ATLAS experiment in 2011, corresponding to an integrated luminosity of 4.5~$\mathrm{fb}^{-1}$. 
The measurement is performed using jets with $\pt{}\ge100$~\GeV{} and $|y|<3$. 
The integrated luminosity of the data used in this paper is more than 100 times larger than that of the previous ATLAS measurement~\cite{Aad:2011fc}, allowing larger kinematic reach, with the jet \pt{} measured up to 2~\TeV{}, corresponding to $\xt=2\pt/\rts\lesssim0.6$.
A precise measurement with full details of uncertainties and their correlations is performed taking advantage of the increased statistical power and improved jet calibration~\cite{Aad:2014bia}. 
A set of NLO pQCD calculations, to which corrections for both non-perturbative QCD effects and electroweak effects are applied, is compared to the results.
The comparison is quantitatively evaluated. 

The outline of the paper is as follows. 
The inclusive jet cross-section is defined in section~\ref{sec:def}. 
A brief description of the ATLAS detector is given in section~\ref{sec:atlas}. 
The Monte Carlo simulations and the theoretical predictions are described in sections~\ref{sec:mc} and~\ref{sec:theo}. 
The event selection is presented in section~\ref{sec:selec}, followed by discussions of the unfolding of detector effects and the systematic uncertainties in the measurement in sections~\ref{sec:unfold} and~\ref{sec:syst}, respectively.
The results are presented in section~\ref{sec:result}, together with a 
quantitative evaluation of the theory predictions in comparison to the measurement.
The conclusions are given in section~\ref{sec:concl}.

%% file: definition.tex
\section{Definition of the cross-section}
\label{sec:def}
Jets are identified using the \antikt algorithm~\cite{Cacciari:2008gp} in the four-momentum recombination scheme, implemented in the \fastjet~\cite{Fastjet, Cacciari200657} software package. 
Two values of the jet radius parameter, $R=0.4$ and $R=0.6$, are considered. 
Inputs to the jet algorithm can be partons in the NLO pQCD calculation, stable particles after the hadronisation process in the Monte Carlo simulations, or energy deposits in the detector. 

Throughout this paper, the jet cross-section refers to the cross-section of jets clustered from stable particles with a proper mean lifetime, $\tau$, given by $c\tau>10$ mm. 
Muons and neutrinos from decaying hadrons are included in this definition. These jets are referred to as particle-level jets in this paper. 

Jets built using partons from NLO pQCD predictions are referred to as parton-level jets. The NLO pQCD predictions with the parton-level jets must be corrected for hadronisation and underlying-event effects in order to be compared to the particle-level measurements. 

The double-differential inclusive jet cross-section, $\mathrm{d}^2\sigma/\mathrm{d}\pt \mathrm{d}y$, is measured in bins of the jet \pt{} and $y$, averaged in each bin. 
The measurement is performed in a kinematic region with $\pt\ge100$~\GeV{} and $|y|<3$.

%% file: atlas.tex
\section{The ATLAS detector}
\label{sec:atlas}
The ATLAS detector consists of a tracking system (inner detector) immersed in a $2\,\mathrm{T}$ axial magnetic field and covering pseudorapidities up to $|\eta|=$~2.5, 
electromagnetic and hadronic sampling calorimeters up to $|\eta|=4.9$, and muon chambers in an azimuthal magnetic field provided by a system of toroidal magnets. 
A detailed description of the ATLAS detector can be found in ref.~\cite{Aad:2008zzm}. 

The inner detector consists of layers of silicon pixel detectors, silicon microstrip detectors and transition radiation tracking detectors. In this analysis, it is used for the reconstruction of vertices from tracks.
Jets are reconstructed using energy deposits in the calo\-ri\-me\-ters, whose granularity and material vary as a function of~$\eta$.
The fine-granularity electromagnetic calorimeter uses lead as absorber and liquid argon (LAr) 
as the active medium.
It consists of a barrel ($|\eta|<1.475$) and two endcap ($1.375<|\eta|<3.2$) regions. 
The hadronic calorimeter is divided into five distinct regions: a barrel region ($|\eta|<0.8$), two extended barrel regions ($0.8<|\eta|<1.7$) and two endcap regions ($1.5<|\eta|<3.2$).
The barrel and extended barrel regions are instrumented with steel/scintillator-tile modules and the endcap regions are instrumented using copper/LAr modules.
Finally, the forward calorimeter ($3.1<|\eta|<4.9$) is instrumented with copper/LAr and tungsten/LAr modules to provide electromagnetic and hadronic energy measurements, respectively. 

%% file: montecarlo.tex
\section{Monte Carlo simulation}
\label{sec:mc}
For the simulation of the detector response to scattered particles in proton--proton collisions, events are generated with the \pythia{}~6.425~\cite{Sjostrand:2006za} generator. 
This utilises leading-order (LO) pQCD matrix elements for $2\rarrow2$ processes, along with a leading-logarithmic (LL) \pt-ordered parton shower~\cite{Sjostrand:2004ef} including photon radiation, underlying-event simulation with multiple parton interactions~\cite{Sjostrand:2004pf}, and hadronisation with the Lund string model~\cite{Andersson:1983ia}.
A sample generated with the \Perugia{} 2011 set of parameter values (tune)~\cite{Skands:2010akv4} and the CTEQ 5L PDF set~\cite{Lai:1999wy} is used for correction of detector effects in this measurement. 

The stable particles from the generated events are passed through the ATLAS
detector simulation~\cite{Aad:2010ah} based on the \geant{} software tool
kit~\cite{Agostinelli:2002hh}.
Effects from multiple proton--proton interactions in the same and neighbouring bunch crossings are included by overlaying minimum-bias events, which consist of single-, double- and non-diffractive collisions generated by the \pythia{}~6.425 generator.
The number of overlaid minimum-bias events 
follows a Poisson distribution with its mean equal to 
the averaged number of interactions per bunch-crossing throughout the analysed data-taking period.

For evaluation of non-perturbative effects, 
the \pythia{}~8.175~\cite{Sjostrand:2007gs} and \herwigpp{}~2.6.3 \cite{Bahr:2008pv,Arnold:2012fq} generators are also employed as described in section~\ref{sec:npcorr}.
The latter utilises LO $2\rarrow2$ matrix elements with an LL angle-ordered parton shower~\cite{Gieseke:2003rz}. 
It implements an underlying-event simulation based on an eikonal model~\cite{Bahr:2008dy} and hadronisation based on a cluster model~\cite{Webber:1983if}. 

%% file: theory.tex
\section{Theoretical predictions}
\label{sec:theo}
Theoretical predictions of the cross-section to be compared to the measurement are obtained from NLO pQCD calculations with corrections for non-perturbative effects. 
Predictions from NLO matrix elements interfaced to a Monte Carlo (MC) simulation of shower partons are also considered.  
In both cases, the predictions are corrected for electroweak effects. 

\subsection{NLO pQCD calculations}
The NLO pQCD predictions are calculated by the \nlojetpp{}~4.1.2 program~\cite{Nagy:2003tz}. The APPLGRID software~\cite{applgrid:2009} is interfaced with \nlojetpp{} for fast and flexible calculations with various PDF sets and various values of the renormalisation and factorisation scales. 
The renormalisation scale, $\mu_\mathrm{R}$, and the factorisation scale, $\mu_\mathrm{F}$, are chosen to be the leading jet transverse momentum, $\pt^\mathrm{max}$, for each event. 
Predictions are made with several NLO PDF sets, namely CT10~\cite{Lai:2010vv}, MSTW2008~\cite{Martin:2009iq}, NNPDF 2.1~\cite{Ball:2010de,Forte:2010ta}, ABM 11 ($n_\mathrm{f}=5$, i.e. for five fixed flavours)~\cite{Alekhin:2012ig} and HERAPDF 1.5~\cite{HERAPDF15}. 
The value of the strong coupling constant, \alphas{}, is set to that assumed in the corresponding PDF set. 

Uncertainties in the PDF sets, the choice of renormalisation and factorisation scales, and the uncertainty in the value of \alphas{} are considered as sources of uncertainties in the NLO pQCD calculations. 
Uncertainties in the PDF sets are propagated through the calculations following the prescription given for each PDF set and the PDF4LHC recommendations~\cite{Botje:arXiv1101.0538}. 
The evaluated uncertainties on the predictions are scaled to the 68\% confidence level for all PDF sets.  
Calculations are redone with varied renormalisation and factorisation scales to estimate the uncertainty due to missing higher-order terms in the pQCD expansion. 
The nominal scales are multiplied by factors of $(f_{\mu_\mathrm{R}}, f_{\mu_\mathrm{F}})=(0.5,0.5), (1,0.5), (0.5,1), (2,1), (1,2), (2,2)$.
The envelope of resulting variations of the prediction is taken as the scale uncertainty. 
The uncertainty reflecting the \alphas{} precision is evaluated following the recommended prescription of the CTEQ group~\cite{Lai:2010nw}, by calculating the cross-sections using a series of PDFs which are derived with various fixed \alphas{} values.

\begin{figure*}
\begin{center}
\subfigure[]{\includegraphics[width=6.9cm]{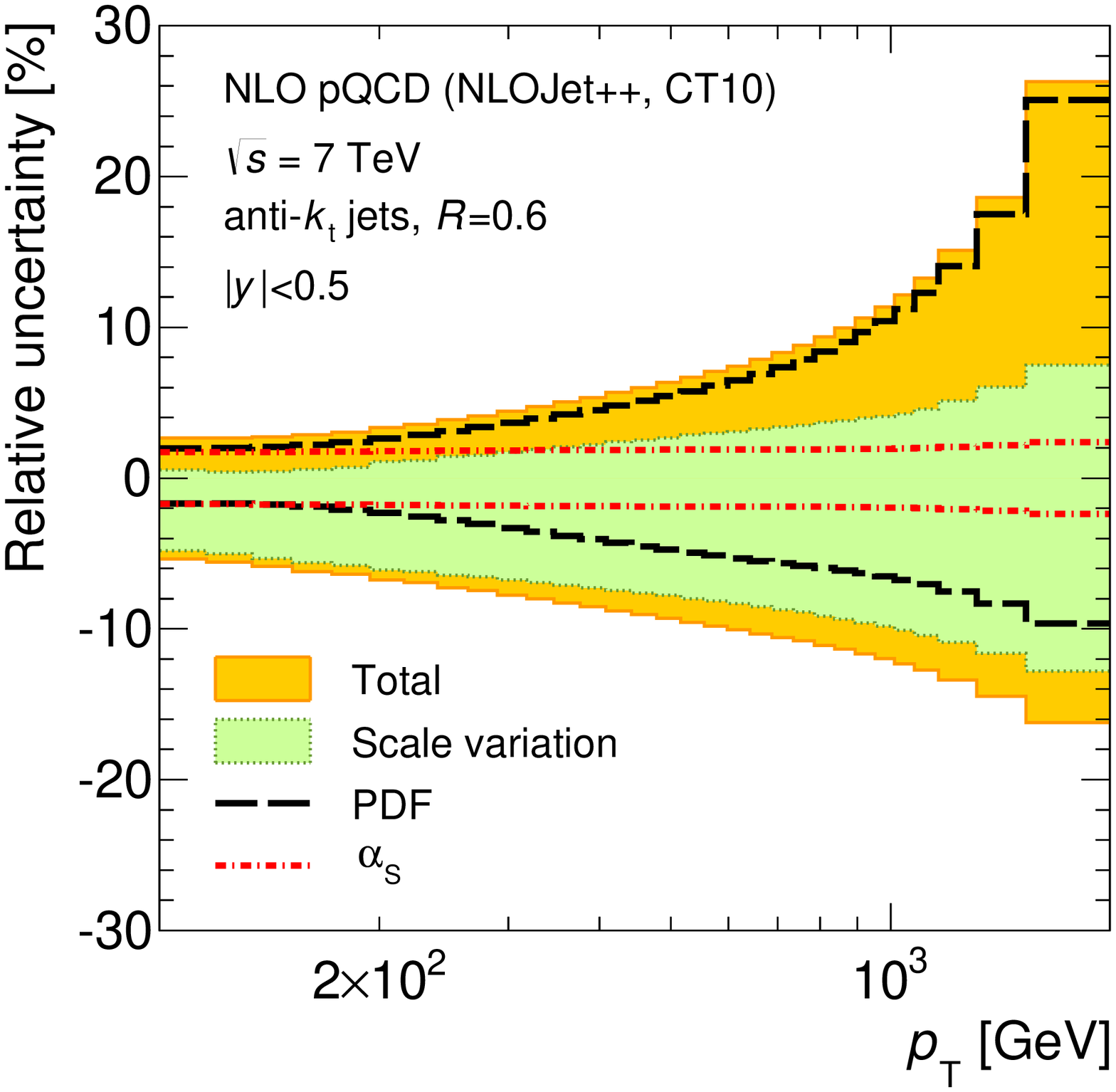}}
\subfigure[]{\includegraphics[width=6.9cm]{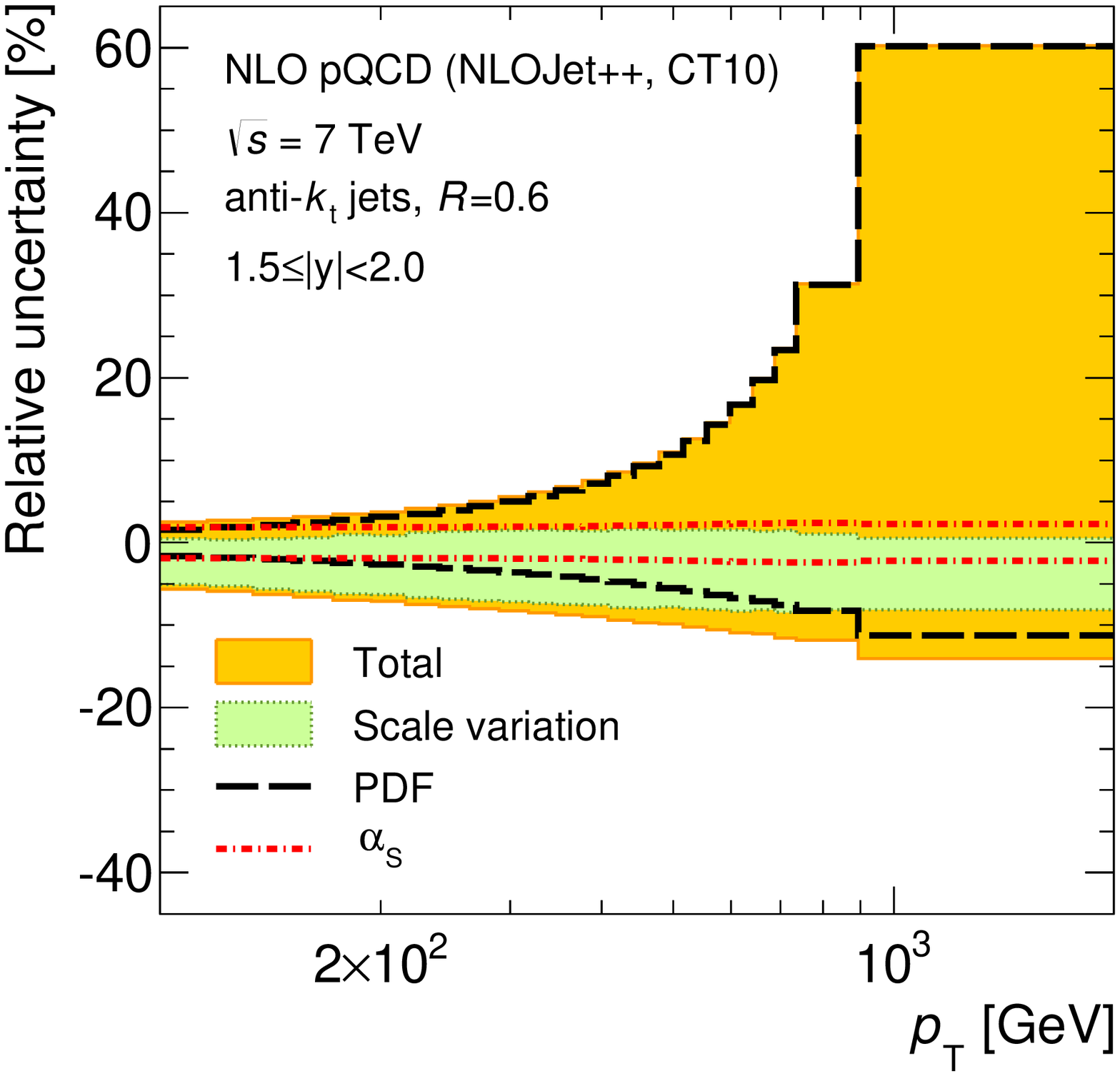}}
\subfigure[]{\includegraphics[width=6.9cm]{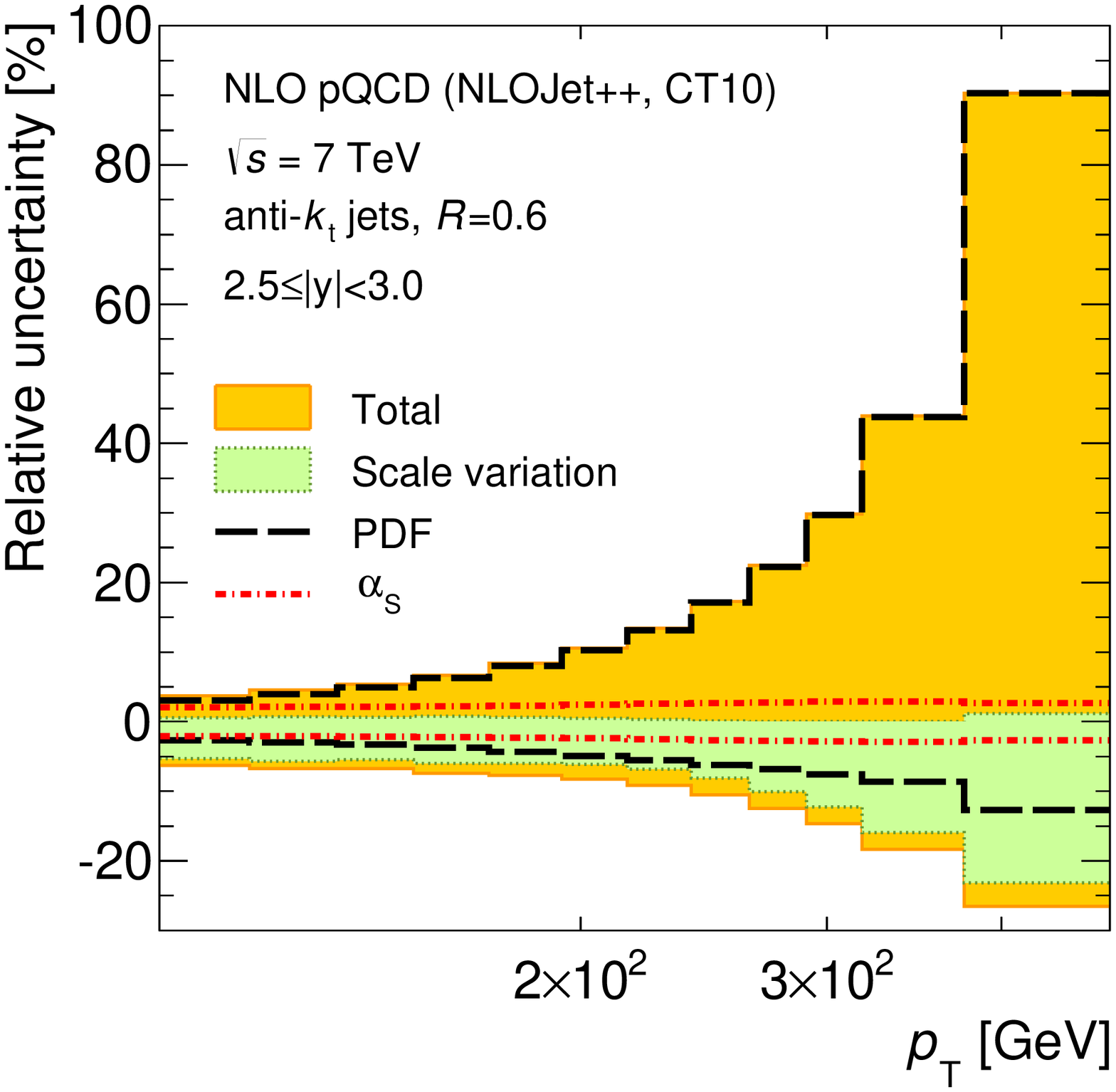}}
\caption{The uncertainty in the NLO pQCD prediction of the inclusive jet cross-section at $\rts=7$~\TeV{}, calculated using \nlojetpp{} with the CT10 PDF set, for \antikt jets with $R=0.6$ shown in three representative rapidity bins (as indicated in the legends), as a function of the jet \pt. In addition to the total uncertainty, the uncertainties from the scale choice, the PDF set and the strong coupling constant are shown.\label{fig:nlojet}}
\end{center}
\end{figure*}

Figure~\ref{fig:nlojet} shows the relative uncertainties in the NLO pQCD
calculations evaluated using the CT10 PDF set for the inclusive jet
cross-section as a function of the jet \pt{}, in representative rapidity bins for jets with $R=0.6$.
The uncertainty is mostly driven by the PDF uncertainty in the region $\pt>0.5$~\TeV{} or in the high-rapidity region. 
The uncertainties for the calculations with $R=0.4$ are similar. 
  
\subsection{Non-perturbative corrections to the NLO pQCD calculations}
\label{sec:npcorr}
Non-perturbative corrections are applied to the parton-level cross-sections from the NLO pQCD calculations. The corrections are derived using LO MC generators complemented by an LL parton shower.
The correction factors are calculated as the bin-by-bin ratio of the MC cross-sections obtained with and
without modelling of hadronisation and the underlying event.
The NLO pQCD calculations are then multiplied by these factors.

The correction factors are evaluated using several generators and tunes:
\pythia~6.427 using the AUET2B~\cite{ATL-PHYS-PUB-2011-009} and Perugia
2011~\cite{Skands:2010akv4} tunes, \herwigpp~2.6.3 using the
UE-EE-3~\cite{Gieseke:2012ft} tune, and \pythia~8.157 using the
4C~\cite{Corke:2010yf} and AU2~\cite{ATL-PHYS-PUB-2012-003} tunes.
The CTEQ6L1 PDF set~\cite{Pumplin:2002vw} is used except for the calculation with the Perugia 2011 tune, where the CTEQ5L PDF set is used. 
The baseline correction is taken from \pythia{} with the Perugia 2011 tune. The envelope of all correction factors is considered as a systematic uncertainty. 

\begin{figure*}
\begin{center}
\subfigure[]{\includegraphics[width=6.9cm]{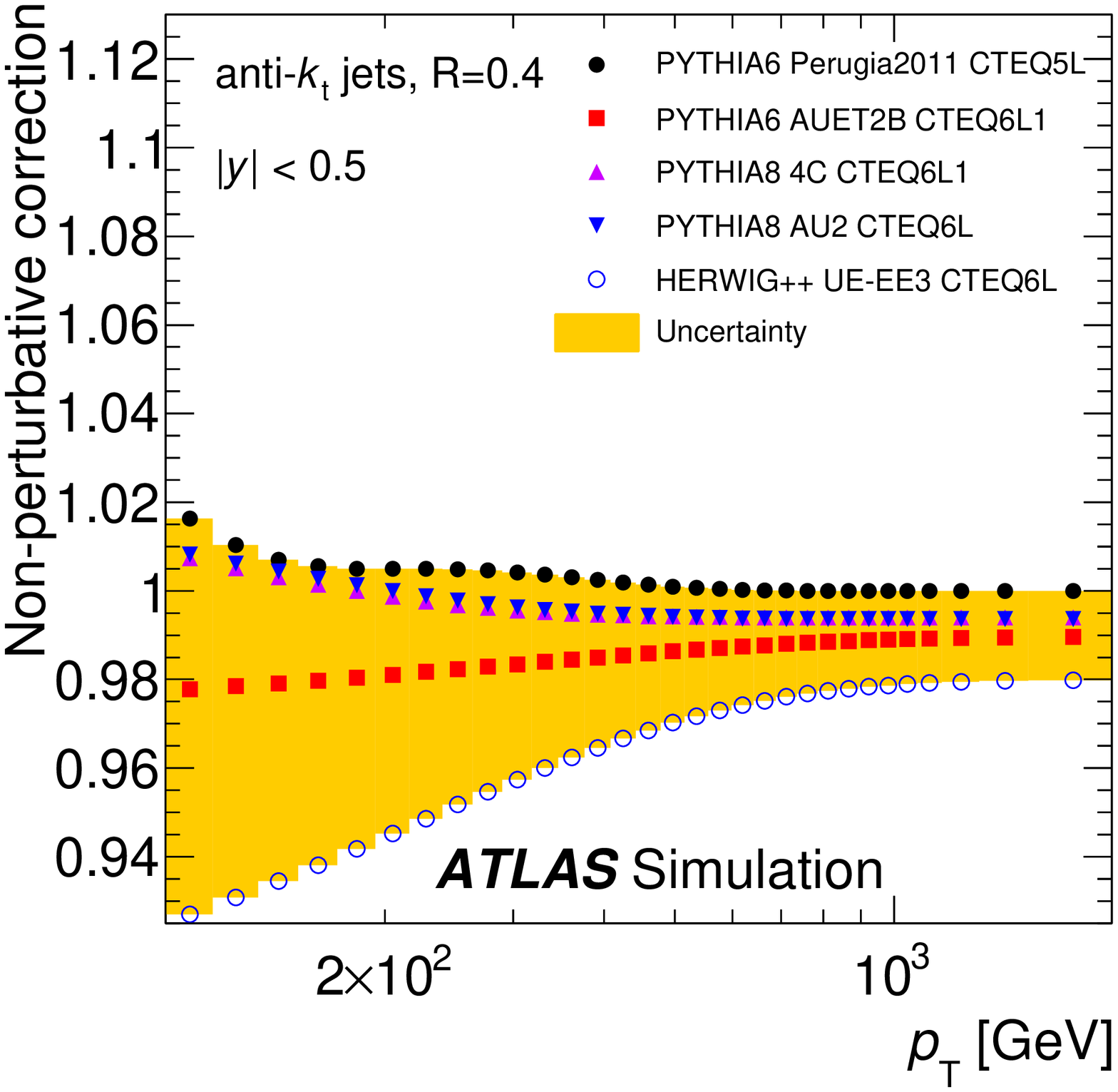}}
\subfigure[]{\includegraphics[width=6.9cm]{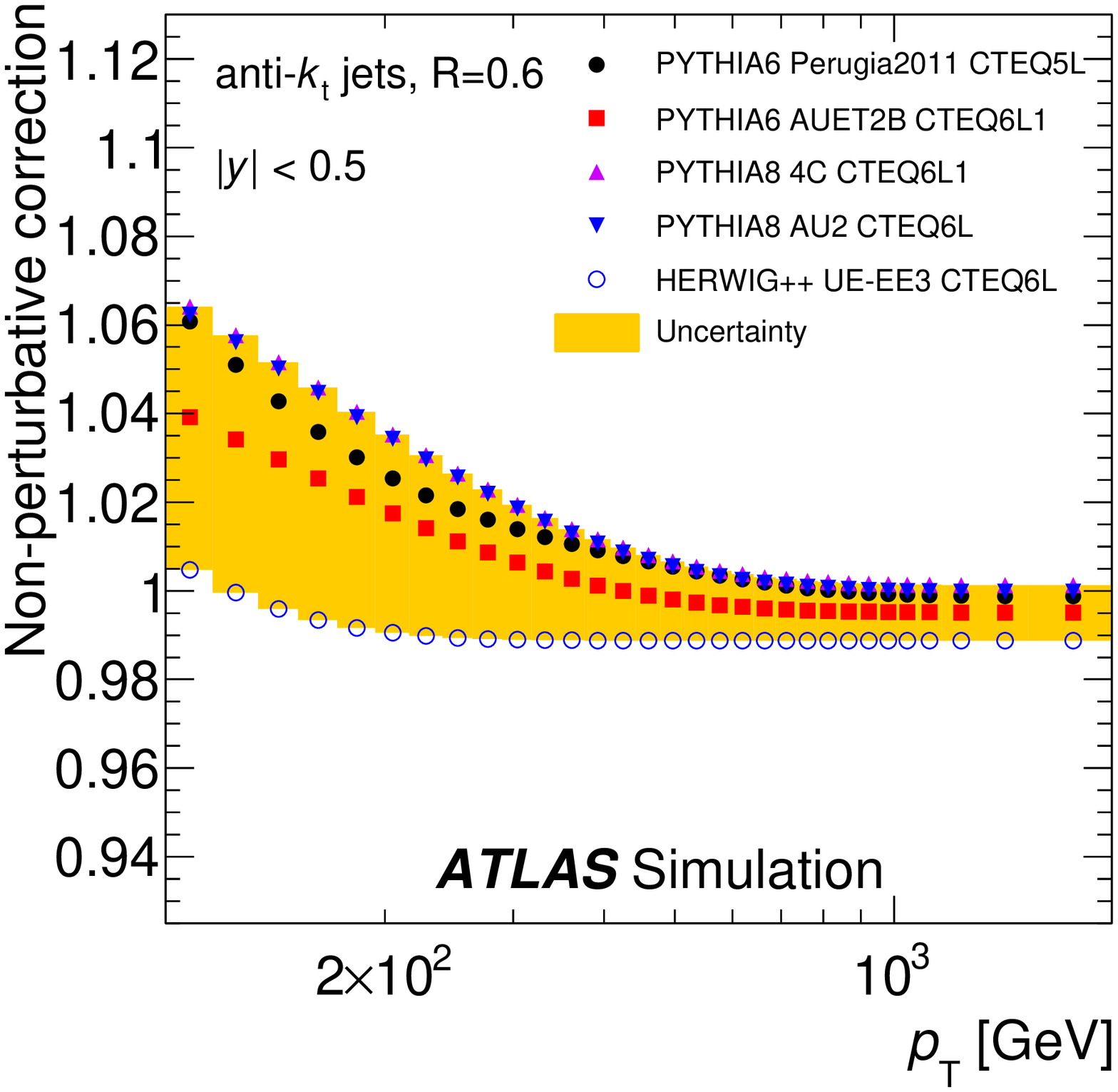}}
\subfigure[]{\includegraphics[width=6.9cm]{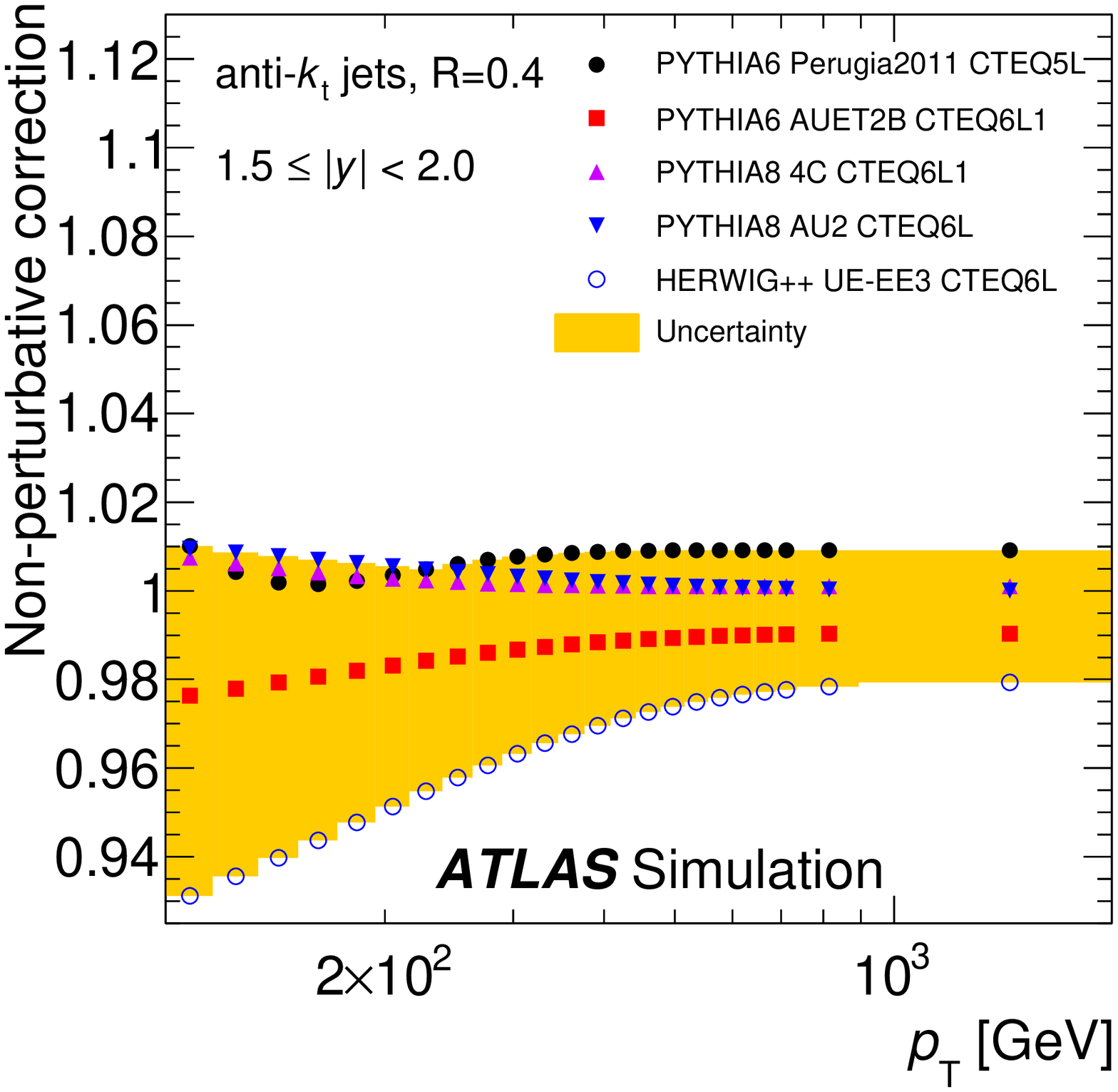}}
\subfigure[]{\includegraphics[width=6.9cm]{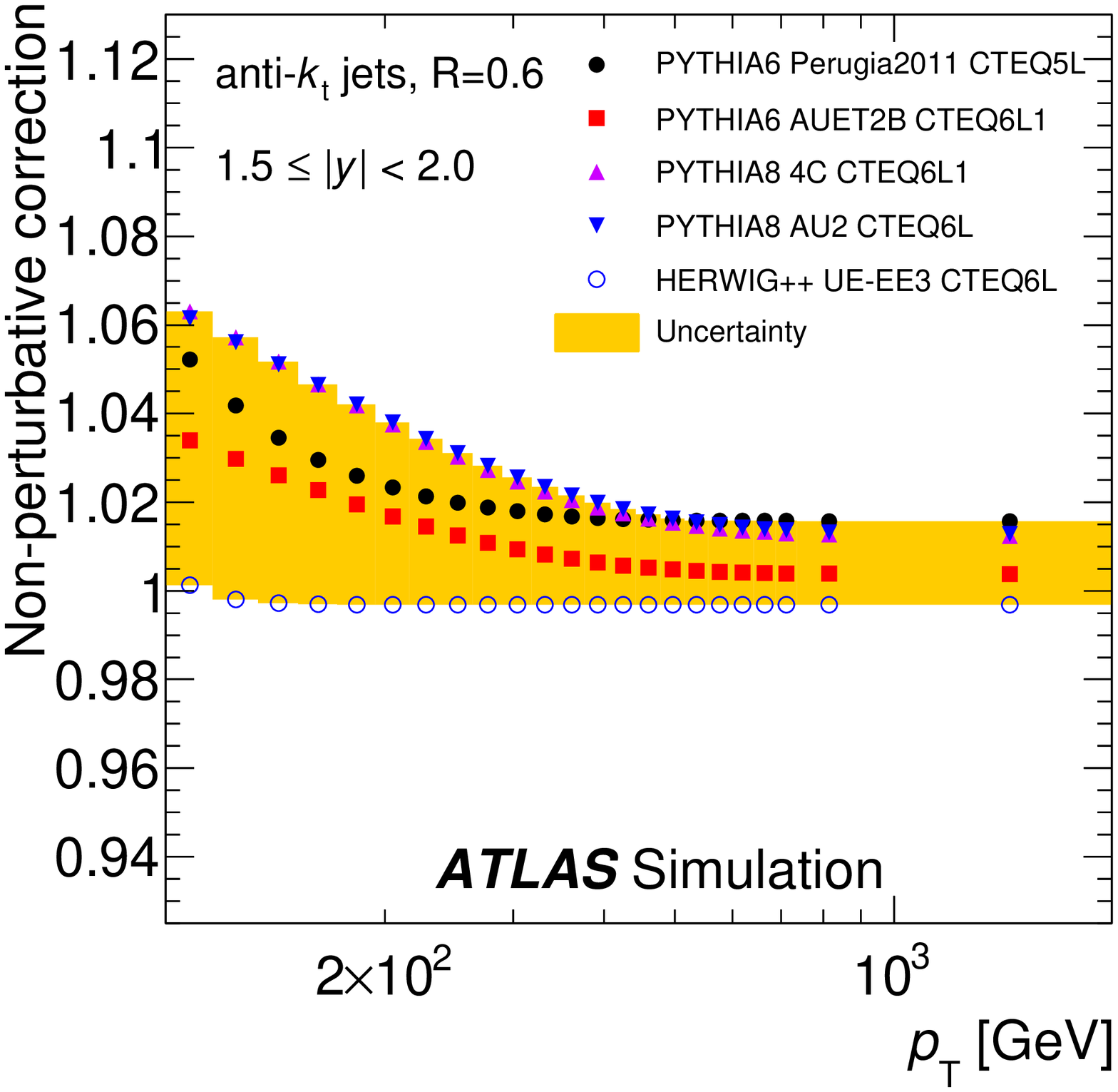}}
\subfigure[]{\includegraphics[width=6.9cm]{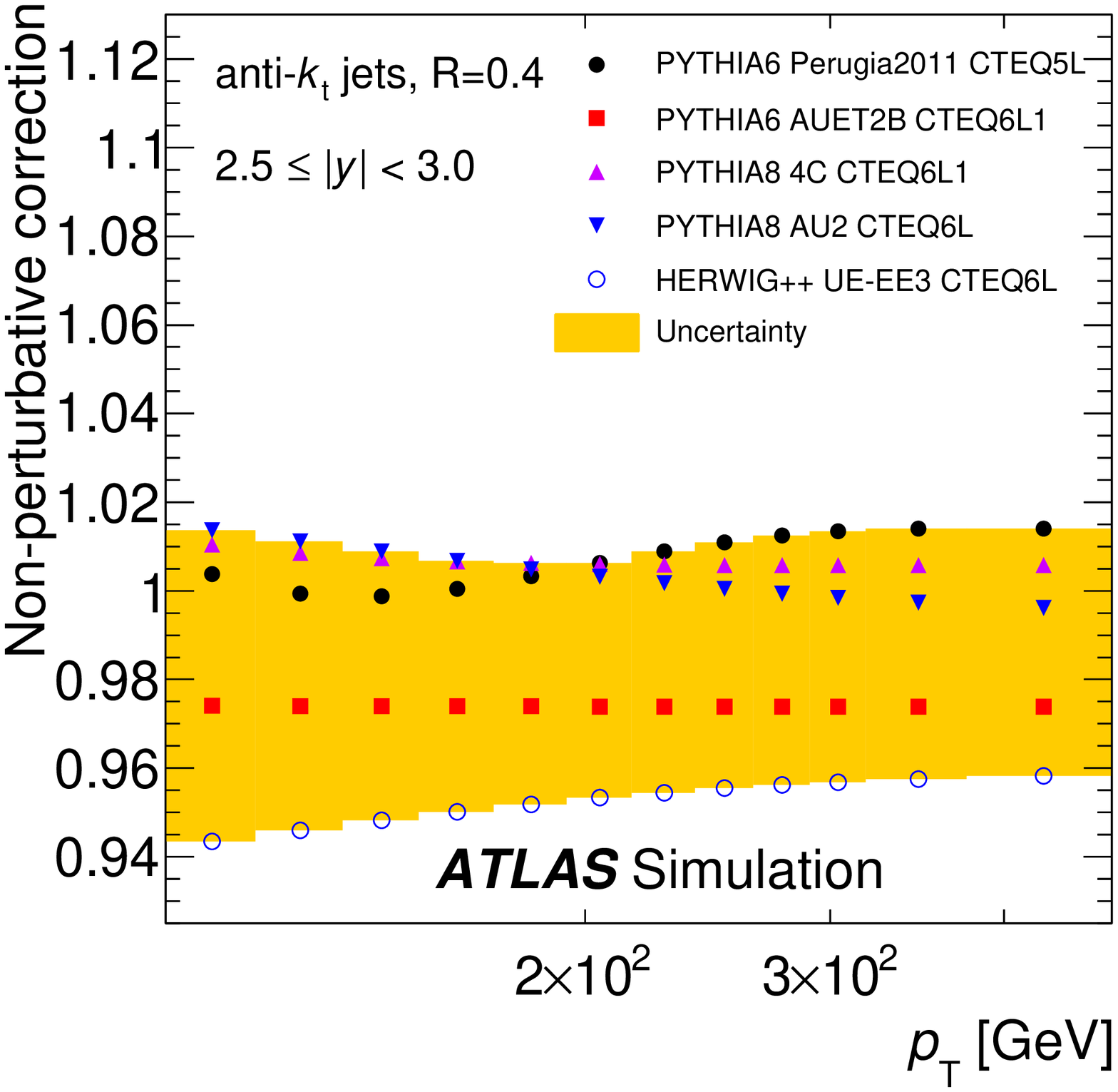}}
\subfigure[]{\includegraphics[width=6.9cm]{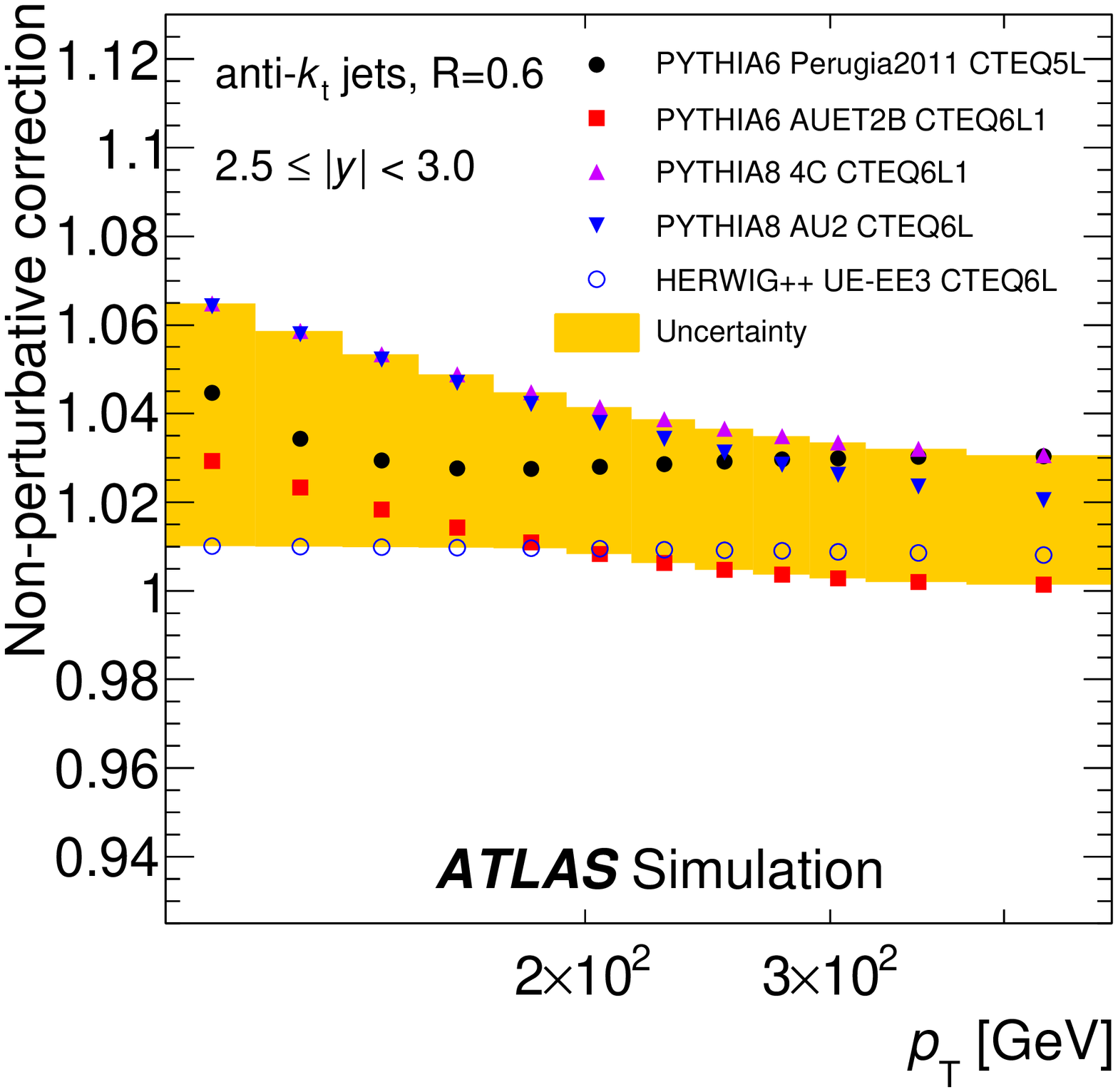}}
\caption{Non-perturbative correction factors applied to fixed order NLO calculations of the inclusive jet cross-section
for \antikt jets, with (a), (c), (e) $R=0.4$ and (b), (d), (f) $R=0.6$ 
in representative rapidity bins (as indicated in the legends),
as a function of the parton-level jet \pt{}, calculated from MC
simulations with various tunes. \label{fig:npcorr}}
\end{center}
\end{figure*}
\afterpage{\clearpage}

The correction factors are shown in figure~\ref{fig:npcorr} in representative rapidity bins for jets with $R=0.4$ and $R=0.6$, as a function of the jet \pt. 
The baseline correction factors for $R=0.4$ have a very weak dependence on
jet \pt{} and are typically 2\% or less from unity. On the other hand the
corrections for $R=0.6$ are up to 6\% at low \pt{}.
These differences between the two jet sizes result from the different interplay of hadronisation and underlying-event effects. 
In the high-rapidity region, the uncertainties are similar in size to those in
the low-rapidity region at low \pt, but do not decrease with the jet \pt{} as rapidly as in the low-rapidity region.

\subsection{Predictions from NLO matrix elements with LL parton showers}
Predictions from \powheg{} dijet production\footnote{
\powheg{} revision 2169~\cite{POWHEGspikes} is used with an option {\tt doublefsr 1} to activate the $q \rightarrow gq$ and $g \rightarrow \bar{q}q$ splitting processes.}
~\cite{Alioli:2010xa} are also compared to the measured cross-sections.
The predictions are made with the \powhegbox{} 1.0 package~\cite{Nason:2004rx,Frixione:2007vw,Alioli:2010xd}.
The \powheg{} generator utilises NLO matrix elements and can be interfaced to different MC programs to simulate parton showers, the underlying event and hadronisation. 

Events are generated for $2\rarrow2$ partonic scattering with the renormalisation and factorisation scales set to $\pt^\mathrm{Born}$, the transverse momentum of the scattered parton. In addition to the hard scatter, the hardest partonic radiation in the event is generated by the \powheg{} generator. 
The event configuration is then passed to the \pythia{} generator to be evolved to the particle level, where the radiative emissions in the parton shower are limited by a matching scale given by \powheg{}.
The predictions are made with the CT10 PDF set using two \pythia{} tunes, AUET2B and Perugia 2011.

The uncertainty in the partonic event generation with the \powheg{} generator is expected to be similar to that in the \nlojetpp{} calculations.
The matching of the \powheg{} generator to the \pythia{} generator can alter the parton shower, the initial-state radiation and the multiple interactions, but the procedure to evaluate the uncertainty on this matching is not well defined.
Therefore, the \powheg{} predictions are used without uncertainties. 

\subsection{Electroweak corrections}
\begin{figure*}
\begin{center}
\subfigure[]{\includegraphics[width=6.8cm]{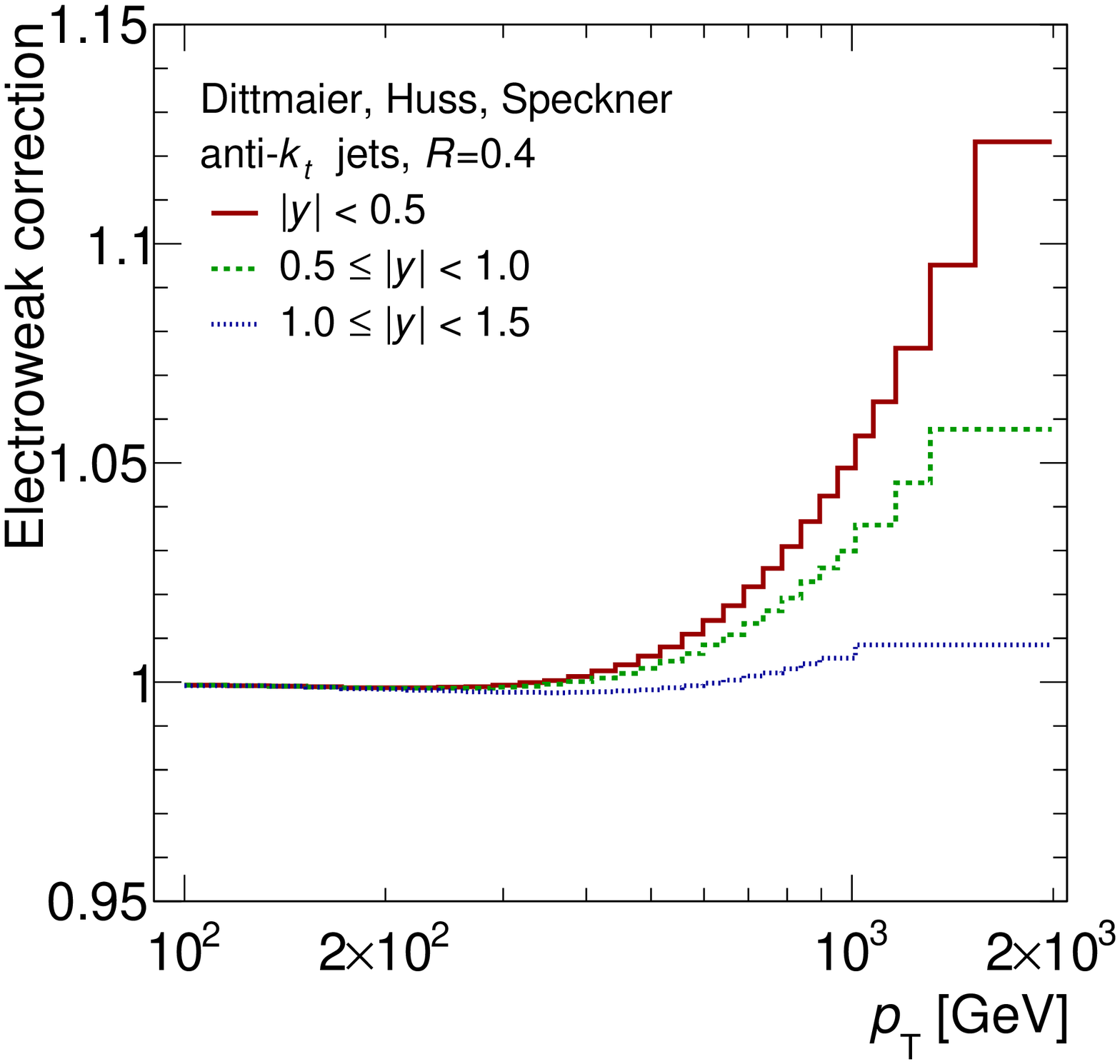}}
\subfigure[]{\includegraphics[width=6.8cm]{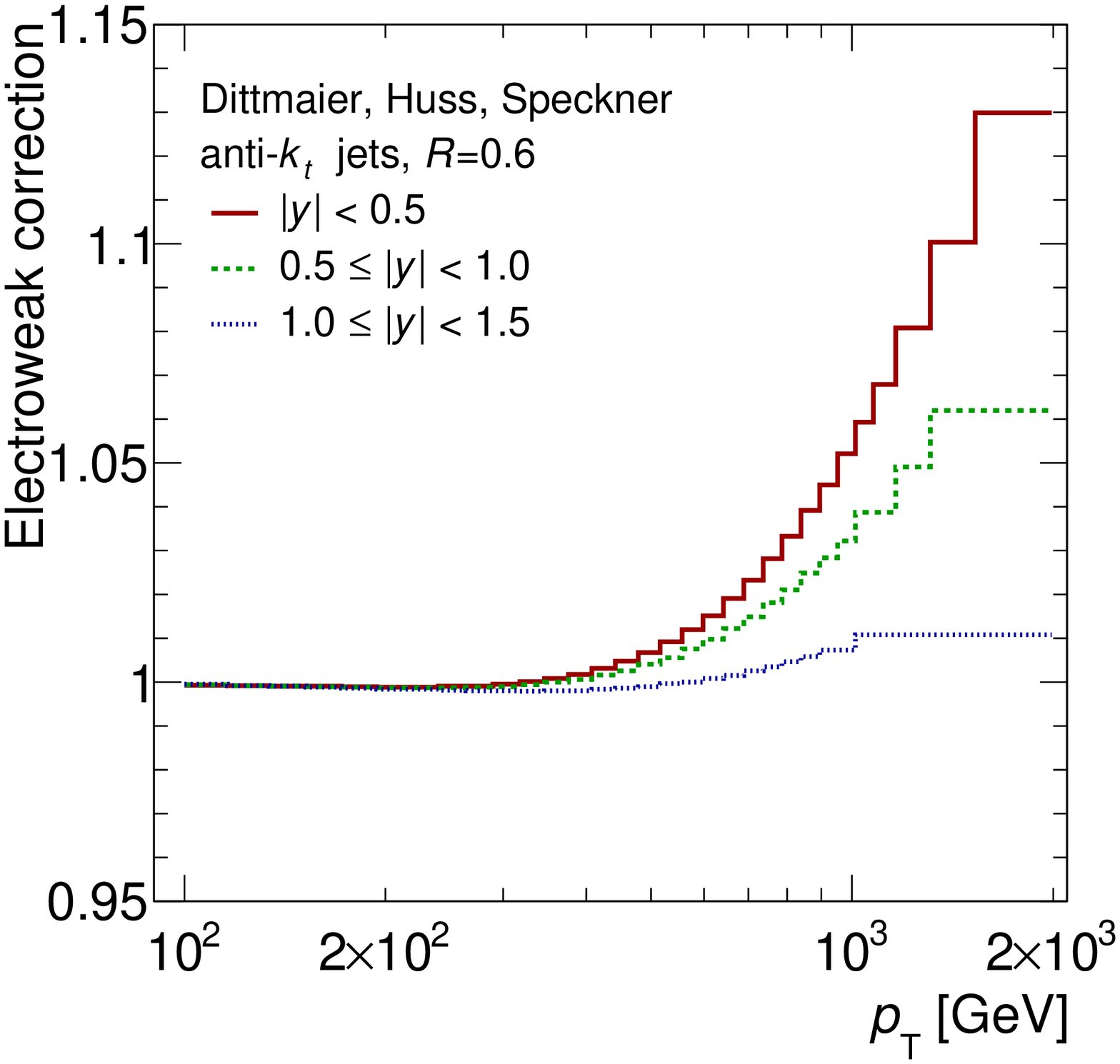}}
\caption{Electroweak correction factors for the inclusive jet cross-section for \antikt jets with (a) $R=0.4$ and (b) $R=0.6$ in the low rapidity bins with $|y|<1.5$ as a function of the jet \pt{}. 
\label{fig:ewcorr}}
\end{center}
\end{figure*}

The electroweak corrections are provided by the authors of ref.~\cite{Dittmaier:2012kx}.
The corrections comprise tree-level effects of $O(\alpha\alpha_\mathrm{S},\alpha^2)$ as well as weak loop effects of $O(\alpha\alpha_\mathrm{S}^2)$ on the cross-section, where $\alpha$ is the electroweak coupling constant.
Effects of photon or \Wboson/\Zboson{} radiation are not included in the corrections, though real \Wboson/\Zboson{} radiation may affect the cross-section by a few percent at $\pt\sim1$~TeV as the calculation at $\rts=14$ TeV in ref.~\cite{Baur:2006sn} shows.
The correction factors are derived by considering NLO electroweak effects on an LO QCD prediction in the phase space considered here.\footnote{Calculations specific to the present measurements are provided by the authors. The numerical values of the parameters are given in ref.~\cite{Dittmaier:2012kx}. The renormalisation and factorisation scales are set to the leading jet \pt.}
Figure~\ref{fig:ewcorr} shows the electroweak corrections for jets with $R=0.6$,
in the lowest rapidity bins.
The correction reaches more than 10\% for $\pt>1$~\TeV{} in the lowest rapidity
bin, but decreases rapidly as the rapidity increases.
It is less than 1\% for jets with $|y|>1$.

\afterpage{\clearpage}

The corrections are multiplicatively applied to the NLO QCD predictions from \nlojetpp{} and \powheg{}. 
Alternatively, the corrections can be applied only to the LO QCD term in the predictions from \nlojetpp{}~\cite{Butterworth:2014efa}.
This alternative procedure results in predictions that are lower by 3\% (4\%)
for jets with $R=0.4$ ($R=0.6$) at most.

%% file: dataselec.tex
\section{Event selection}
\label{sec:selec}
\subsection{Data set}
The measurement is made using proton--proton collision data at $\rts=7$~TeV collected by the ATLAS detector during the data-taking period of the LHC in 2011. 
The total integrated luminosity corresponds to 4.5~\ifb~\cite{Aad:2013ucp}.
Due to the increasing instantaneous luminosity at the LHC, the average number of proton--proton interactions per bunch crossing, $\langle\mu\rangle$, increased from $\langle\mu\rangle\sim5$ at the beginning of the data-taking period to $\langle\mu\rangle\sim18$ at the end.
The number of colliding bunches increased in 2011 with respect to the previous ATLAS measurements~\cite{Aad:2011fc} with a minimum bunch spacing of 50 ns.

The overlay of multiple proton--proton interactions in the same and neighbouring bunch crossings are called in-time and out-of-time pile-up, respectively. 
They affect the energy  measurement due to additional energy deposits in the calorimeter and residual electronic signals in the readout system. 
This is corrected in the jet calibration. 
As a consequence of the pile-up, an event may have additional low-\pt{} jets which do not originate from the hardest interaction.  
Their contribution is negligible in the kinematic region of this measurement. 
Several checks are done as described in section~\ref{sec:checks}.

\subsection{Trigger and offline event selection}
The ATLAS trigger system is composed of three consecutive levels: level 1, level 2 and the event filter, with progressively larger processing time available per event, finer granularity and access to more detector systems.
Online event selection was done 
using a set of single-jet triggers.
Each single-jet trigger selects events that contain a jet with transverse momentum above a certain threshold at the electromagnetic scale\footnote{
The electromagnetic scale is the basic signal scale 
to which the ATLAS calorimeters are calibrated. It does not take into account the lower response to hadrons.
} 
in the region $|\eta|<3.2$. 
Online jet reconstruction uses the \antikt algorithm with a jet radius parameter of $R=0.4$ at the event filter.
Depending on its output rate, a single-jet trigger may be suppressed by recording only a predefined fraction of events. 
Since the jet-production rate falls steeply with the jet \pt,
triggers with different \pt{} thresholds are considered in this measurement.
The triggers with low \pt{} thresholds are highly suppressed.
For a given offline jet \pt{} value, the least suppressed trigger whose efficiency is greater than 99\% is used.

All events used in this measurement were collected during stable beams conditions. 
They are required to pass data-quality requirements from the relevant detector systems for jet reconstruction.
In addition, events are required to have at least one well-reconstructed vertex, which must have at least two associated tracks with $\pt{}> 400$ MeV and be consistent with the proton-proton collision region.
 The number of vertices which fulfil these criteria is used in studies of pile-up effects and is denoted by $N_\mathrm{PV}$. 

\subsection{Jet reconstruction and calibration}
Jets are reconstructed with the \antikt algorithm using topological cell clusters~\cite{topoclusters} in the calorimeter as input objects.
These clusters are constructed from calorimeter cells at the electromagnetic scale and are then calibrated using local hadronic calibration weights (LCW)~\cite{Aad:2011he}.
The LCW correct for the non-compensating response of the ATLAS calorimeters, for
energy losses in inactive regions of the detector,
and for signal losses due to the clustering itself.

Reconstructed jets require the following corrections~\cite{Aad:2014bia}.
Additional energy due to pile-up is subtracted by applying a correction derived from MC simulation as a function of $N_\mathrm{PV}$, $\langle\mu\rangle$ in bins of the jet $\eta$ and \pt{}. 
The jet direction is corrected under the assumption that the jet originates from the hardest event vertex, which is the vertex with the highest $\sum\pt^2$ of associated tracks. 
The jet energy and direction are further corrected to account for instrumental effects which cannot be corrected at the level of the topological cell clusters. 
These corrections are derived from MC simulations.
Finally, jets reconstructed in data are corrected based on in-situ \pt-balance measurements, to account for residual differences between MC simulation and data.

Jets are required to pass jet-quality selections to reject fake jets reconstructed from non-collision signals, such as beam-related background, cosmic rays or detector noise. 
The ``Medium'' selection described in ref.~\cite{ATLAS-CONF-2012-020} is applied, which gives an efficiency larger than 99\% for jets with $\pt\geq100$~\GeV. 

Part of the data-taking period was affected by a read-out problem in a region of the LAr calorimeter, causing jets in this region to be poorly reconstructed. 
In order to avoid any bias in the measurement, jets reconstructed in the region $-0.1<\eta<1.5$ and $-0.88<\phi<-0.50$ were rejected in both data and MC simulation, regardless of the data-taking period. 
The unfolding procedure described in section~\ref{sec:unfold} corrects for the corresponding inefficiency.

All selected jets with $\pt\geq100$~\GeV, $|y|<3$, and a positive decision from the trigger in the corresponding kinematic region are considered in this analysis. 

\subsection{Validity and consistency checks of the analysis}\label{sec:checks}
The following checks are performed on the selected jet distributions to confirm the validity and consistency of the analysis. 
The distributions of the jet $\eta$ and $\phi$ are well described by the MC simulation.
The simulation reproduces the effects of the energy corrections for 
time-dependent calorimeter defects.
To ascertain the robustness against pile-up, jet \pt{} distributions in the data are extracted separately in bins of $N_\mathrm{PV}$ and $\langle\mu\rangle$.
No statistically significant deviation compared with the effect from the pile-up component of the uncertainty in the jet energy scale (JES) is observed.
The stability of the jet yield over time shows no significant variations, indicating stability against the increase of pile-up at the LHC and against time-dependent calorimeter defects.
Track information is used to verify that the selected jets come from the hardest event vertex. 
The events containing the highest-\pt{} jets in each rapidity region are visually scanned, assuring no contamination of the events by fake jets.

%% file: unfold.tex
\section{Unfolding of detector effects}
\label{sec:unfold}
Cross-sections are measured in six rapidity bins as a function of the jet \pt{}. 
The definition of the \pt{} bins is chosen to ensure that the statistical uncertainty in each bin is less than 40\% of the systematic uncertainty discussed in section~\ref{sec:syst}.
Furthermore, according to the MC simulation, at least half of the jets reconstructed in each bin of the measurement must be generated in the same bin at the particle level.
In addition, correlations due to bin-to-bin migration between adjacent bins are required to be less than 80\%. 

The data distribution is unfolded to correct for detector inefficiencies and resolution effects to obtain the particle-level cross-section. 
The Iterative, Dynamically Stabilised (IDS) unfolding method~\cite{Malaescu:2009dm}, a modified Bayesian technique, is used.
This method takes into account the migrations of events across the bins and uses a data-driven regularisation. 
It is performed separately for each rapidity bin, since the migrations across rapidity bins are small while those across jet \pt{} bins are significant. 
The migrations across rapidity bins are taken into account in bin-by-bin corrections.

A transfer matrix which relates the \pt{} of the jet at the particle level and that after the reconstruction is used in the unfolding process.  
It is derived by matching a particle-level jet with a reconstructed jet 
in MC simulations, when both are 
closer to one another than to any other jet and lie
within a radius of $\Delta R_{jj}=0.3$, where $\Delta R_{jj}$ is the distance between two jets in the $(\eta, \phi)$-plane.
If jets migrate to other rapidity bins, they are unmatched. 
The \pt{} spectra of unmatched reconstructed jets are used to determine the sample purity, $r_\mathcal{P}$, which is defined as 
the fraction of reconstructed jets that are matched. 
The analysis efficiency, $r_\mathcal{E}$, is derived from the \pt{} spectra of matched particle-level jets and is defined 
as the fraction of particle-level jets that are matched. 
The migrations across \pt{} bins are irrelevant to the definition of $r_\mathcal{P}$ and $r_\mathcal{E}$.

The data are unfolded to the particle level in a three-step procedure. 
First, they are corrected for the sample impurities, followed by unfolding for the \pt{} migration.
Finally, the data are corrected for the analysis inefficiencies. 
The final result is given by
\begin{equation}
N^\mathrm{part}_i=\sum_jN^\mathrm{reco}_j\cdot r_{\mathcal{P},j}\cdot A_{ij}\ /\ r_{\mathcal{E},i},
\end{equation}
where $i$ and $j$ are the particle-level and reconstructed bin indices, respectively, $N^\mathrm{part}_k$ and $N^\mathrm{reco}_k$ are the number of particle-level jets and the number of reconstructed jets in bin $k$, and 
$A_{ij}$~is an unfolding matrix extracted from the transfer matrix. 
This unfolding matrix gives the probability for a reconstructed jet in \pt{} bin $j$ to originate from particle-level \pt{} bin $i$.
The number of iterations in the IDS unfolding method is chosen such that the bias in the closure test described below is small, at most at the percent level in bins with a statistical uncertainty of less than 20\%.
In this measurement, this is achieved after one iteration. 

The precision of the unfolding technique is studied using a data-driven closure test. 
In this study, the particle-level \pt{} spectrum in the MC simulation is reweighted in the transfer matrix,
such that significantly improved agreement between the resulting reconstructed spectrum and the data is obtained. 
The reconstructed spectrum in this reweighted MC simulation is then unfolded using the same procedure as for the data.
Comparison of the spectrum obtained from the unfolding procedure with the original reweighted particle-level spectrum provides an estimate of the unfolding bias, which is interpreted as the associated systematic uncertainty. 

As an estimate of further systematic uncertainties, the unfolding procedure is repeated using different transfer matrices created with tighter and looser matching criteria of $\Delta R_{jj}=0.2$ and $\Delta R_{jj}=0.4$. 
The deviations of the results from the nominal unfolding result are considered as an additional uncertainty. 
They are found to be smaller than 0.05\%.

The statistical uncertainties are propagated through the unfolding procedure by performing pseudo-experiments.
An ensemble of pseudo-experiments is created in which a weight is applied to each event in both the data and the MC sample, using a Poisson distribution with expectation value equal to one.
This procedure takes into account the correlation between jets produced in the same event. 
For a combination of this measurement with other results using the same data set, the pseudo-random Poisson distribution is seeded uniquely for each event based on the event number and the run number in the ATLAS experiment. 
The fluctuation of the MC sample is also done in a similar way, where both the transfer matrix and the efficiency corrections are modified. 
The unfolding is performed in each pseudo-experiment and a set of results from the ensemble is used to calculate a covariance matrix. 
The total statistical uncertainty is obtained from the covariance matrix, where bin-to-bin correlations are also encoded. 
The separate contributions from the data and from the MC statistics can be obtained from the same procedure by fluctuating only either the data or the MC samples.  

The unfolding procedure is repeated for the propagation of the uncertainties in the jet energy and angle measurements, as described in the next section.

%% file: syst.tex
\section{Experimental systematic uncertainties}
\label{sec:syst}
The sources of systematic uncertainty considered in this measurement are those associated with 
the jet reconstruction and calibration, the unfolding procedure, and the luminosity measurement. 
Uncertainties related to the trigger efficiency are found to be negligible and are not considered. 

The uncertainty in the JES is the dominant source of uncertainty in the inclusive jet cross-section measurement. 
The full description of the JES uncertainty can be found in ref.~\cite{Aad:2014bia} and a brief description is given in appendix~\ref{sec:AppXsec}.
The total size of the JES uncertainty is below $2\%$ in the central region and increases to $4\%$ in the forward region for jets with $\pt\sim O(100)$~\GeV. 
The correlations among the components of the JES uncertainty are described by 63 nuisance parameters which are treated as independent. 
Each corresponding uncertainty component in the JES is assumed to have a Gaussian uncertainty which is fully correlated across the jet \pt{} and rapidity ranges.

An uncertainty component is added specifically for this measurement, to take into account that the MC sample used in the unfolding is generated with a tune different from that used in the derivation of the jet calibration. 
This component is derived from a comparison of jet-\pt{} responses, which are ratios of the reconstructed jet \pt{} to the particle-level jet \pt{}, between the two MC samples. 
Its size is $O(0.1)$\% for central jets, with a maximum value of 3\% for the jets with the highest pseudorapidity in this measurement.
 
The JES uncertainty is propagated to the measured cross-section. 
For each component of the JES uncertainty, the jet energies are scaled up and down by one standard deviation in the MC simulation. 
The resulting \pt{} spectra are unfolded using the nominal unfolding matrix.
The original MC \pt{} spectra are also unfolded and the difference is taken as the uncertainty on the cross-section measurement from the given component.

Since the knowledge of the correlations between the experimental components of the JES uncertainty is limited, two different configurations of nuisance parameters are considered. 
They are constructed with different assumptions on the correlations of the components and have ``stronger'' and ``weaker'' correlations with respect to the nominal configuration of the uncertainties.  
The uncertainties in the cross-section using these two configurations are available in HEPDATA, providing access to the influence of the assumed correlation. 
The total uncertainty in the measured cross-section due to the JES does not change with these different configurations. 

Usually uncertainties in experimental measurements are treated as having Gaussian distributions.
A test is performed to see the shape of the probability density functions of the cross-section due to the JES uncertainty.
The test is performed for large components of the JES uncertainty, with an assumption that each component has a Gaussian shape before the propagation to the cross-section.   
In the test, quantiles of the probability density functions of the cross-section after the propagation of the corresponding component are determined experimentally. 
They are evaluated by shifting the jet energies by $\pm1\sigma$, $\pm2\sigma$, $\pm3\sigma$, $\pm4\sigma$, and $\pm5\sigma$ for a given uncertainty
in the MC simulation. 
The shifts are propagated to the cross-section by the procedure described above. 
The determined experimental cross-section quantiles are compared with the expected quantile positions obtained from Gaussian and log-normal shape assumptions. 
The expectations are derived from the nominal cross-section and the experimental $1\sigma$ quantile.
For the components giving $O(10)$\% uncertainties, the experimental quantiles deviate from the quantiles expected with the Gaussian assumption and better descriptions are given by the log-normal assumption.

The jet energy resolution (JER) is determined using the in-situ techniques described in ref.~\cite{Aad:2012ag} and the JER difference between data and the MC simulations is considered as its uncertainty. 
The effect of this uncertainty in the cross-section measurements is evaluated by smearing the energy of reconstructed jets in the MC simulation such that the resolution is worsened by the size of its uncertainty. 
A new transfer matrix is constructed using this smeared sample and used to unfold the data spectra.
The resulting deviations from the measured cross-sections unfolded using the nominal transfer matrix are taken as the uncertainty in the measurement, applied symmetrically as upward and downward uncertainties.

The jet angular resolution is estimated from comparisons of the polar angles of a reconstructed jet and the matched particle-level jet using the MC simulation.  
No bias is found in the angular reconstruction and the resolution is 0.035 radians at most in the sample with high pile-up ($10\leq N_\mathrm{PV}\leq12$) for jets with energy $E\ge100$~\GeV. 
An uncertainty is assigned to the resolution to account for possible differences between data and the MC simulation.
It is propagated to the cross-section in the same way as for the JER.

The jet reconstruction efficiency is evaluated using jets reconstructed from tracks following the technique described in ref.~\cite{Aad:2011he}. 
Inefficiency is only seen for jets with very low \pt, well below the kinematic region for this measurement. 
No uncertainty is considered for the jet reconstruction efficiency. 

Estimating the efficiency of the jet-quality selections shows agreement between data and the MC simulations for the ``Medium'' criteria at the level of 0.25\%~\cite{ATLAS-CONF-2012-020}. 
A corresponding systematic uncertainty is assigned to the measurement. 

The uncertainties associated with the unfolding procedure are described in section~\ref{sec:unfold}. 
The closure test quantifies the impact of a possible mis-modelling in the MC simulation. 
The variations of the matching criterion in the construction of the transfer matrix are checked. 

The uncertainty in the luminosity measurement is 1.8\%~\cite{Aad:2013ucp}. Due to changes in the hardware of the detector and the algorithm used in the luminosity measurement, the uncertainty is not correlated with that for the 2010 data set. 

The systematic uncertainties propagated through the unfolding are evaluated using a set of pseudo-experiments for each component, as in the evaluation of the statistical uncertainties.
Remaining statistical fluctuations of the systematic uncertainties are minimised using a smoothing procedure. 
For each component, the \pt{} bins are combined until the propagated uncertainty value in the bin has a Poisson statistical significance larger than two standard deviations. 
Then a Gaussian kernel smoothing is performed to regain the original fine bins.

\begin{figure*}
  \centering
 \subfigure[]{\includegraphics[width=6.9cm]{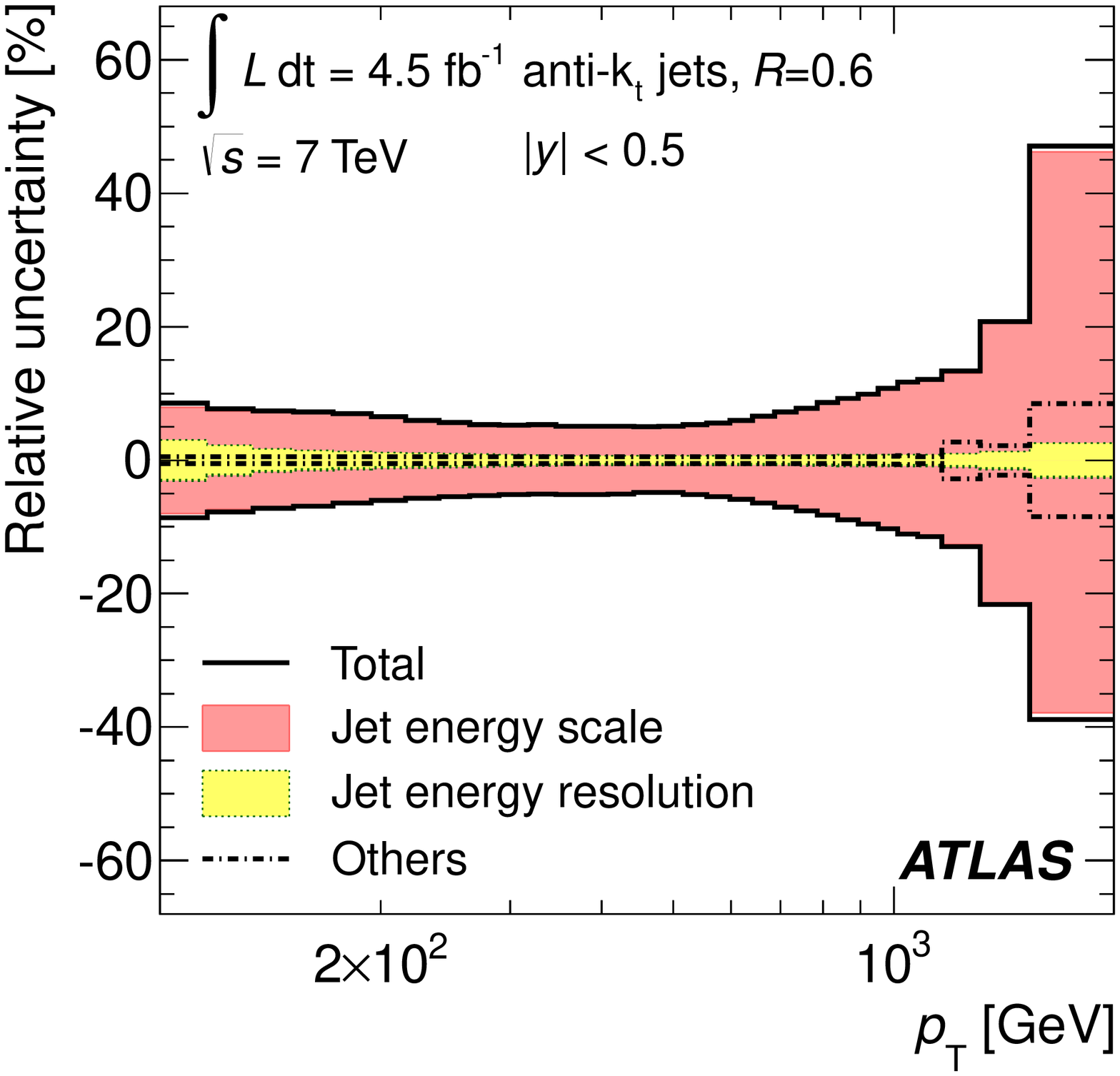}}
 \subfigure[]{\includegraphics[width=6.9cm]{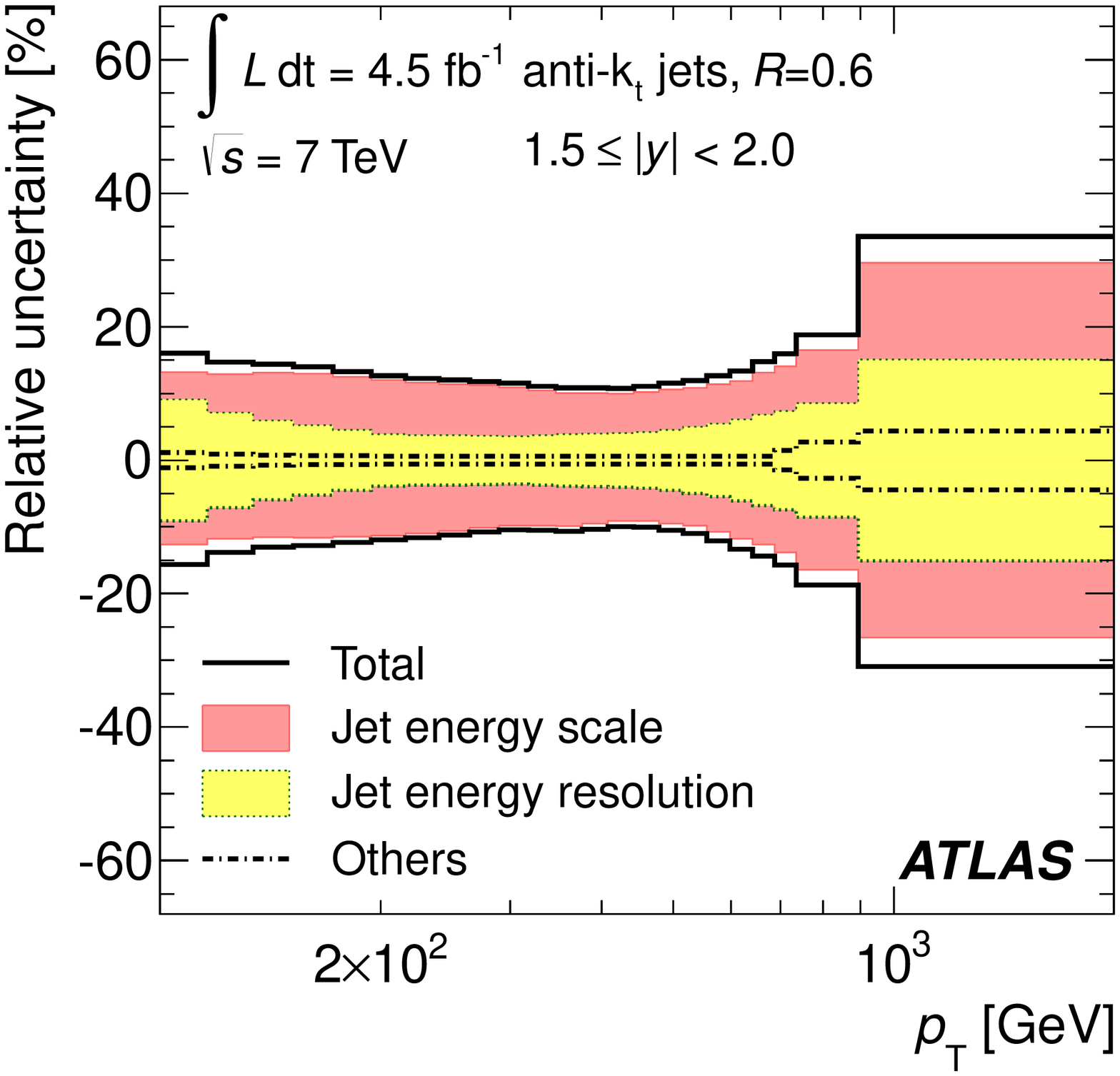}}
 \subfigure[]{\includegraphics[width=6.9cm]{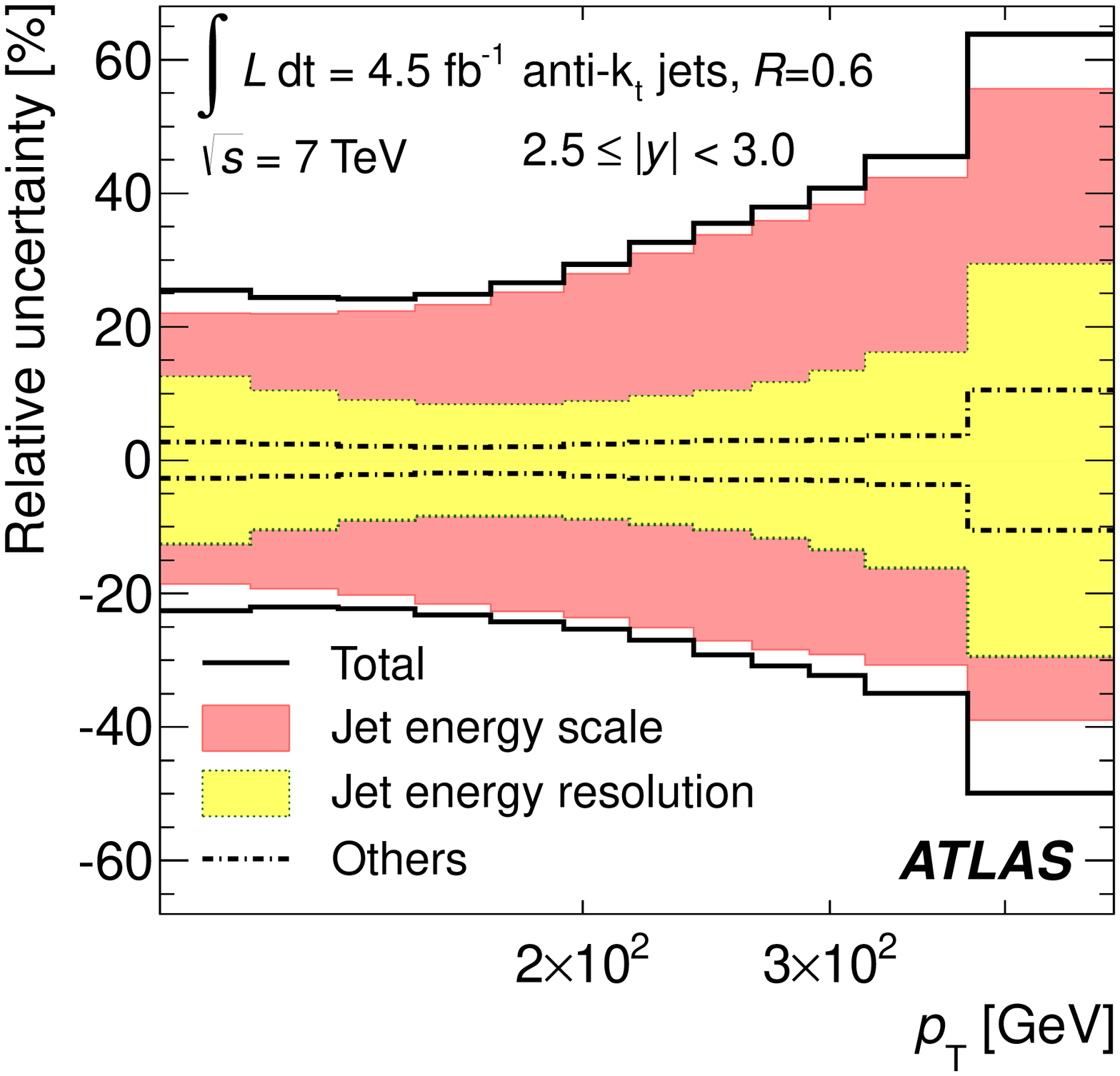}}
  \caption{Experimental systematic uncertainties in the inclusive jet cross-section measurement for \antikt jets with $R=0.6$ in three representative rapidity bins, as a function of the jet \pt. In addition to the total uncertainty, the uncertainties from the jet energy scale (JES), the jet energy resolution (JER) and other systematic sources are shown separately. The 1.8\% uncertainty from the luminosity measurement is not shown. 
  \label{fig:systr06}}
\end{figure*}

Uncertainties from individual sources are treated as uncorrelated with each other and added in quadrature.  
The evaluated systematic uncertainties on the cross-section measurement are shown in figure~\ref{fig:systr06} for representative rapidity bins for jets with $R=0.6$.
The uncertainties for the measurement using jets with $R=0.4$ yield similar total uncertainties, with smaller contributions from the JER and larger contributions from the JES. 
The systematic uncertainty in this measurement is dominated by the uncertainties in the JES. 
The large uncertainty in the highest \pt{} bin is caused by the JES uncertainty associated with the single-hadron energy response measurement~\cite{Aad:2014bia}. 
The increased uncertainty in the high-rapidity region is mainly due to the modelling of the additional parton radiation, which gives the largest uncertainty in the calibration technique using the \pt{} balance between a central jet and a forward jet. 

\begin{figure}
\begin{center}
\includegraphics[width=14.5cm]{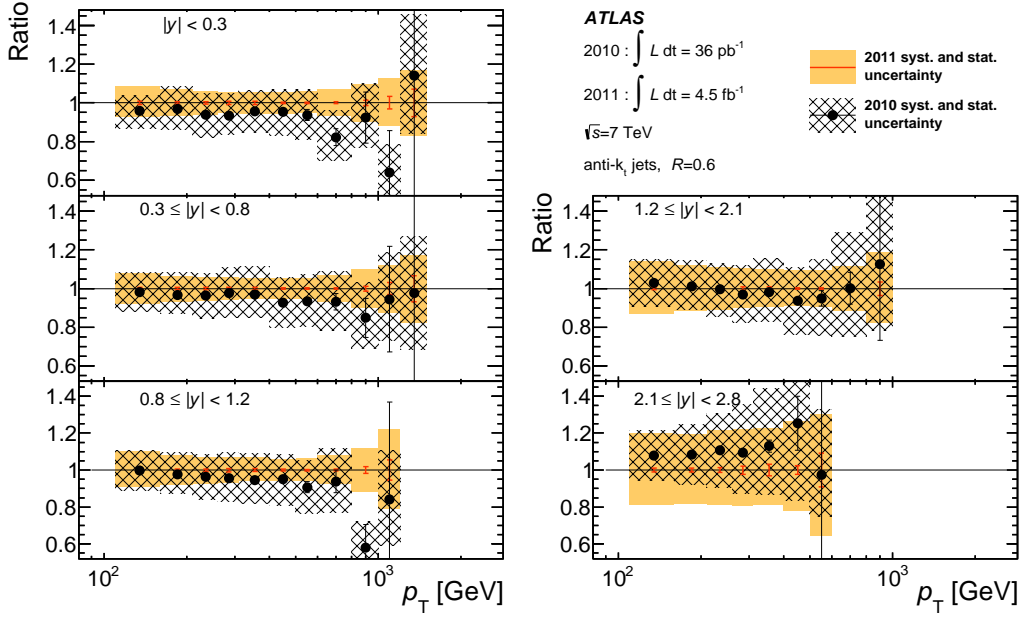}
\caption{\label{fig:ratio20102011r06}
Ratios of inclusive jet cross-sections 
using 2010 data~\cite{Aad:2011fc} to the measurement using 2011 data, both in the same binning, as a function of the jet \pt{} in bins of rapidity, for \antikt jets with $R=0.6$. 
The comparison is done in the common phase space only. 
The statistical uncertainties on the measurement are indicated by the error bars and the systematic uncertainties on each measurement are shown by the bands. 
The uncertainties from the luminosity measurements are not included. 
}
\end{center}
\end{figure}

In order to compare the results of this measurement with those previously published using data collected by ATLAS in 2010~\cite{Aad:2011fc}, the measurement is repeated with the same binning as used in that measurement. 
Figure~\ref{fig:ratio20102011r06} shows the cross-section ratio of the published measurement\footnote{
The cross-sections are multiplied by a factor of 1.0187 to take into account the updated value of the integrated luminosity for the ATLAS 2010 data-taking period. See ref.~\cite{Aad:2013ucp} for more details.} 
using the 2010 data set to that repeated using the 2011 data set.  
The central values of the ratio are in most bins contained within the size of the systematic uncertainties of either measurement. 
As expected, the statistical uncertainties are smaller for the 2011 data. 
The systematic uncertainties are also smaller in most of the common phase space, especially in the low-rapidity region.

\clearpage

%% file: result.tex
\section{Results}\label{sec:result}
\begin{figure}
\begin{center}
\includegraphics[width=12.3cm]{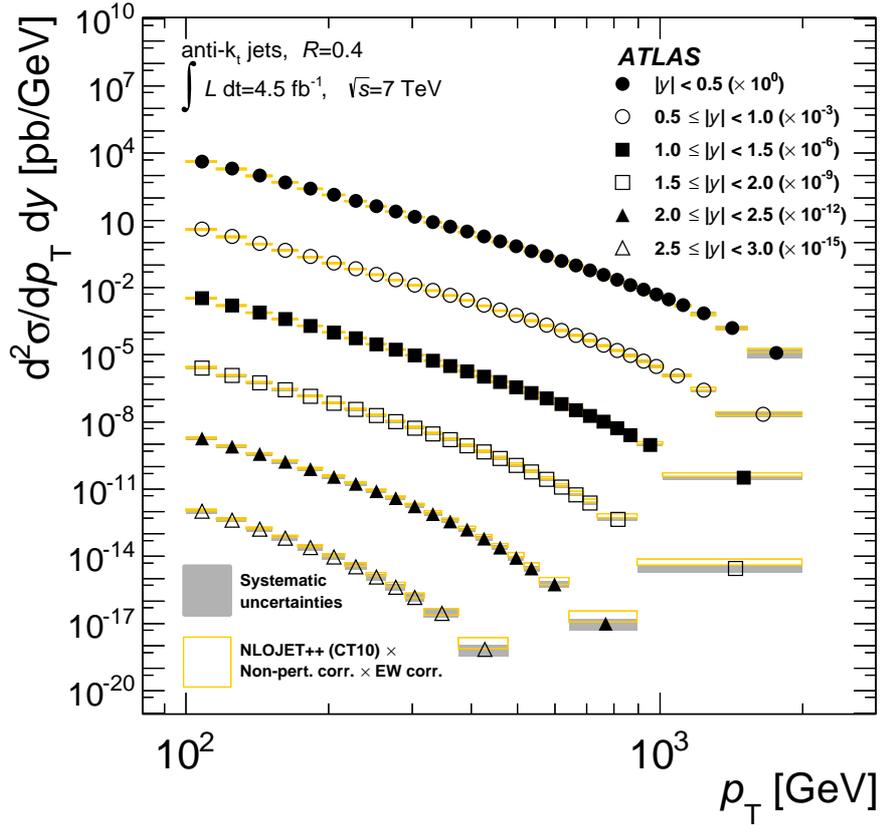}
\caption{\label{fig:xsecr04}
Double-differential inclusive jet cross-sections as a function of the jet \pt{} in bins of rapidity, for \antikt jets with $R=0.4$. For presentation, the cross-sections are multiplied by the factors indicated in the legend. 
The statistical uncertainties are smaller than the size of the symbols used to plot the cross-section values.
The shaded areas indicate the experimental systematic uncertainties. 
The data are compared to NLO pQCD predictions calculated using \nlojetpp{} with the CT10 NLO PDF set, to which non-perturbative corrections and electroweak corrections are applied. The open boxes indicate the predictions with their uncertainties. 
The 1.8\% uncertainty from the luminosity measurement is not shown.
}
\end{center}
\end{figure}

\begin{figure}
\begin{center}
\includegraphics[width=12.3cm]{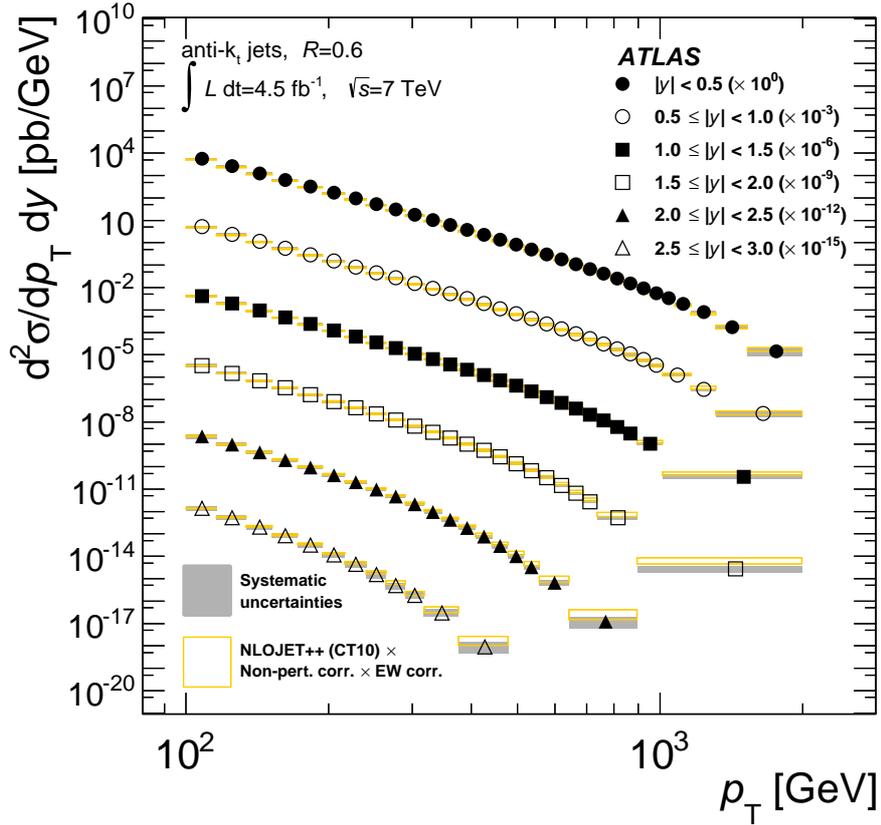}
\caption{Double-differential inclusive jet cross-sections as a function of the jet \pt{} in bins of rapidity, for \antikt jets with $R=0.6$. For presentation, the cross-sections are multiplied by the factors indicated in the legend. 
The statistical uncertainties are smaller than the size of the symbols used to plot the cross-section values.
The shaded areas indicate the experimental systematic uncertainties. 
The data are compared to NLO pQCD predictions calculated using \nlojetpp{} with the CT10 NLO PDF set, to which non-perturbative corrections and electroweak corrections are applied. The open boxes indicate the predictions with their uncertainties. 
The 1.8\% uncertainty from the luminosity measurement is not shown.\label{fig:xsecr06}
}
\end{center}
\end{figure}

The double-differential inclusive jet cross-sections are shown in figures~\ref{fig:xsecr04} and \ref{fig:xsecr06} for jets reconstructed using the \antikt algorithm with $R=0.4$ and $R=0.6$, respectively. 
The measurement extends over jet transverse momenta from 100~GeV to 2~TeV in the rapidity region $|y|<3$. 
The NLO pQCD predictions calculated with \nlojetpp{} using the CT10 PDF set with corrections for non-perturbative effects and electroweak effects applied are compared to the measurement. 
The figures show that the NLO pQCD predictions reproduce the measured cross-sections, which range over eight orders of magnitude in the six rapidity bins.

\begin{figure}
\begin{center}
\includegraphics[width=14.5cm]{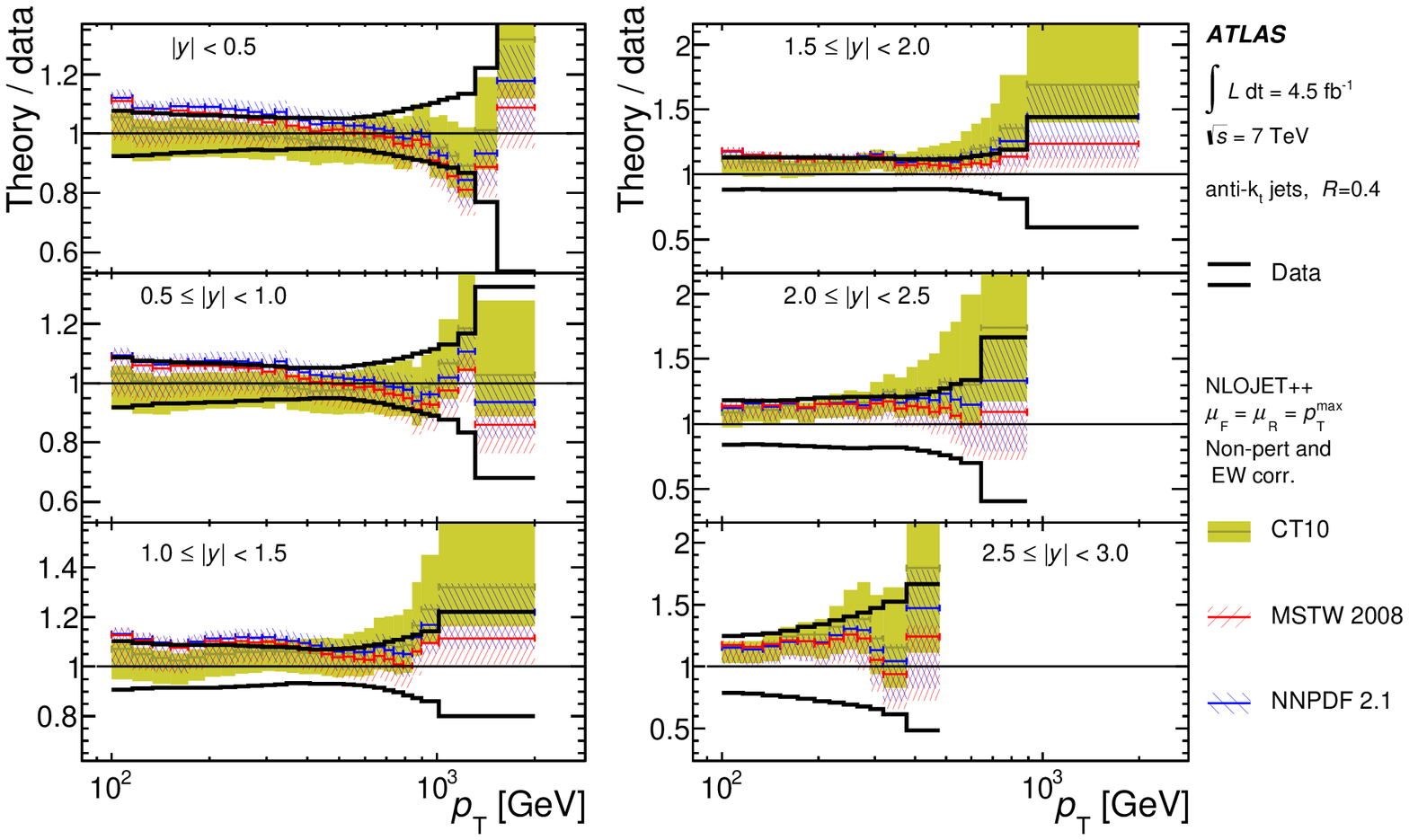}
\caption{Ratio of NLO pQCD predictions to the measured double-differential inclusive jet cross-section, shown as a function of the jet \pt{} in bins of the jet rapidity, for \antikt jets with $R=0.4$. The predictions are calculated using \nlojetpp{} with different NLO PDF sets, namely CT10, MSTW2008 and NNPDF 2.1. Non-perturbative corrections and electroweak corrections are applied to the predictions. Their uncertainties are shown by the bands, including all the uncertainties discussed in section~\ref{sec:theo}. 
The data lines show the total uncertainty except the 1.8\% uncertainty from the luminosity measurement. 
\label{fig:ratior04}
}
\end{center}
\end{figure}

\begin{figure}
\begin{center}
\includegraphics[width=14.5cm]{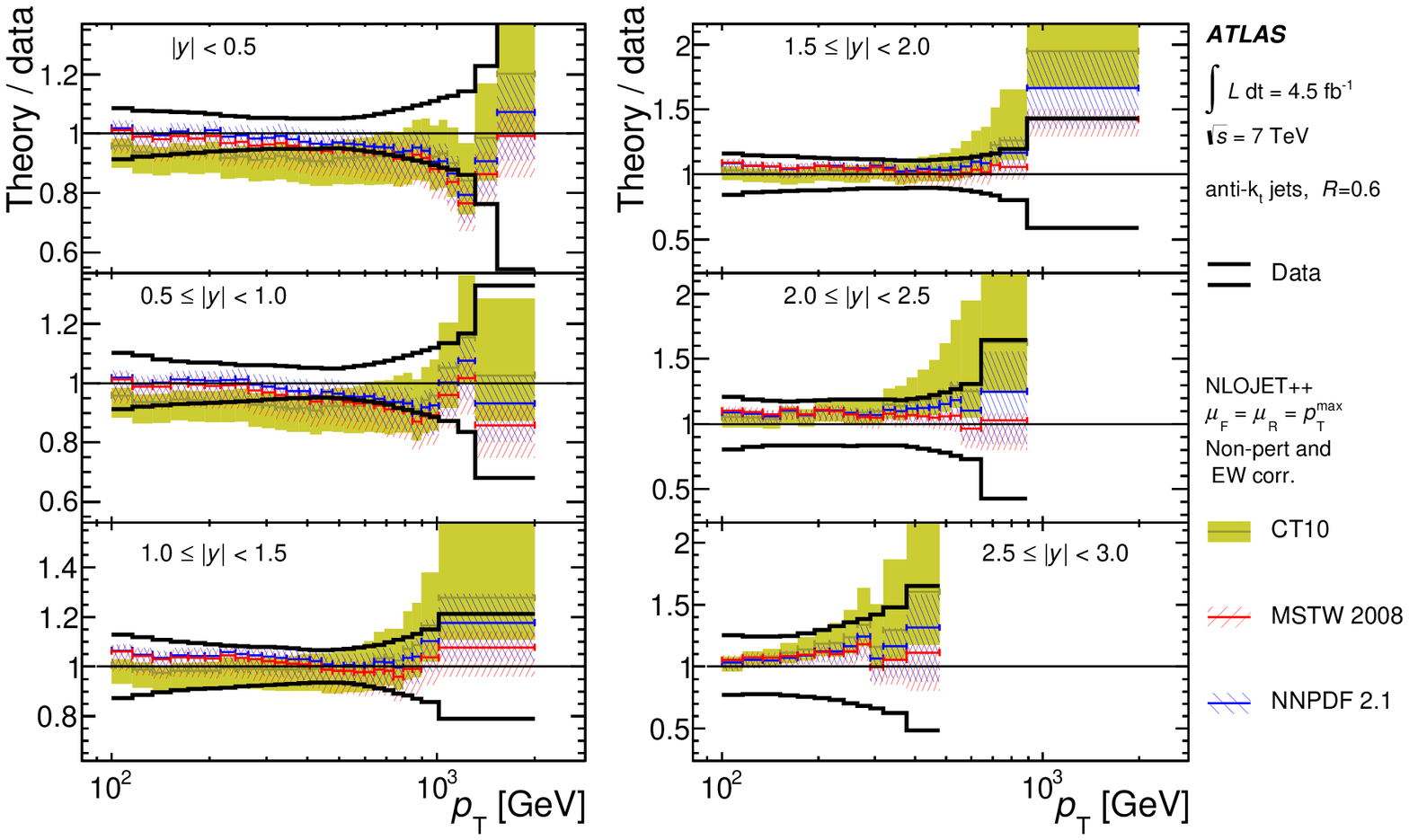}
\caption{Ratio of NLO pQCD predictions to the measured double-differential inclusive jet cross-section, shown as a function of the jet \pt{} in bins of the jet rapidity, for \antikt jets with $R=0.6$. The predictions are calculated using \nlojetpp{} with different NLO PDF sets, namely CT10, MSTW2008 and NNPDF 2.1. Non-perturbative corrections and electroweak corrections are applied to the predictions. Their uncertainties are shown by the bands, including all the uncertainties discussed in section~\ref{sec:theo}. 
The data lines show the total uncertainty except the 1.8\% uncertainty from the luminosity measurement. 
\label{fig:ratior06}
}
\end{center}
\end{figure}

\begin{figure}
\begin{center}
\includegraphics[width=14.5cm]{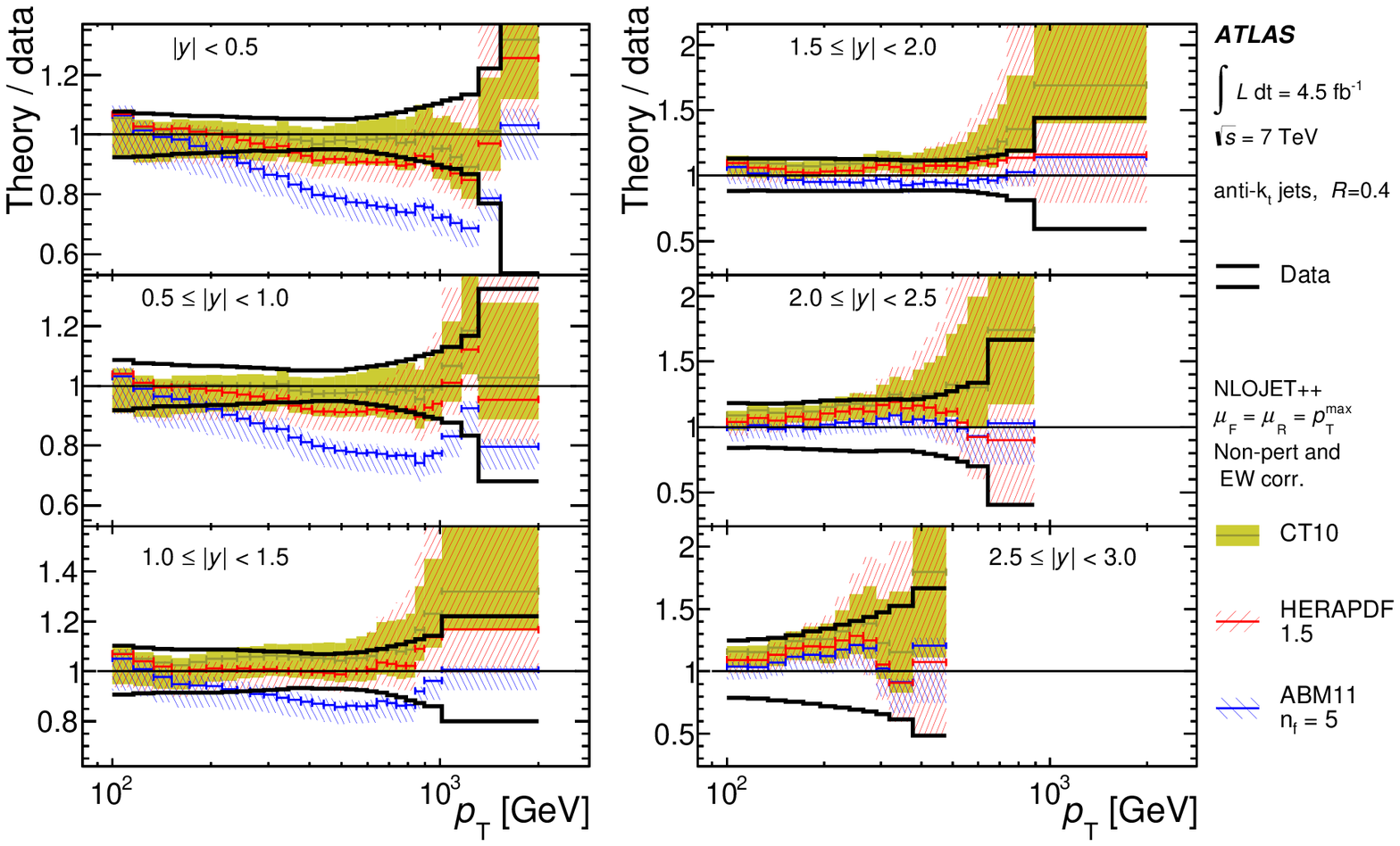}
\caption{Ratio of NLO pQCD predictions to the measured double-differential inclusive jet cross-section, shown as a function of the jet \pt{} in bins of the jet rapidity, for \antikt jets with $R=0.4$. The predictions are calculated using \nlojetpp{} with different NLO PDF sets, namely CT10, HERAPDF 1.5 and ABM11. Non-perturbative corrections and electroweak corrections are applied to the predictions. Their uncertainties are shown by the bands, including all the uncertainties discussed in section~\ref{sec:theo}. 
The data lines show the total uncertainty except the 1.8\% uncertainty from the luminosity measurement. 
\label{fig:ratio2r04}
}
\end{center}
\end{figure}

\begin{figure}
\begin{center}
\includegraphics[width=14.5cm]{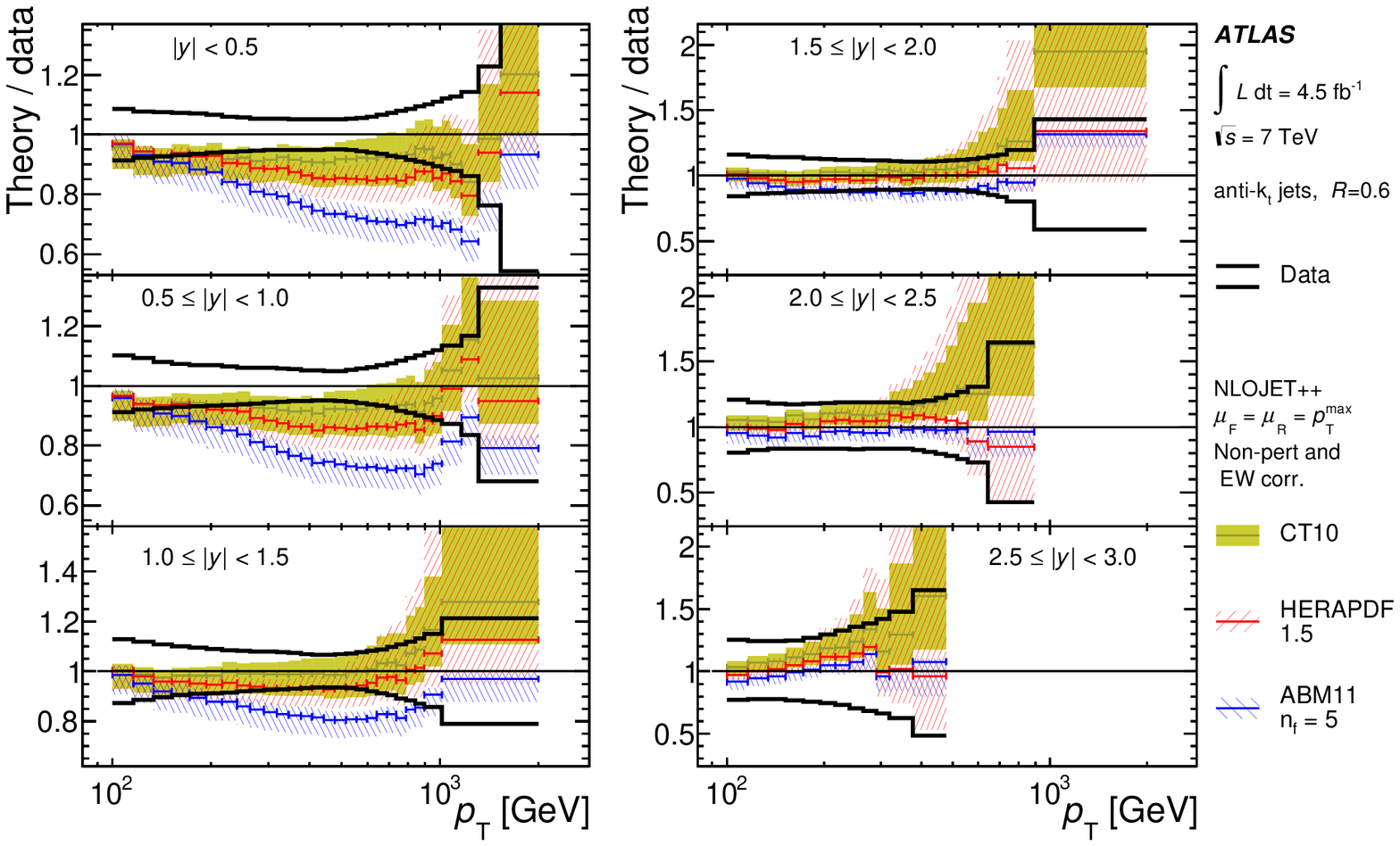}
\caption{Ratio of NLO pQCD predictions to the measured double-differential inclusive jet cross-section, shown as a function of the jet \pt{} in bins of the jet rapidity, for \antikt jets with $R=0.6$. The predictions are calculated using \nlojetpp{} with different NLO PDF sets, namely CT10, HERAPDF 1.5 and ABM11. Non-perturbative corrections and electroweak corrections are applied to the predictions. Their uncertainties are shown by the bands, including all the uncertainties discussed in section~\ref{sec:theo}. 
The data lines show the total uncertainty except the 1.8\% uncertainty from the luminosity measurement. 
\label{fig:ratio2r06}}
\end{center}
\end{figure}

The ratios of the NLO pQCD predictions to the measured cross-sections are presented in figures~\ref{fig:ratior04}--\ref{fig:ratio2r06}. 
The comparison is shown for the predictions using the NLO PDF sets CT10, MSTW 2008, NNPDF 2.1, HERAPDF1.5 and ABM 11 ($n_\mathrm{f}=5$). 
The predictions are generally consistent with the measured cross-sections for jets with both radius parameter values, though the level of consistency varies among the predictions with the different PDF sets.  

\afterpage{\clearpage}

A quantitative comparison of the theoretical predictions to the measurement is performed using a frequentist method. 
The employed method is fully described in ref.~\cite{Aad:2013tea} for the ATLAS dijet cross-section measurement. 
It uses a generalised definition of $\chi^2$ which takes into account the asymmetry of the uncertainties. 
A large set of pseudo-experiments is generated by fluctuating the theoretical predictions according to the full set of experimental and theoretical uncertainties. 
The asymmetries and the correlations of these uncertainties are taken into account.
The $\chi^2$ value is computed between each pseudo-experimental data set and the theoretical predictions, and a $\chi^2$ distribution is constructed. 
The observed $\chi^2$ value, $\chi^2_\mathrm{obs}$, is calculated from the measured points and the theoretical prediction. 
The observed p-value, $P_\mathrm{obs}$, which is defined as the fractional area of the $\chi^2$ distribution with $\chi^2>\chi^2_\mathrm{obs}$, is obtained. 
Tables~\ref{tab:pvaluer04}~and~\ref{tab:pvaluer06} show the evaluated values of $P_\mathrm{obs}$ for the NLO pQCD predictions with non-perturbative and electroweak corrections applied.
The predictions generally show agreement with the measured cross-sections, with a few exceptions. 
The predictions using the ABM11 NLO PDF set fail to describe the measured cross-sections in the low-rapidity region but show good agreement in the high-rapidity region.

\begin{table}
\centering
\small
\begin{tabular}{c|cccccc}
\hline\hline
$y$ ranges &   &\multicolumn{5}{c}{$P_\mathrm{obs}$ } \\
         &{\scriptsize NLO PDF set:} & CT10 & MSTW2008 & NNPDF2.1 & HERAPDF1.5 & ABM11 \\
\hline
$|y|<0.5$        && 81\% & 60\% & 70\% & 58\% & <0.1\% \\
$0.5\leq|y|<1.0$ && 90\% & 92\% & 88\% & 50\% & <0.1\%\\
$1.0\leq|y|<1.5$ && 87\% & 87\% & 84\% & 92\% & 3.5\%\\
$1.5\leq|y|<2.0$ && 91\% & 88\% & 90\% & 72\% & 60\%\\
$2.0\leq|y|<2.5$ && 89\% & 82\% & 85\% & 25\% & 54\%\\
$2.5\leq|y|<3.0$ && 95\% & 92\% & 96\% & 83\% & 87\%\\
\hline\hline
\end{tabular}
\caption{\label{tab:pvaluer04}Observed p-values, $P_\mathrm{obs}$, evaluated for the NLO pQCD predictions with corrections for non-perturbative and electroweak effects, in comparison to the measured cross-section of \antikt{} jets with $R=0.4$. The values are given for the predictions using the NLO PDF sets of CT10, MSTW2008, NNPDF2.1, HERAPDF1.5 and ABM11, for each rapidity bin. 
}
\end{table}

\begin{table}
\centering
\small
\begin{tabular}{c|cccccc}
\hline\hline
$y$ ranges &   &\multicolumn{5}{c}{$P_\mathrm{obs}$ } \\
         &{\scriptsize NLO PDF set:} & CT10 & MSTW2008 & NNPDF2.1 & HERAPDF1.5 & ABM11 \\
\hline
$|y|<0.5$        && 60\% & 52\% & 65\% & 29\% & <0.1\% \\
$0.5\leq|y|<1.0$ && 37\% & 54\% & 48\% & 6.0\% & <0.1\% \\
$1.0\leq|y|<1.5$ && 96\% & 94\% & 92\%& 94\% & 3.3\% \\
$1.5\leq|y|<2.0$ && 90\% & 84\% & 86\%& 93\% & 56\%\\
$2.0\leq|y|<2.5$ && 87\% & 86\% & 89\%& 49\% & 74\% \\
$2.5\leq|y|<3.0$ && 92\% & 99\% & 98\%& 80\% & 80\% \\
\hline\hline
\end{tabular}
\caption{\label{tab:pvaluer06}Observed p-values, $P_\mathrm{obs}$, evaluated for the NLO pQCD predictions with corrections for non-perturbative and electroweak effects, in comparison to the measured cross-section of \antikt{} jets with $R=0.6$. The values are given for the predictions using the NLO PDF sets of CT10, MSTW2008, NNPDF2.1, HERAPDF1.5 and ABM11, for each rapidity bin.
}
\end{table}

\begin{figure}
\begin{center}
\includegraphics[width=14.5cm]{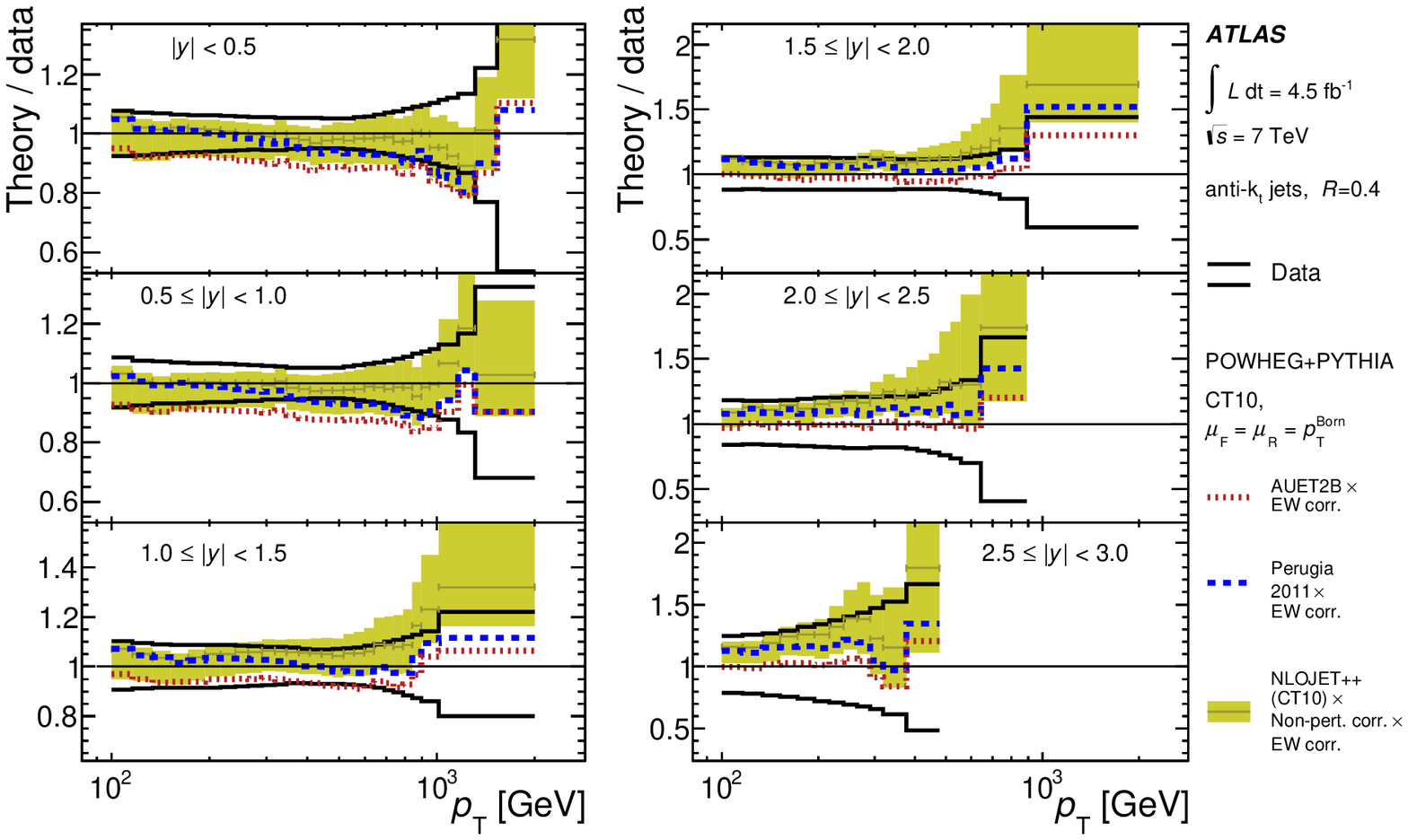}
\caption{Ratio of predictions from \powheg{} to the measured double-differential inclusive jet cross-section, shown as a function of the jet \pt{} in bins of jet rapidity, for \antikt jets with $R=0.4$. The figure also shows the NLO pQCD prediction using \nlojetpp{} with the CT10 NLO PDF set, corrected for non-perturbative effects and electroweak effects. The \powheg{} predictions use \pythia{} for the simulation of parton showers, hadronisation, and the underlying event with the AUET2B tune and the Perugia 2011 tune. Electroweak corrections are applied to the predictions. 
The data lines show the total uncertainty except the 1.8\% uncertainty from the luminosity measurement. 
\label{fig:powratior04}}
\end{center}
\end{figure}

\begin{figure}
\begin{center}
\includegraphics[width=14.5cm]{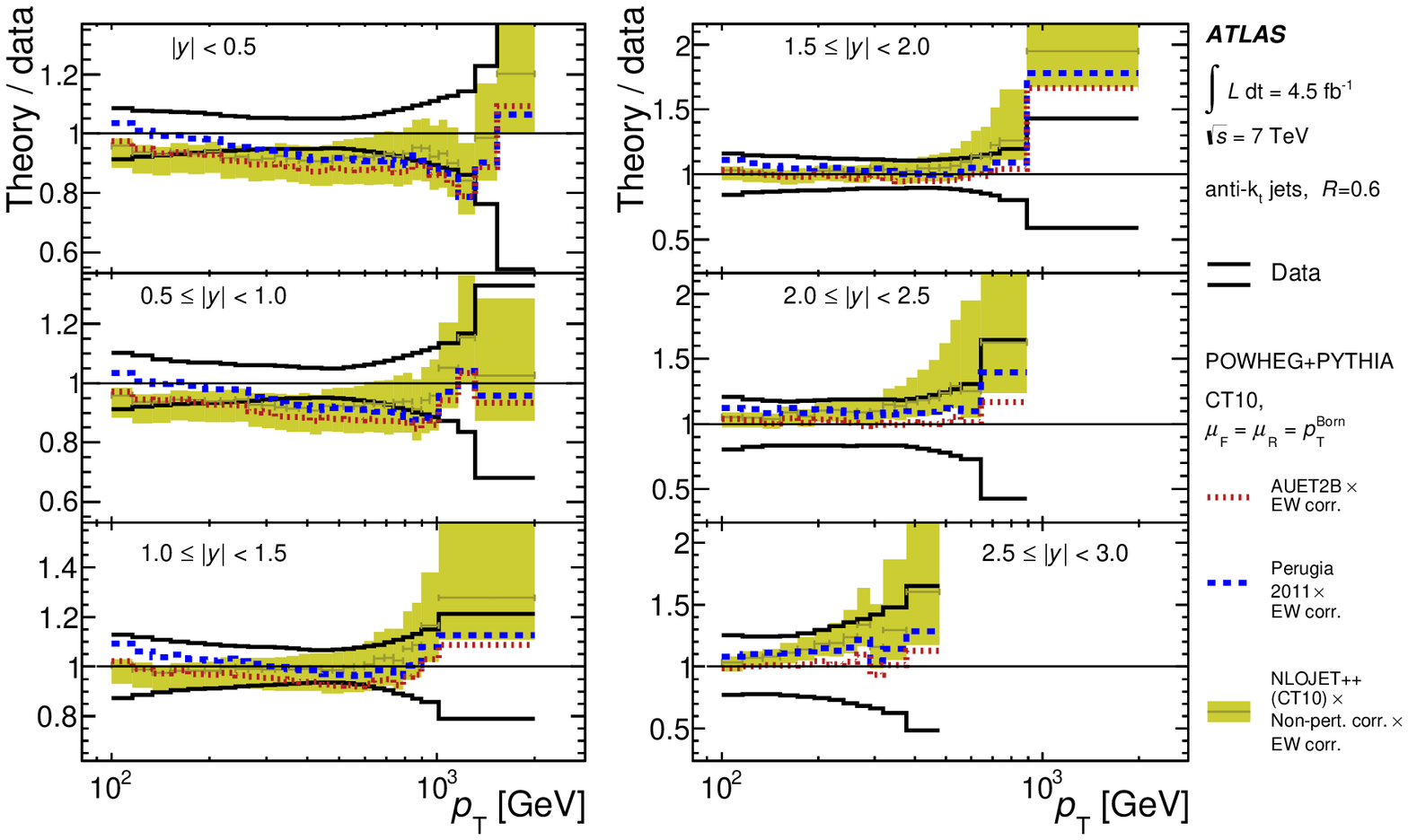}
\caption{Ratio of predictions from \powheg{} to the measured double-differential inclusive jet cross-section, shown as a function of the jet \pt{} in bins of jet rapidity, for \antikt jets with $R=0.6$. The figure also shows the NLO pQCD prediction using \nlojetpp{} with the CT10 NLO PDF set, corrected for non-perturbative effects and electroweak effects. The \powheg{} predictions use \pythia{} for the simulation of parton showers, hadronisation, and the underlying event with the AUET2B tune and the Perugia 2011 tune. Electroweak corrections are applied to the predictions. 
The data lines show the total uncertainty except the 1.8\% uncertainty from the luminosity measurement. 
\label{fig:powratior06}}
\end{center}
\end{figure}

The comparisons of the \powheg{} predictions with the measurement for jets with $R=0.4$ and $R=0.6$ are shown in figures \ref{fig:powratior04} and \ref{fig:powratior06}, respectively, as a function of the jet \pt{} in bins of the jet rapidity. 
The NLO pQCD prediction with the CT10 PDF set is also shown. 
In general, the \powheg{} predictions are found to be in agreement with the measurement. 
In the high-rapidity region, the shape of the measured cross-section is very well reproduced by the \powheg{} predictions, while the predictions tend to be slightly smaller than the measurement for high \pt{} in the low-rapidity region.
As seen in previous measurements~\cite{Aad:2011fc,Aad:2013lpa}, the \Perugia{}~2011 tune gives a consistently larger prediction than the AUET2B tune. 

\FloatBarrier

%% file: conclusion.tex
\section{Conclusions}
\label{sec:concl}
The inclusive jet cross-section in proton--proton collisions at $\rts=7$~\TeV{} is measured for jets reconstructed with the \antikt algorithm with jet radius parameter values of $R=0.4$ and $R=0.6$ in the kinematic region $\pt\geq100$~\GeV{} and $|y|<3$. 
The measurement is based on the data collected with the ATLAS detector during LHC operation in 2011, corresponding to an integrated luminosity of 4.5~\ifb.
The cross-sections are measured double differentially in the jet transverse momentum and rapidity.

The measurement extends up to 2~\TeV{} in jet transverse momentum. 
Compared to the previous measurement using the data collected in 2010, this measurement has a finer binning in \pt, thus giving more precise information on the \pt-dependence of the inclusive jet cross-section.
Full details of uncertainties and their correlations are provided.
The dominant systematic uncertainty arises from the jet energy calibration. 

Fixed-order NLO perturbative QCD calculations, to which corrections for both non-perturbative effects and electroweak effects are applied, are compared to the measurement. 
Several NLO PDF sets are used in the predictions for the comparisons. 
Based on a quantitative evaluation, most of the NLO pQCD predictions are in good agreement with the measurement, confirming that perturbative QCD can describe jet production up to a jet transverse momentum of 2~\TeV. 
The measurement is also well described by the predictions from an NLO matrix element MC generator with matched parton showers and with electroweak correction applied.

%% file: Acknowledgement.tex

\section*{Acknowledgements}

We thank CERN for the very successful operation of the LHC, as well as the
support staff from our institutions without whom ATLAS could not be
operated efficiently.

We acknowledge the support of ANPCyT, Argentina; YerPhI, Armenia; ARC,
Australia; BMWFW and FWF, Austria; ANAS, Azerbaijan; SSTC, Belarus; CNPq and FAPESP,
Brazil; NSERC, NRC and CFI, Canada; CERN; CONICYT, Chile; CAS, MOST and NSFC,
China; COLCIENCIAS, Colombia; MSMT CR, MPO CR and VSC CR, Czech Republic;
DNRF, DNSRC and Lundbeck Foundation, Denmark; EPLANET, ERC and NSRF, European Union;
IN2P3-CNRS, CEA-DSM/IRFU, France; GNSF, Georgia; BMBF, DFG, HGF, MPG and AvH
Foundation, Germany; GSRT and NSRF, Greece; ISF, MINERVA, GIF, I-CORE and Benoziyo Center,
Israel; INFN, Italy; MEXT and JSPS, Japan; CNRST, Morocco; FOM and NWO,
Netherlands; BRF and RCN, Norway; MNiSW and NCN, Poland; GRICES and FCT, Portugal; MNE/IFA, Romania; MES of Russia and ROSATOM, Russian Federation; JINR; MSTD,
Serbia; MSSR, Slovakia; ARRS and MIZ\v{S}, Slovenia; DST/NRF, South Africa;
MINECO, Spain; SRC and Wallenberg Foundation, Sweden; SER, SNSF and Cantons of
Bern and Geneva, Switzerland; NSC, Taiwan; TAEK, Turkey; STFC, the Royal
Society and Leverhulme Trust, United Kingdom; DOE and NSF, United States of
America.

The crucial computing support from all WLCG partners is acknowledged
gratefully, in particular from CERN and the ATLAS Tier-1 facilities at
TRIUMF (Canada), NDGF (Denmark, Norway, Sweden), CC-IN2P3 (France),
KIT/GridKA (Germany), INFN-CNAF (Italy), NL-T1 (Netherlands), PIC (Spain),
ASGC (Taiwan), RAL (UK) and BNL (USA) and in the Tier-2 facilities
worldwide.

%% file: appxsec.tex
\clearpage
\section{Tables of the measured cross-sections}
\label{sec:AppXsec}
The measured inclusive jet cross-sections are shown in tables~\ref{tab:xsecr04y0}--\ref{tab:xsecr04y5} and \ref{tab:xsecr06y0}--\ref{tab:xsecr06y5} for jets with $R=0.4$ and $R=0.6$, respectively.
The correction factors for non-perturbative effects and electroweak effects, which are applied to the NLO pQCD predictions, are also shown in the same table. 

The uncertainties due to the JES uncertainty are separated into four categories, {\it in-situ}, {\it pile-up}, {\it close-by} and {\it flavour}.
The {\it in-situ} category shows the uncertainties from the components of the JES uncertainty given by in-situ calibration techniques.  These techniques are based on the transverse momentum balance between a jet and a well-calibrated reference object, such as the balance between a central jet and a forward jet in a dijet system, the balance between a jet and a \Zboson{} boson or a photon, and the balance between a recoil system of jets and a photon or a high-\pt{} jet. 
For jets with $\pt\gtrsim1$~\TeV, where the techniques employing \pt{} balance are limited by sample size, the uncertainty is estimated from a study of the calorimeter response to single hadrons.
The {\it pile-up} category shows the uncertainties from 
the JES due to the subtraction of pile-up energy in the calibration.
These uncertainties are evaluated from in-situ studies based on the $N_\mathrm{PV}$ and $\langle\mu\rangle$ values. 
The  {\it close-by} category shows the uncertainty from the JES due to the event topology, i.e. the presence of close-by jets. 
Finally, the {\it flavour} category shows the uncertainty from the JES due to the assumption of the fraction of jets originating from a quark or a gluon, which are likely to have different fragmentation.
Further description can be found in ref.~\cite{Aad:2014bia}.
Due to improvements in the jet calibration technique in 2011, the correlation to the JES uncertainty in 2010 is not available.

\input{sys_R04_Eta1}

\input{sys_R04_Eta2}

\input{sys_R04_Eta3}

\input{sys_R04_Eta4}

\input{sys_R04_Eta5}

\input{sys_R04_Eta6}

\input{sys_R06_Eta1}

\input{sys_R06_Eta2}

\input{sys_R06_Eta3}

\input{sys_R06_Eta4}

\input{sys_R06_Eta5}

\input{sys_R06_Eta6}

\renewcommand{\arraystretch}{1.}

%% file: sys_R04_Eta1.tex
\renewcommand{\arraystretch}{1.8}
\begin{table*}[!ht]
\tiny
\centering
{\fontfamily{ptm}\selectfont
\begin{tabular}{|@{}c@{}@{}c@{}@{}c@{}@{}r@{}@{}r@{}@{}r@{}@{}r@{}@{}r@{}@{}r@{}@{}r@{}@{}r@{}|@{}r@{}@{}r@{}|@{}r@{}@{}r@{}@{}r@{}|}
\hline
 \ \ \pt range \ \ &
 \ \ $\sigma$ \ \ &
 \ \ $\delta_{\mathrm{stat}}^{\mathrm{data}}$ \ \ &
 \ \ $\delta_{\mathrm{stat}}^{\mathrm{MC}}$ \ \ &
 \ \ $u_{\operatorname{in-situ}}$ \ \ &
 \ \ $u_{\operatorname{pile-up}}$ \ \ &
 \ \ $u_{\operatorname{close-by}}$ \ \ &
\ \ $u_{\mathrm{flavour}}$ \ \ &
 \ \ $u_{\mathrm{JER}}$ \ \ &
 \ \ $u_{\mathrm{JAR}}$ \ \ &
 \ \ $u_{\mathrm{unfold}}$ \ \ &
 \ \ $u_{\mathrm{qual.}}$ \ \ &
 \ \ $u_{\mathrm{lumi}}$ \ \ &
 \ \ NPC \ \ &
 \ \ $u_{\mathrm{NP}}$ \ \ &
 \ \ EWC \ \ \\ \relax
\ \ [GeV] \ \ & \ \ [pb/GeV] \ \ &\ \  \% \ \ & \ \ \% \ \ & \ \  \% \ \ &
 \ \ \% \ \ & \ \ \% \ \ & \ \ \% \ \ & \ \ \% \ \ &
 \ \ \% \ \ & \ \ \% \ \ & \ \ \% \ \ & \ \ \% \ \ & \ \ \ \ \ & \ \ \% \ \ & \ \ \ \ \ \\
\hline
 \ \ $100$--$116$ \ \ & \ \ $4.23\cdot 10^{3}$ \ \ & \ \ $0.55$ \ \ & \ \ $0.69$ \ \ & \ \ $^{+3.9}_{-3.9}$ \ \ & \ \ $^{+1.4}_{-1.3}$ \ \ & \ \ $^{+0.8}_{-0.9}$ \ \ & \ \ $^{+5.8}_{-5.5}$ \ \ & \ \ $3.0$ \ \ & \ \ $0.0$ \ \ & \ \ $0.1$ \ \ & \multirow{31}{*}{\ \ $0.25$ \ \ } & \multirow{31}{*}{\ \ $1.8$ \ \ } & \ \ $1.02$ \ \ & \ \ $^{+0}_{-9}$ \ \ & \ \ $1.00$ \ \ \\ 
 \ \ $116$--$134$ \ \ & \ \ $2.02\cdot 10^{3}$ \ \ & \ \ $0.75$ \ \ & \ \ $0.48$ \ \ & \ \ $^{+3.8}_{-3.9}$ \ \ & \ \ $^{+1.1}_{-1.2}$ \ \ & \ \ $^{+0.8}_{-0.8}$ \ \ & \ \ $^{+5.3}_{-5.3}$ \ \ & \ \ $2.4$ \ \ & \ \ $0.0$ \ \ & \ \ $0.1$ \ \ & & & \ \ $1.01$ \ \ & \ \ $^{+0}_{-8}$ \ \ & \ \ $1.00$ \ \ \\ 
 \ \ $134$--$152$ \ \ & \ \ $9.88\cdot 10^{2}$ \ \ & \ \ $0.88$ \ \ & \ \ $0.39$ \ \ & \ \ $^{+3.9}_{-4.0}$ \ \ & \ \ $^{+0.9}_{-1.1}$ \ \ & \ \ $^{+0.7}_{-0.7}$ \ \ & \ \ $^{+5.0}_{-5.0}$ \ \ & \ \ $1.9$ \ \ & \ \ $0.0$ \ \ & \ \ $0.0$ \ \ & & & \ \ $1.01$ \ \ & \ \ $^{+0}_{-7}$ \ \ & \ \ $1.00$ \ \ \\ 
 \ \ $152$--$172$ \ \ & \ \ $5.02\cdot 10^{2}$ \ \ & \ \ $0.71$ \ \ & \ \ $0.43$ \ \ & \ \ $^{+4.0}_{-4.1}$ \ \ & \ \ $^{+0.9}_{-0.9}$ \ \ & \ \ $^{+0.7}_{-0.7}$ \ \ & \ \ $^{+4.7}_{-4.6}$ \ \ & \ \ $1.6$ \ \ & \ \ $0.0$ \ \ & \ \ $0.0$ \ \ & & & \ \ $1.01$ \ \ & \ \ $^{+0}_{-7}$ \ \ & \ \ $1.00$ \ \ \\ 
 \ \ $172$--$194$ \ \ & \ \ $2.58\cdot 10^{2}$ \ \ & \ \ $0.56$ \ \ & \ \ $0.36$ \ \ & \ \ $^{+4.1}_{-4.2}$ \ \ & \ \ $^{+0.8}_{-0.9}$ \ \ & \ \ $^{+0.8}_{-0.7}$ \ \ & \ \ $^{+4.3}_{-4.2}$ \ \ & \ \ $1.4$ \ \ & \ \ $0.0$ \ \ & \ \ $0.0$ \ \ & & & \ \ $1.01$ \ \ & \ \ $^{+0}_{-6}$ \ \ & \ \ $1.00$ \ \ \\ 
 \ \ $194$--$216$ \ \ & \ \ $1.37\cdot 10^{2}$ \ \ & \ \ $0.72$ \ \ & \ \ $0.36$ \ \ & \ \ $^{+4.3}_{-4.3}$ \ \ & \ \ $^{+0.8}_{-0.8}$ \ \ & \ \ $^{+0.8}_{-0.8}$ \ \ & \ \ $^{+4.0}_{-3.9}$ \ \ & \ \ $1.3$ \ \ & \ \ $0.0$ \ \ & \ \ $0.0$ \ \ & & & \ \ $1.00$ \ \ & \ \ $^{+0}_{-6}$ \ \ & \ \ $1.00$ \ \ \\ 
 \ \ $216$--$240$ \ \ & \ \ $7.55\cdot 10^{1}$ \ \ & \ \ $0.52$ \ \ & \ \ $0.34$ \ \ & \ \ $^{+4.3}_{-4.3}$ \ \ & \ \ $^{+0.7}_{-0.7}$ \ \ & \ \ $^{+0.9}_{-0.9}$ \ \ & \ \ $^{+3.6}_{-3.5}$ \ \ & \ \ $1.2$ \ \ & \ \ $0.0$ \ \ & \ \ $0.0$ \ \ & & & \ \ $1.00$ \ \ & \ \ $^{+0}_{-6}$ \ \ & \ \ $1.00$ \ \ \\ 
 \ \ $240$--$264$ \ \ & \ \ $4.28\cdot 10^{1}$ \ \ & \ \ $0.67$ \ \ & \ \ $0.37$ \ \ & \ \ $^{+4.3}_{-4.2}$ \ \ & \ \ $^{+0.7}_{-0.6}$ \ \ & \ \ $^{+1.0}_{-1.0}$ \ \ & \ \ $^{+3.3}_{-3.2}$ \ \ & \ \ $1.1$ \ \ & \ \ $0.0$ \ \ & \ \ $0.0$ \ \ & & & \ \ $1.00$ \ \ & \ \ $^{+0}_{-5}$ \ \ & \ \ $1.00$ \ \ \\ 
 \ \ $264$--$290$ \ \ & \ \ $2.47\cdot 10^{1}$ \ \ & \ \ $0.86$ \ \ & \ \ $0.34$ \ \ & \ \ $^{+4.2}_{-4.1}$ \ \ & \ \ $^{+0.6}_{-0.6}$ \ \ & \ \ $^{+1.1}_{-1.0}$ \ \ & \ \ $^{+3.0}_{-3.0}$ \ \ & \ \ $1.1$ \ \ & \ \ $0.0$ \ \ & \ \ $0.0$ \ \ & & & \ \ $1.00$ \ \ & \ \ $^{+0}_{-5}$ \ \ & \ \ $1.00$ \ \ \\ 
 \ \ $290$--$318$ \ \ & \ \ $1.44\cdot 10^{1}$ \ \ & \ \ $1.0$ \ \ & \ \ $0.32$ \ \ & \ \ $^{+4.2}_{-4.2}$ \ \ & \ \ $^{+0.6}_{-0.7}$ \ \ & \ \ $^{+1.1}_{-1.1}$ \ \ & \ \ $^{+2.9}_{-2.9}$ \ \ & \ \ $1.1$ \ \ & \ \ $0.0$ \ \ & \ \ $0.0$ \ \ & & & \ \ $1.00$ \ \ & \ \ $^{+0}_{-5}$ \ \ & \ \ $1.00$ \ \ \\ 
 \ \ $318$--$346$ \ \ & \ \ $8.40\cdot 10^{0}$ \ \ & \ \ $0.98$ \ \ & \ \ $0.46$ \ \ & \ \ $^{+4.1}_{-4.3}$ \ \ & \ \ $^{+0.6}_{-0.7}$ \ \ & \ \ $^{+1.1}_{-1.0}$ \ \ & \ \ $^{+2.8}_{-2.8}$ \ \ & \ \ $1.0$ \ \ & \ \ $0.0$ \ \ & \ \ $0.0$ \ \ & & & \ \ $1.00$ \ \ & \ \ $^{+0}_{-4}$ \ \ & \ \ $1.00$ \ \ \\ 
 \ \ $346$--$376$ \ \ & \ \ $5.14\cdot 10^{0}$ \ \ & \ \ $0.54$ \ \ & \ \ $0.58$ \ \ & \ \ $^{+4.2}_{-4.1}$ \ \ & \ \ $^{+0.8}_{-0.8}$ \ \ & \ \ $^{+1.0}_{-0.9}$ \ \ & \ \ $^{+2.7}_{-2.7}$ \ \ & \ \ $0.9$ \ \ & \ \ $0.1$ \ \ & \ \ $0.0$ \ \ & & & \ \ $1.00$ \ \ & \ \ $^{+0}_{-4}$ \ \ & \ \ $1.00$ \ \ \\ 
 \ \ $376$--$408$ \ \ & \ \ $3.11\cdot 10^{0}$ \ \ & \ \ $0.30$ \ \ & \ \ $0.49$ \ \ & \ \ $^{+4.4}_{-4.1}$ \ \ & \ \ $^{+0.8}_{-0.7}$ \ \ & \ \ $^{+0.9}_{-0.8}$ \ \ & \ \ $^{+2.7}_{-2.6}$ \ \ & \ \ $0.9$ \ \ & \ \ $0.1$ \ \ & \ \ $0.0$ \ \ & & & \ \ $1.00$ \ \ & \ \ $^{+0}_{-4}$ \ \ & \ \ $1.00$ \ \ \\ 
 \ \ $408$--$442$ \ \ & \ \ $1.89\cdot 10^{0}$ \ \ & \ \ $0.33$ \ \ & \ \ $0.40$ \ \ & \ \ $^{+4.5}_{-4.0}$ \ \ & \ \ $^{+0.7}_{-0.6}$ \ \ & \ \ $^{+0.6}_{-0.6}$ \ \ & \ \ $^{+2.5}_{-2.4}$ \ \ & \ \ $0.9$ \ \ & \ \ $0.1$ \ \ & \ \ $0.0$ \ \ & & & \ \ $1.00$ \ \ & \ \ $^{+0}_{-4}$ \ \ & \ \ $1.00$ \ \ \\ 
 \ \ $442$--$478$ \ \ & \ \ $1.13\cdot 10^{0}$ \ \ & \ \ $0.33$ \ \ & \ \ $0.38$ \ \ & \ \ $^{+4.3}_{-4.3}$ \ \ & \ \ $^{+0.4}_{-0.4}$ \ \ & \ \ $^{+0.4}_{-0.3}$ \ \ & \ \ $^{+2.2}_{-2.1}$ \ \ & \ \ $0.9$ \ \ & \ \ $0.0$ \ \ & \ \ $0.0$ \ \ & & & \ \ $1.00$ \ \ & \ \ $^{+0}_{-3}$ \ \ & \ \ $1.00$ \ \ \\ 
 \ \ $478$--$516$ \ \ & \ \ $6.83\cdot 10^{-1}$ \ \ & \ \ $0.27$ \ \ & \ \ $0.28$ \ \ & \ \ $^{+4.5}_{-4.4}$ \ \ & \ \ $^{+0.2}_{-0.1}$ \ \ & \ \ $^{+0.2}_{-0.2}$ \ \ & \ \ $^{+2.0}_{-1.9}$ \ \ & \ \ $0.8$ \ \ & \ \ $0.0$ \ \ & \ \ $0.0$ \ \ & & & \ \ $1.00$ \ \ & \ \ $^{+0}_{-3}$ \ \ & \ \ $1.01$ \ \ \\ 
 \ \ $516$--$556$ \ \ & \ \ $4.19\cdot 10^{-1}$ \ \ & \ \ $0.34$ \ \ & \ \ $0.23$ \ \ & \ \ $^{+4.7}_{-4.6}$ \ \ & \ \ $^{+0.1}_{-0.0}$ \ \ & \ \ $^{+0.1}_{-0.1}$ \ \ & \ \ $^{+1.8}_{-1.7}$ \ \ & \ \ $0.8$ \ \ & \ \ $0.0$ \ \ & \ \ $0.0$ \ \ & & & \ \ $1.00$ \ \ & \ \ $^{+0}_{-3}$ \ \ & \ \ $1.01$ \ \ \\ 
 \ \ $556$--$598$ \ \ & \ \ $2.53\cdot 10^{-1}$ \ \ & \ \ $0.42$ \ \ & \ \ $0.21$ \ \ & \ \ $^{+5.1}_{-5.0}$ \ \ & \ \ $^{+0.1}_{-0.1}$ \ \ & \ \ $^{+0.0}_{-0.0}$ \ \ & \ \ $^{+1.7}_{-1.7}$ \ \ & \ \ $0.8$ \ \ & \ \ $0.0$ \ \ & \ \ $0.0$ \ \ & & & \ \ $1.00$ \ \ & \ \ $^{+0}_{-3}$ \ \ & \ \ $1.01$ \ \ \\ 
 \ \ $598$--$642$ \ \ & \ \ $1.55\cdot 10^{-1}$ \ \ & \ \ $0.53$ \ \ & \ \ $0.20$ \ \ & \ \ $^{+5.5}_{-5.5}$ \ \ & \ \ $^{+0.1}_{-0.1}$ \ \ & \ \ $^{+0.0}_{-0.0}$ \ \ & \ \ $^{+1.6}_{-1.6}$ \ \ & \ \ $0.9$ \ \ & \ \ $0.0$ \ \ & \ \ $0.0$ \ \ & & & \ \ $1.00$ \ \ & \ \ $^{+0}_{-3}$ \ \ & \ \ $1.01$ \ \ \\ 
 \ \ $642$--$688$ \ \ & \ \ $9.48\cdot 10^{-2}$ \ \ & \ \ $0.67$ \ \ & \ \ $0.22$ \ \ & \ \ $^{+6.1}_{-6.0}$ \ \ & \ \ $^{+0.1}_{-0.1}$ \ \ & \ \ $^{+0.0}_{-0.0}$ \ \ & \ \ $^{+1.6}_{-1.6}$ \ \ & \ \ $0.9$ \ \ & \ \ $0.0$ \ \ & \ \ $0.0$ \ \ & & & \ \ $1.00$ \ \ & \ \ $^{+0}_{-2}$ \ \ & \ \ $1.02$ \ \ \\ 
 \ \ $688$--$736$ \ \ & \ \ $5.80\cdot 10^{-2}$ \ \ & \ \ $0.83$ \ \ & \ \ $0.22$ \ \ & \ \ $^{+6.8}_{-6.6}$ \ \ & \ \ $^{+0.1}_{-0.1}$ \ \ & \ \ $^{+0.0}_{-0.0}$ \ \ & \ \ $^{+1.7}_{-1.7}$ \ \ & \ \ $0.9$ \ \ & \ \ $0.0$ \ \ & \ \ $0.0$ \ \ & & & \ \ $1.00$ \ \ & \ \ $^{+0}_{-2}$ \ \ & \ \ $1.02$ \ \ \\ 
 \ \ $736$--$786$ \ \ & \ \ $3.61\cdot 10^{-2}$ \ \ & \ \ $1.0$ \ \ & \ \ $0.22$ \ \ & \ \ $^{+7.2}_{-7.2}$ \ \ & \ \ $^{+0.1}_{-0.1}$ \ \ & \ \ $^{+0.0}_{-0.0}$ \ \ & \ \ $^{+1.7}_{-1.6}$ \ \ & \ \ $0.9$ \ \ & \ \ $0.0$ \ \ & \ \ $0.0$ \ \ & & & \ \ $1.00$ \ \ & \ \ $^{+0}_{-2}$ \ \ & \ \ $1.03$ \ \ \\ 
 \ \ $786$--$838$ \ \ & \ \ $2.22\cdot 10^{-2}$ \ \ & \ \ $1.3$ \ \ & \ \ $0.22$ \ \ & \ \ $^{+8.0}_{-7.8}$ \ \ & \ \ $^{+0.1}_{-0.1}$ \ \ & \ \ $^{+0.0}_{-0.0}$ \ \ & \ \ $^{+1.6}_{-1.6}$ \ \ & \ \ $0.9$ \ \ & \ \ $0.0$ \ \ & \ \ $0.0$ \ \ & & & \ \ $1.00$ \ \ & \ \ $^{+0}_{-2}$ \ \ & \ \ $1.03$ \ \ \\ 
 \ \ $838$--$894$ \ \ & \ \ $1.31\cdot 10^{-2}$ \ \ & \ \ $1.6$ \ \ & \ \ $0.23$ \ \ & \ \ $^{+8.6}_{-8.4}$ \ \ & \ \ $^{+0.1}_{-0.0}$ \ \ & \ \ $^{+0.0}_{-0.0}$ \ \ & \ \ $^{+1.5}_{-1.5}$ \ \ & \ \ $1.0$ \ \ & \ \ $0.0$ \ \ & \ \ $0.0$ \ \ & & & \ \ $1.00$ \ \ & \ \ $^{+0}_{-2}$ \ \ & \ \ $1.04$ \ \ \\ 
 \ \ $894$--$952$ \ \ & \ \ $7.94\cdot 10^{-3}$ \ \ & \ \ $2.1$ \ \ & \ \ $0.26$ \ \ & \ \ $^{+9.3}_{-8.9}$ \ \ & \ \ $^{+0.1}_{-0.0}$ \ \ & \ \ $^{+0.0}_{-0.0}$ \ \ & \ \ $^{+1.4}_{-1.4}$ \ \ & \ \ $1.0$ \ \ & \ \ $0.0$ \ \ & \ \ $0.1$ \ \ & & & \ \ $1.00$ \ \ & \ \ $^{+0}_{-2}$ \ \ & \ \ $1.04$ \ \ \\ 
 \ \ $952$--$1012$ \ \ & \ \ $4.98\cdot 10^{-3}$ \ \ & \ \ $2.5$ \ \ & \ \ $0.27$ \ \ & \ \ $^{+9.9}_{-9.7}$ \ \ & \ \ $^{+0.1}_{-0.0}$ \ \ & \ \ $^{+0.0}_{-0.0}$ \ \ & \ \ $^{+1.3}_{-1.3}$ \ \ & \ \ $1.0$ \ \ & \ \ $0.0$ \ \ & \ \ $0.1$ \ \ & & & \ \ $1.00$ \ \ & \ \ $^{+0}_{-2}$ \ \ & \ \ $1.05$ \ \ \\ 
 \ \ $1012$--$1076$ \ \ & \ \ $2.97\cdot 10^{-3}$ \ \ & \ \ $3.3$ \ \ & \ \ $0.37$ \ \ & \ \ $^{+11}_{-10}$ \ \ & \ \ $^{+0.1}_{-0.0}$ \ \ & \ \ $^{+0.0}_{-0.0}$ \ \ & \ \ $^{+1.2}_{-1.2}$ \ \ & \ \ $1.0$ \ \ & \ \ $0.0$ \ \ & \ \ $0.1$ \ \ & & & \ \ $1.00$ \ \ & \ \ $^{+0}_{-2}$ \ \ & \ \ $1.06$ \ \ \\ 
 \ \ $1076$--$1162$ \ \ & \ \ $1.67\cdot 10^{-3}$ \ \ & \ \ $3.9$ \ \ & \ \ $0.32$ \ \ & \ \ $^{+11}_{-11}$ \ \ & \ \ $^{+0.1}_{-0.0}$ \ \ & \ \ $^{+0.0}_{-0.0}$ \ \ & \ \ $^{+1.1}_{-1.1}$ \ \ & \ \ $1.1$ \ \ & \ \ $0.0$ \ \ & \ \ $0.0$ \ \ & & & \ \ $1.00$ \ \ & \ \ $^{+0}_{-2}$ \ \ & \ \ $1.06$ \ \ \\ 
 \ \ $1162$--$1310$ \ \ & \ \ $7.00\cdot 10^{-4}$ \ \ & \ \ $5.3$ \ \ & \ \ $0.25$ \ \ & \ \ $^{+12}_{-12}$ \ \ & \ \ $^{+0.1}_{-0.0}$ \ \ & \ \ $^{+0.0}_{-0.0}$ \ \ & \ \ $^{+1.0}_{-0.9}$ \ \ & \ \ $1.3$ \ \ & \ \ $0.0$ \ \ & \ \ $1.0$ \ \ & & & \ \ $1.00$ \ \ & \ \ $^{+0}_{-2}$ \ \ & \ \ $1.08$ \ \ \\ 
 \ \ $1310$--$1530$ \ \ & \ \ $1.55\cdot 10^{-4}$ \ \ & \ \ $10$ \ \ & \ \ $0.27$ \ \ & \ \ $^{+20}_{-21}$ \ \ & \ \ $^{+0.1}_{-0.0}$ \ \ & \ \ $^{+0.0}_{-0.0}$ \ \ & \ \ $^{+0.9}_{-0.8}$ \ \ & \ \ $1.8$ \ \ & \ \ $0.0$ \ \ & \ \ $0.2$ \ \ & & & \ \ $1.00$ \ \ & \ \ $^{+0}_{-2}$ \ \ & \ \ $1.10$ \ \ \\ 
 \ \ $1530$--$1992$ \ \ & \ \ $1.17\cdot 10^{-5}$ \ \ & \ \ $25$ \ \ & \ \ $0.42$ \ \ & \ \ $^{+47}_{-39}$ \ \ & \ \ $^{+0.0}_{-0.0}$ \ \ & \ \ $^{+0.0}_{-0.0}$ \ \ & \ \ $^{+0.8}_{-0.8}$ \ \ & \ \ $3.0$ \ \ & \ \ $0.0$ \ \ & \ \ $5.5$ \ \ & & & \ \ $1.00$ \ \ & \ \ $^{+0}_{-2}$ \ \ & \ \ $1.12$ \ \ \\ 
\hline
\end{tabular}
}
\caption{
Measured double-differential inclusive jet cross-sections for jets with $R=0.4$ in the rapidity bin $|y| < 0.5$.
Here, $\sigma$ is the measured double differential cross-section $\mathrm{d}^2\sigma/\mathrm{d}\pt \mathrm{d}y$, averaged in each bin.
All uncertainties are given in \%.
The variable $\delta_\mathrm{stat}^\mathrm{data}$ ($\delta_\mathrm{stat}^\mathrm{MC}$) is the statistical uncertainty from the data (MC simulation).
The $u$ components show the uncertainties due to the jet energy calibration from the in-situ, pile-up, close-by jet, and flavour components.
The uncertainty due to the jet energy and angular resolution, the unfolding, 
the quality selection, and the integrated luminosity are also shown by the $u$ components.
While all columns are uncorrelated with each other, the in-situ, pile-up, and flavour uncertainties shown here are the sum in quadrature of multiple uncorrelated components.
In the last three columns, the correction factors for non-perturbative effects (NPC) with their uncertainties ($u_\mathrm{NP}$) and electroweak effects (EWC) are shown. 
}
\label{tab:xsecr04y0}
\end{table*}

%% file: sys_R04_Eta2.tex
\renewcommand{\arraystretch}{1.8}
\begin{table*}[!ht]
\tiny
\centering
{\fontfamily{ptm}\selectfont
\begin{tabular}{|@{}c@{}@{}c@{}@{}c@{}@{}r@{}@{}r@{}@{}r@{}@{}r@{}@{}r@{}@{}r@{}@{}r@{}@{}r@{}|@{}r@{}@{}r@{}|@{}r@{}@{}r@{}@{}r@{}|}
\hline
 \ \ \pt range \ \ &
 \ \ $\sigma$ \ \ &
 \ \ $\delta_{\mathrm{stat}}^{\mathrm{data}}$ \ \ &
 \ \ $\delta_{\mathrm{stat}}^{\mathrm{MC}}$ \ \ &
 \ \ $u_{\operatorname{in-situ}}$ \ \ &
 \ \ $u_{\operatorname{pile-up}}$ \ \ &
 \ \ $u_{\operatorname{close-by}}$ \ \ &
\ \ $u_{\mathrm{flavour}}$ \ \ &
 \ \ $u_{\mathrm{JER}}$ \ \ &
 \ \ $u_{\mathrm{JAR}}$ \ \ &
 \ \ $u_{\mathrm{unfold}}$ \ \ &
 \ \ $u_{\mathrm{qual.}}$ \ \ &
 \ \ $u_{\mathrm{lumi}}$ \ \ &
 \ \ NPC \ \ &
 \ \ $u_{\mathrm{NP}}$ \ \ &
 \ \ EWC \ \ \\ \relax
\ \ [GeV] \ \ & \ \ [pb/GeV] \ \ &\ \  \% \ \ & \ \ \% \ \ & \ \  \% \ \ &
 \ \ \% \ \ & \ \ \% \ \ & \ \ \% \ \ & \ \ \% \ \ &
 \ \ \% \ \ & \ \ \% \ \ & \ \ \% \ \ & \ \ \% \ \ & \ \ \ \ \ & \ \ \% \ \ & \ \ \ \ \ \\
\hline
 \ \ $100$--$116$ \ \ & \ \ $4.03\cdot 10^{3}$ \ \ & \ \ $0.55$ \ \ & \ \ $0.91$ \ \ & \ \ $^{+4.4}_{-4.5}$ \ \ & \ \ $^{+1.4}_{-1.2}$ \ \ & \ \ $^{+0.8}_{-0.8}$ \ \ & \ \ $^{+6.4}_{-5.9}$ \ \ & \ \ $3.0$ \ \ & \ \ $0.2$ \ \ & \ \ $0.1$ \ \ & \multirow{29}{*}{\ \ $0.25$ \ \ } & \multirow{29}{*}{\ \ $1.8$ \ \ } & \ \ $1.02$ \ \ & \ \ $^{+0}_{-9}$ \ \ & \ \ $1.00$ \ \ \\ 
 \ \ $116$--$134$ \ \ & \ \ $1.90\cdot 10^{3}$ \ \ & \ \ $0.77$ \ \ & \ \ $0.52$ \ \ & \ \ $^{+4.2}_{-4.2}$ \ \ & \ \ $^{+1.1}_{-1.1}$ \ \ & \ \ $^{+0.8}_{-0.8}$ \ \ & \ \ $^{+5.6}_{-5.3}$ \ \ & \ \ $2.5$ \ \ & \ \ $0.1$ \ \ & \ \ $0.1$ \ \ & & & \ \ $1.01$ \ \ & \ \ $^{+0}_{-8}$ \ \ & \ \ $1.00$ \ \ \\ 
 \ \ $134$--$152$ \ \ & \ \ $9.31\cdot 10^{2}$ \ \ & \ \ $0.92$ \ \ & \ \ $0.40$ \ \ & \ \ $^{+4.3}_{-4.2}$ \ \ & \ \ $^{+1.1}_{-1.0}$ \ \ & \ \ $^{+0.8}_{-0.7}$ \ \ & \ \ $^{+5.1}_{-4.8}$ \ \ & \ \ $2.2$ \ \ & \ \ $0.1$ \ \ & \ \ $0.0$ \ \ & & & \ \ $1.01$ \ \ & \ \ $^{+0}_{-7}$ \ \ & \ \ $1.00$ \ \ \\
\hline
\end{tabular}
}
\caption{
Measured double-differential inclusive jet cross-sections for jets with $R=0.4$ in the rapidity bin $0.5\leq|y| < 1.0$. See caption of table~\ref{tab:xsecr04y0} for details.
}
\label{tab:xsecr04y1}
\end{table*}

%% file: sys_R04_Eta3.tex
\renewcommand{\arraystretch}{1.8}
\begin{table*}[!ht]
\tiny
\centering
{\fontfamily{ptm}\selectfont
\begin{tabular}{|@{}c@{}@{}c@{}@{}c@{}@{}r@{}@{}r@{}@{}r@{}@{}r@{}@{}r@{}@{}r@{}@{}r@{}@{}r@{}|@{}r@{}@{}r@{}|@{}r@{}@{}r@{}@{}r@{}|}
\hline
 \ \ \pt range \ \ &
 \ \ $\sigma$ \ \ &
 \ \ $\delta_{\mathrm{stat}}^{\mathrm{data}}$ \ \ &
 \ \ $\delta_{\mathrm{stat}}^{\mathrm{MC}}$ \ \ &
 \ \ $u_{\operatorname{in-situ}}$ \ \ &
 \ \ $u_{\operatorname{pile-up}}$ \ \ &
 \ \ $u_{\operatorname{close-by}}$ \ \ &
\ \ $u_{\mathrm{flavour}}$ \ \ &
 \ \ $u_{\mathrm{JER}}$ \ \ &
 \ \ $u_{\mathrm{JAR}}$ \ \ &
 \ \ $u_{\mathrm{unfold}}$ \ \ &
 \ \ $u_{\mathrm{qual.}}$ \ \ &
 \ \ $u_{\mathrm{lumi}}$ \ \ &
 \ \ NPC \ \ &
 \ \ $u_{\mathrm{NP}}$ \ \ &
 \ \ EWC \ \ \\ \relax
\ \ [GeV] \ \ & \ \ [pb/GeV] \ \ &\ \  \% \ \ & \ \ \% \ \ & \ \  \% \ \ &
 \ \ \% \ \ & \ \ \% \ \ & \ \ \% \ \ & \ \ \% \ \ &
 \ \ \% \ \ & \ \ \% \ \ & \ \ \% \ \ & \ \ \% \ \ & \ \ \ \ \ & \ \ \% \ \ & \ \ \ \ \ \\
\hline
 \ \ $100$--$116$ \ \ & \ \ $3.33\cdot 10^{3}$ \ \ & \ \ $0.61$ \ \ & \ \ $0.73$ \ \ & \ \ $^{+7.1}_{-6.3}$ \ \ & \ \ $^{+1.7}_{-1.2}$ \ \ & \ \ $^{+0.8}_{-0.8}$ \ \ & \ \ $^{+6.5}_{-5.8}$ \ \ & \ \ $3.4$ \ \ & \ \ $0.4$ \ \ & \ \ $0.1$ \ \ & \multirow{26}{*}{\ \ $0.25$ \ \ } & \multirow{26}{*}{\ \ $1.8$ \ \ } & \ \ $1.01$ \ \ & \ \ $^{+0}_{-8}$ \ \ & \ \ $1.00$ \ \ \\ 
 \ \ $116$--$134$ \ \ & \ \ $1.56\cdot 10^{3}$ \ \ & \ \ $0.85$ \ \ & \ \ $0.56$ \ \ & \ \ $^{+6.6}_{-6.2}$ \ \ & \ \ $^{+1.2}_{-1.1}$ \ \ & \ \ $^{+0.8}_{-0.8}$ \ \ & \ \ $^{+5.9}_{-5.3}$ \ \ & \ \ $2.9$ \ \ & \ \ $0.3$ \ \ & \ \ $0.1$ \ \ & & & \ \ $1.01$ \ \ & \ \ $^{+0}_{-7}$ \ \ & \ \ $1.00$ \ \ \\ 
 \ \ $134$--$152$ \ \ & \ \ $7.58\cdot 10^{2}$ \ \ & \ \ $1.0$ \ \ & \ \ $0.42$ \ \ & \ \ $^{+6.5}_{-6.4}$ \ \ & \ \ $^{+1.0}_{-1.0}$ \ \ & \ \ $^{+0.7}_{-0.7}$ \ \ & \ \ $^{+5.4}_{-4.9}$ \ \ & \ \ $2.4$ \ \ & \ \ $0.1$ \ \ & \ \ $0.0$ \ \ & & & \ \ $1.00$ \ \ & \ \ $^{+0}_{-7}$ \ \ & \ \ $1.00$ \ \ \\
\hline
\end{tabular}
}
\caption{
Measured double-differential inclusive jet cross-sections for jets with $R=0.4$ in the rapidity bin $1.0\leq|y| < 1.5$. See caption of table~\ref{tab:xsecr04y0} for details.
}
\label{tab:xsecr04y2}
\end{table*}

%% file: sys_R04_Eta4.tex
\renewcommand{\arraystretch}{1.8}
\begin{table*}[!ht]
\tiny
\centering
{\fontfamily{ptm}\selectfont
\begin{tabular}{|@{}c@{}@{}c@{}@{}c@{}@{}r@{}@{}r@{}@{}r@{}@{}r@{}@{}r@{}@{}r@{}@{}r@{}@{}r@{}|@{}r@{}@{}r@{}|@{}r@{}@{}r@{}@{}r@{}|}
\hline
 \ \ \pt range \ \ &
 \ \ $\sigma$ \ \ &
 \ \ $\delta_{\mathrm{stat}}^{\mathrm{data}}$ \ \ &
 \ \ $\delta_{\mathrm{stat}}^{\mathrm{MC}}$ \ \ &
 \ \ $u_{\operatorname{in-situ}}$ \ \ &
 \ \ $u_{\operatorname{pile-up}}$ \ \ &
 \ \ $u_{\operatorname{close-by}}$ \ \ &
\ \ $u_{\mathrm{flavour}}$ \ \ &
 \ \ $u_{\mathrm{JER}}$ \ \ &
 \ \ $u_{\mathrm{JAR}}$ \ \ &
 \ \ $u_{\mathrm{unfold}}$ \ \ &
 \ \ $u_{\mathrm{qual.}}$ \ \ &
 \ \ $u_{\mathrm{lumi}}$ \ \ &
 \ \ NPC \ \ &
 \ \ $u_{\mathrm{NP}}$ \ \ &
 \ \ EWC \ \ \\ \relax
\ \ [GeV] \ \ & \ \ [pb/GeV] \ \ &\ \  \% \ \ & \ \ \% \ \ & \ \  \% \ \ &
 \ \ \% \ \ & \ \ \% \ \ & \ \ \% \ \ & \ \ \% \ \ &
 \ \ \% \ \ & \ \ \% \ \ & \ \ \% \ \ & \ \ \% \ \ & \ \ \ \ \ & \ \ \% \ \ & \ \ \ \ \ \\
\hline
 \ \ $100$--$116$ \ \ & \ \ $2.54\cdot 10^{3}$ \ \ & \ \ $0.74$ \ \ & \ \ $0.87$ \ \ & \ \ $^{+11}_{-9.6}$ \ \ & \ \ $^{+1.7}_{-1.2}$ \ \ & \ \ $^{+1.0}_{-0.8}$ \ \ & \ \ $^{+6.0}_{-5.3}$ \ \ & \ \ $3.5$ \ \ & \ \ $0.5$ \ \ & \ \ $0.1$ \ \ & \multirow{23}{*}{\ \ $0.25$ \ \ } & \multirow{23}{*}{\ \ $1.8$ \ \ } & \ \ $1.01$ \ \ & \ \ $^{+0}_{-8}$ \ \ & \ \ $1.00$ \ \ \\ 
 \ \ $116$--$134$ \ \ & \ \ $1.16\cdot 10^{3}$ \ \ & \ \ $1.0$ \ \ & \ \ $0.67$ \ \ & \ \ $^{+11}_{-9.6}$ \ \ & \ \ $^{+1.5}_{-1.1}$ \ \ & \ \ $^{+0.9}_{-0.7}$ \ \ & \ \ $^{+5.9}_{-5.1}$ \ \ & \ \ $3.0$ \ \ & \ \ $0.4$ \ \ & \ \ $0.1$ \ \ & & & \ \ $1.00$ \ \ & \ \ $^{+0}_{-7}$ \ \ & \ \ $1.00$ \ \ \\ 
 \ \ $134$--$152$ \ \ & \ \ $5.52\cdot 10^{2}$ \ \ & \ \ $1.3$ \ \ & \ \ $0.47$ \ \ & \ \ $^{+11}_{-10}$ \ \ & \ \ $^{+1.3}_{-1.0}$ \ \ & \ \ $^{+0.8}_{-0.7}$ \ \ & \ \ $^{+5.5}_{-4.8}$ \ \ & \ \ $2.6$ \ \ & \ \ $0.3$ \ \ & \ \ $0.0$ \ \ & & & \ \ $1.00$ \ \ & \ \ $^{+1}_{-6}$ \ \ & \ \ $1.00$ \ \ \\
\hline
\end{tabular}
}
\caption{
Measured double-differential inclusive jet cross-sections for jets with $R=0.4$ in the rapidity bin $1.5\leq|y| < 2.0$. See caption of table~\ref{tab:xsecr04y0} for details.
}
\label{tab:xsecr04y3}
\end{table*}

%% file: sys_R04_Eta5.tex
\renewcommand{\arraystretch}{1.8}
\begin{table*}[!ht]
\tiny
\centering
{\fontfamily{ptm}\selectfont
\begin{tabular}{|@{}c@{}@{}c@{}@{}c@{}@{}r@{}@{}r@{}@{}r@{}@{}r@{}@{}r@{}@{}r@{}@{}r@{}@{}r@{}|@{}r@{}@{}r@{}|@{}r@{}@{}r@{}@{}r@{}|}
\hline
 \ \ \pt range \ \ &
 \ \ $\sigma$ \ \ &
 \ \ $\delta_{\mathrm{stat}}^{\mathrm{data}}$ \ \ &
 \ \ $\delta_{\mathrm{stat}}^{\mathrm{MC}}$ \ \ &
 \ \ $u_{\operatorname{in-situ}}$ \ \ &
 \ \ $u_{\operatorname{pile-up}}$ \ \ &
 \ \ $u_{\operatorname{close-by}}$ \ \ &
\ \ $u_{\mathrm{flavour}}$ \ \ &
 \ \ $u_{\mathrm{JER}}$ \ \ &
 \ \ $u_{\mathrm{JAR}}$ \ \ &
 \ \ $u_{\mathrm{unfold}}$ \ \ &
 \ \ $u_{\mathrm{qual.}}$ \ \ &
 \ \ $u_{\mathrm{lumi}}$ \ \ &
 \ \ NPC \ \ &
 \ \ $u_{\mathrm{NP}}$ \ \ &
 \ \ EWC \ \ \\ \relax
\ \ [GeV] \ \ & \ \ [pb/GeV] \ \ &\ \  \% \ \ & \ \ \% \ \ & \ \  \% \ \ &
 \ \ \% \ \ & \ \ \% \ \ & \ \ \% \ \ & \ \ \% \ \ &
 \ \ \% \ \ & \ \ \% \ \ & \ \ \% \ \ & \ \ \% \ \ & \ \ \ \ \ & \ \ \% \ \ & \ \ \ \ \ \\
\hline
 \ \ $100$--$116$ \ \ & \ \ $1.87\cdot 10^{3}$ \ \ & \ \ $0.88$ \ \ & \ \ $0.79$ \ \ & \ \ $^{+17}_{-14}$ \ \ & \ \ $^{+1.7}_{-1.6}$ \ \ & \ \ $^{+0.7}_{-0.7}$ \ \ & \ \ $^{+4.4}_{-4.2}$ \ \ & \ \ $4.8$ \ \ & \ \ $0.8$ \ \ & \ \ $0.0$ \ \ & \multirow{19}{*}{\ \ $0.25$ \ \ } & \multirow{19}{*}{\ \ $1.8$ \ \ } & \ \ $1.01$ \ \ & \ \ $^{+1}_{-7}$ \ \ & \ \ $1.00$ \ \ \\ 
 \ \ $116$--$134$ \ \ & \ \ $7.83\cdot 10^{2}$ \ \ & \ \ $1.2$ \ \ & \ \ $0.96$ \ \ & \ \ $^{+17}_{-15}$ \ \ & \ \ $^{+1.3}_{-1.3}$ \ \ & \ \ $^{+0.7}_{-0.7}$ \ \ & \ \ $^{+3.9}_{-3.7}$ \ \ & \ \ $3.7$ \ \ & \ \ $0.7$ \ \ & \ \ $0.0$ \ \ & & & \ \ $1.00$ \ \ & \ \ $^{+1}_{-6}$ \ \ & \ \ $1.00$ \ \ \\ 
 \ \ $134$--$152$ \ \ & \ \ $3.63\cdot 10^{2}$ \ \ & \ \ $1.6$ \ \ & \ \ $0.52$ \ \ & \ \ $^{+17}_{-15}$ \ \ & \ \ $^{+1.3}_{-1.2}$ \ \ & \ \ $^{+0.6}_{-0.6}$ \ \ & \ \ $^{+3.5}_{-3.2}$ \ \ & \ \ $3.2$ \ \ & \ \ $0.6$ \ \ & \ \ $0.0$ \ \ & & & \ \ $1.00$ \ \ & \ \ $^{+1}_{-5}$ \ \ & \ \ $1.00$ \ \ \\
\hline
\end{tabular}
}
\caption{
Measured double-differential inclusive jet cross-sections for jets with $R=0.4$ in the rapidity bin $2.0\leq|y| < 2.5$. See caption of table~\ref{tab:xsecr04y0} for details.
}
\label{tab:xsecr04y4}
\end{table*}

%% file: sys_R04_Eta6.tex
\renewcommand{\arraystretch}{1.8}
\begin{table*}[!ht]
\tiny
\centering
{\fontfamily{ptm}\selectfont
\begin{tabular}{|@{}c@{}@{}c@{}@{}c@{}@{}r@{}@{}r@{}@{}r@{}@{}r@{}@{}r@{}@{}r@{}@{}r@{}@{}r@{}|@{}r@{}@{}r@{}|@{}r@{}@{}r@{}@{}r@{}|}
\hline
 \ \ \pt range \ \ &
 \ \ $\sigma$ \ \ &
 \ \ $\delta_{\mathrm{stat}}^{\mathrm{data}}$ \ \ &
 \ \ $\delta_{\mathrm{stat}}^{\mathrm{MC}}$ \ \ &
 \ \ $u_{\operatorname{in-situ}}$ \ \ &
 \ \ $u_{\operatorname{pile-up}}$ \ \ &
 \ \ $u_{\operatorname{close-by}}$ \ \ &
\ \ $u_{\mathrm{flavour}}$ \ \ &
 \ \ $u_{\mathrm{JER}}$ \ \ &
 \ \ $u_{\mathrm{JAR}}$ \ \ &
 \ \ $u_{\mathrm{unfold}}$ \ \ &
 \ \ $u_{\mathrm{qual.}}$ \ \ &
 \ \ $u_{\mathrm{lumi}}$ \ \ &
 \ \ NPC \ \ &
 \ \ $u_{\mathrm{NP}}$ \ \ &
 \ \ EWC \ \ \\ \relax
\ \ [GeV] \ \ & \ \ [pb/GeV] \ \ &\ \  \% \ \ & \ \ \% \ \ & \ \  \% \ \ &
 \ \ \% \ \ & \ \ \% \ \ & \ \ \% \ \ & \ \ \% \ \ &
 \ \ \% \ \ & \ \ \% \ \ & \ \ \% \ \ & \ \ \% \ \ & \ \ \ \ \ & \ \ \% \ \ & \ \ \ \ \ \\
\hline
 \ \ $100$--$116$ \ \ & \ \ $1.04\cdot 10^{3}$ \ \ & \ \ $1.2$ \ \ & \ \ $1.1$ \ \ & \ \ $^{+24}_{-20}$ \ \ & \ \ $^{+3.1}_{-2.1}$ \ \ & \ \ $^{+0.8}_{-0.7}$ \ \ & \ \ $^{+3.9}_{-3.5}$ \ \ & \ \ $5.3$ \ \ & \ \ $1.5$ \ \ & \ \ $0.1$ \ \ & \multirow{12}{*}{\ \ $0.25$ \ \ } & \multirow{12}{*}{\ \ $1.8$ \ \ } & \ \ $1.00$ \ \ & \ \ $^{+1}_{-6}$ \ \ & \ \ $1.00$ \ \ \\ 
 \ \ $116$--$134$ \ \ & \ \ $4.16\cdot 10^{2}$ \ \ & \ \ $1.7$ \ \ & \ \ $0.95$ \ \ & \ \ $^{+25}_{-21}$ \ \ & \ \ $^{+2.4}_{-1.9}$ \ \ & \ \ $^{+0.7}_{-0.6}$ \ \ & \ \ $^{+3.3}_{-3.1}$ \ \ & \ \ $5.0$ \ \ & \ \ $1.5$ \ \ & \ \ $0.1$ \ \ & & & \ \ $1.00$ \ \ & \ \ $^{+1}_{-5}$ \ \ & \ \ $1.00$ \ \ \\ 
 \ \ $134$--$152$ \ \ & \ \ $1.60\cdot 10^{2}$ \ \ & \ \ $2.4$ \ \ & \ \ $0.86$ \ \ & \ \ $^{+26}_{-22}$ \ \ & \ \ $^{+1.9}_{-2.0}$ \ \ & \ \ $^{+0.7}_{-0.6}$ \ \ & \ \ $^{+2.6}_{-2.8}$ \ \ & \ \ $4.6$ \ \ & \ \ $1.3$ \ \ & \ \ $0.1$ \ \ & & & \ \ $1.00$ \ \ & \ \ $^{+1}_{-5}$ \ \ & \ \ $1.00$ \ \ \\
\hline
\end{tabular}
}
\caption{
Measured double-differential inclusive jet cross-sections for jets with $R=0.4$ in the rapidity bin $2.5\leq|y| < 3.0$. See caption of table~\ref{tab:xsecr04y0} for details.
}
\label{tab:xsecr04y5}
\end{table*}

%% file: sys_R06_Eta1.tex
\renewcommand{\arraystretch}{1.8}
\begin{table*}[!ht]
\tiny
\centering
{\fontfamily{ptm}\selectfont
\begin{tabular}{|@{}c@{}@{}c@{}@{}c@{}@{}r@{}@{}r@{}@{}r@{}@{}r@{}@{}r@{}@{}r@{}@{}r@{}@{}r@{}|@{}r@{}@{}r@{}|@{}r@{}@{}r@{}@{}r@{}|}
\hline
 \ \ \pt range \ \ &
 \ \ $\sigma$ \ \ &
 \ \ $\delta_{\mathrm{stat}}^{\mathrm{data}}$ \ \ &
 \ \ $\delta_{\mathrm{stat}}^{\mathrm{MC}}$ \ \ &
 \ \ $u_{\operatorname{in-situ}}$ \ \ &
 \ \ $u_{\operatorname{pile-up}}$ \ \ &
 \ \ $u_{\operatorname{close-by}}$ \ \ &
\ \ $u_{\mathrm{flavour}}$ \ \ &
 \ \ $u_{\mathrm{JER}}$ \ \ &
 \ \ $u_{\mathrm{JAR}}$ \ \ &
 \ \ $u_{\mathrm{unfold}}$ \ \ &
 \ \ $u_{\mathrm{qual.}}$ \ \ &
 \ \ $u_{\mathrm{lumi}}$ \ \ &
 \ \ NPC \ \ &
 \ \ $u_{\mathrm{NP}}$ \ \ &
 \ \ EWC \ \ \\ \relax
\ \ [GeV] \ \ & \ \ [pb/GeV] \ \ &\ \  \% \ \ & \ \ \% \ \ & \ \  \% \ \ &
 \ \ \% \ \ & \ \ \% \ \ & \ \ \% \ \ & \ \ \% \ \ &
 \ \ \% \ \ & \ \ \% \ \ & \ \ \% \ \ & \ \ \% \ \ & \ \ \ \ \ & \ \ \% \ \ & \ \ \ \ \ \\
\hline
 \ \ $100$--$116$ \ \ & \ \ $5.52\cdot 10^{3}$ \ \ & \ \ $0.54$ \ \ & \ \ $0.73$ \ \ & \ \ $^{+4.8}_{-5.1}$ \ \ & \ \ $^{+1.5}_{-1.7}$ \ \ & \ \ $^{+1.3}_{-1.3}$ \ \ & \ \ $^{+6.0}_{-5.8}$ \ \ & \ \ $3.1$ \ \ & \ \ $0.1$ \ \ & \ \ $0.1$ \ \ & \multirow{31}{*}{\ \ $0.25$ \ \ } & \multirow{31}{*}{\ \ $1.8$ \ \ } & \ \ $1.06$ \ \ & \ \ $^{+0}_{-6}$ \ \ & \ \ $1.00$ \ \ \\ 
 \ \ $116$--$134$ \ \ & \ \ $2.60\cdot 10^{3}$ \ \ & \ \ $0.67$ \ \ & \ \ $0.49$ \ \ & \ \ $^{+4.4}_{-4.6}$ \ \ & \ \ $^{+1.2}_{-1.4}$ \ \ & \ \ $^{+1.1}_{-1.1}$ \ \ & \ \ $^{+5.7}_{-5.5}$ \ \ & \ \ $2.2$ \ \ & \ \ $0.1$ \ \ & \ \ $0.1$ \ \ & & & \ \ $1.05$ \ \ & \ \ $^{+1}_{-5}$ \ \ & \ \ $1.00$ \ \ \\ 
 \ \ $134$--$152$ \ \ & \ \ $1.26\cdot 10^{3}$ \ \ & \ \ $0.75$ \ \ & \ \ $0.39$ \ \ & \ \ $^{+4.3}_{-4.3}$ \ \ & \ \ $^{+1.1}_{-1.1}$ \ \ & \ \ $^{+1.0}_{-1.0}$ \ \ & \ \ $^{+5.5}_{-5.3}$ \ \ & \ \ $1.7$ \ \ & \ \ $0.1$ \ \ & \ \ $0.0$ \ \ & & & \ \ $1.04$ \ \ & \ \ $^{+1}_{-5}$ \ \ & \ \ $1.00$ \ \ \\
\hline
\end{tabular}
}
\caption{
Measured double-differential inclusive jet cross-sections for jets with $R=0.6$ in the rapidity bin $|y| < 0.5$.
Here, $\sigma$ is the measured double differential cross-section $\mathrm{d}^2\sigma/\mathrm{d}\pt \mathrm{d}y$, averaged in each bin.
All uncertainties are given in \%.
The variable $\delta_\mathrm{stat}^\mathrm{data}$ ($\delta_\mathrm{stat}^\mathrm{MC}$) is the statistical uncertainty from the data (MC simulation).
The $u$ components show the uncertainties due to the jet energy calibration from the in-situ, pile-up, close-by jet, and flavour components.
The uncertainty due to the jet energy and angular resolution, the unfolding, 
the quality selection, and the integrated luminosity are also shown by the $u$ components.
While all columns are uncorrelated with each other, the in-situ, pile-up, and flavour uncertainties shown here are the sum in quadrature of multiple uncorrelated components.
In the last three columns, the correction factors for non-perturbative effects (NPC) with their uncertainties ($u_\mathrm{NP}$) and electroweak effects (EWC) are shown. 
}
\label{tab:xsecr06y0}
\end{table*}

%% file: sys_R06_Eta2.tex
\renewcommand{\arraystretch}{1.8}
\begin{table*}[!ht]
\tiny
\centering
{\fontfamily{ptm}\selectfont
\begin{tabular}{|@{}c@{}@{}c@{}@{}c@{}@{}r@{}@{}r@{}@{}r@{}@{}r@{}@{}r@{}@{}r@{}@{}r@{}@{}r@{}|@{}r@{}@{}r@{}|@{}r@{}@{}r@{}@{}r@{}|}
\hline
 \ \ \pt range \ \ &
 \ \ $\sigma$ \ \ &
 \ \ $\delta_{\mathrm{stat}}^{\mathrm{data}}$ \ \ &
 \ \ $\delta_{\mathrm{stat}}^{\mathrm{MC}}$ \ \ &
 \ \ $u_{\operatorname{in-situ}}$ \ \ &
 \ \ $u_{\operatorname{pile-up}}$ \ \ &
 \ \ $u_{\operatorname{close-by}}$ \ \ &
\ \ $u_{\mathrm{flavour}}$ \ \ &
 \ \ $u_{\mathrm{JER}}$ \ \ &
 \ \ $u_{\mathrm{JAR}}$ \ \ &
 \ \ $u_{\mathrm{unfold}}$ \ \ &
 \ \ $u_{\mathrm{qual.}}$ \ \ &
 \ \ $u_{\mathrm{lumi}}$ \ \ &
 \ \ NPC \ \ &
 \ \ $u_{\mathrm{NP}}$ \ \ &
 \ \ EWC \ \ \\ \relax
\ \ [GeV] \ \ & \ \ [pb/GeV] \ \ &\ \  \% \ \ & \ \ \% \ \ & \ \  \% \ \ &
 \ \ \% \ \ & \ \ \% \ \ & \ \ \% \ \ & \ \ \% \ \ &
 \ \ \% \ \ & \ \ \% \ \ & \ \ \% \ \ & \ \ \% \ \ & \ \ \ \ \ & \ \ \% \ \ & \ \ \ \ \ \\
\hline
 \ \ $100$--$116$ \ \ & \ \ $5.14\cdot 10^{3}$ \ \ & \ \ $0.57$ \ \ & \ \ $0.75$ \ \ & \ \ $^{+5.7}_{-4.8}$ \ \ & \ \ $^{+1.9}_{-1.4}$ \ \ & \ \ $^{+1.6}_{-1.2}$ \ \ & \ \ $^{+6.7}_{-5.8}$ \ \ & \ \ $4.1$ \ \ & \ \ $0.6$ \ \ & \ \ $0.1$ \ \ & \multirow{29}{*}{\ \ $0.25$ \ \ } & \multirow{29}{*}{\ \ $1.8$ \ \ } & \ \ $1.06$ \ \ & \ \ $^{+0}_{-6}$ \ \ & \ \ $1.00$ \ \ \\ 
 \ \ $116$--$134$ \ \ & \ \ $2.41\cdot 10^{3}$ \ \ & \ \ $0.70$ \ \ & \ \ $0.55$ \ \ & \ \ $^{+5.6}_{-4.7}$ \ \ & \ \ $^{+1.7}_{-1.3}$ \ \ & \ \ $^{+1.3}_{-1.1}$ \ \ & \ \ $^{+6.1}_{-5.3}$ \ \ & \ \ $3.2$ \ \ & \ \ $0.5$ \ \ & \ \ $0.1$ \ \ & & & \ \ $1.05$ \ \ & \ \ $^{+1}_{-5}$ \ \ & \ \ $1.00$ \ \ \\ 
 \ \ $134$--$152$ \ \ & \ \ $1.16\cdot 10^{3}$ \ \ & \ \ $0.79$ \ \ & \ \ $0.45$ \ \ & \ \ $^{+5.0}_{-4.6}$ \ \ & \ \ $^{+1.4}_{-1.2}$ \ \ & \ \ $^{+1.0}_{-0.9}$ \ \ & \ \ $^{+5.5}_{-5.0}$ \ \ & \ \ $2.3$ \ \ & \ \ $0.3$ \ \ & \ \ $0.0$ \ \ & & & \ \ $1.04$ \ \ & \ \ $^{+1}_{-5}$ \ \ & \ \ $1.00$ \ \ \\ 
 \ \ $152$--$172$ \ \ & \ \ $5.75\cdot 10^{2}$ \ \ & \ \ $0.62$ \ \ & \ \ $0.37$ \ \ & \ \ $^{+4.6}_{-4.6}$ \ \ & \ \ $^{+1.2}_{-1.1}$ \ \ & \ \ $^{+0.9}_{-0.8}$ \ \ & \ \ $^{+5.0}_{-4.9}$ \ \ & \ \ $1.8$ \ \ & \ \ $0.2$ \ \ & \ \ $0.0$ \ \ & & & \ \ $1.03$ \ \ & \ \ $^{+1}_{-4}$ \ \ & \ \ $1.00$ \ \ \\ 
 \ \ $172$--$194$ \ \ & \ \ $2.93\cdot 10^{2}$ \ \ & \ \ $0.51$ \ \ & \ \ $0.37$ \ \ & \ \ $^{+4.7}_{-4.6}$ \ \ & \ \ $^{+1.1}_{-1.0}$ \ \ & \ \ $^{+0.9}_{-0.8}$ \ \ & \ \ $^{+4.7}_{-4.6}$ \ \ & \ \ $1.5$ \ \ & \ \ $0.2$ \ \ & \ \ $0.0$ \ \ & & & \ \ $1.03$ \ \ & \ \ $^{+1}_{-4}$ \ \ & \ \ $1.00$ \ \ \\ 
 \ \ $194$--$216$ \ \ & \ \ $1.54\cdot 10^{2}$ \ \ & \ \ $0.63$ \ \ & \ \ $0.37$ \ \ & \ \ $^{+4.8}_{-4.7}$ \ \ & \ \ $^{+1.0}_{-1.0}$ \ \ & \ \ $^{+0.9}_{-0.9}$ \ \ & \ \ $^{+4.4}_{-4.2}$ \ \ & \ \ $1.4$ \ \ & \ \ $0.1$ \ \ & \ \ $0.0$ \ \ & & & \ \ $1.03$ \ \ & \ \ $^{+1}_{-3}$ \ \ & \ \ $1.00$ \ \ \\ 
 \ \ $216$--$240$ \ \ & \ \ $8.33\cdot 10^{1}$ \ \ & \ \ $0.48$ \ \ & \ \ $0.34$ \ \ & \ \ $^{+4.7}_{-4.7}$ \ \ & \ \ $^{+0.9}_{-0.9}$ \ \ & \ \ $^{+1.0}_{-1.0}$ \ \ & \ \ $^{+3.9}_{-3.8}$ \ \ & \ \ $1.2$ \ \ & \ \ $0.1$ \ \ & \ \ $0.0$ \ \ & & & \ \ $1.02$ \ \ & \ \ $^{+1}_{-3}$ \ \ & \ \ $1.00$ \ \ \\ 
 \ \ $240$--$264$ \ \ & \ \ $4.63\cdot 10^{1}$ \ \ & \ \ $0.63$ \ \ & \ \ $0.33$ \ \ & \ \ $^{+4.6}_{-4.7}$ \ \ & \ \ $^{+0.8}_{-0.9}$ \ \ & \ \ $^{+1.0}_{-1.0}$ \ \ & \ \ $^{+3.5}_{-3.5}$ \ \ & \ \ $1.0$ \ \ & \ \ $0.1$ \ \ & \ \ $0.0$ \ \ & & & \ \ $1.02$ \ \ & \ \ $^{+1}_{-3}$ \ \ & \ \ $1.00$ \ \ \\ 
 \ \ $264$--$290$ \ \ & \ \ $2.69\cdot 10^{1}$ \ \ & \ \ $0.80$ \ \ & \ \ $0.34$ \ \ & \ \ $^{+4.7}_{-4.6}$ \ \ & \ \ $^{+0.8}_{-0.8}$ \ \ & \ \ $^{+1.1}_{-1.0}$ \ \ & \ \ $^{+3.2}_{-3.1}$ \ \ & \ \ $1.0$ \ \ & \ \ $0.1$ \ \ & \ \ $0.0$ \ \ & & & \ \ $1.02$ \ \ & \ \ $^{+1}_{-3}$ \ \ & \ \ $1.00$ \ \ \\ 
 \ \ $290$--$318$ \ \ & \ \ $1.54\cdot 10^{1}$ \ \ & \ \ $1.0$ \ \ & \ \ $0.37$ \ \ & \ \ $^{+4.7}_{-4.4}$ \ \ & \ \ $^{+0.9}_{-0.8}$ \ \ & \ \ $^{+1.0}_{-1.0}$ \ \ & \ \ $^{+2.9}_{-2.8}$ \ \ & \ \ $1.0$ \ \ & \ \ $0.1$ \ \ & \ \ $0.0$ \ \ & & & \ \ $1.01$ \ \ & \ \ $^{+1}_{-2}$ \ \ & \ \ $1.00$ \ \ \\ 
 \ \ $318$--$346$ \ \ & \ \ $9.08\cdot 10^{0}$ \ \ & \ \ $0.94$ \ \ & \ \ $0.46$ \ \ & \ \ $^{+4.8}_{-4.2}$ \ \ & \ \ $^{+1.0}_{-0.9}$ \ \ & \ \ $^{+0.9}_{-0.9}$ \ \ & \ \ $^{+2.6}_{-2.5}$ \ \ & \ \ $1.0$ \ \ & \ \ $0.1$ \ \ & \ \ $0.0$ \ \ & & & \ \ $1.01$ \ \ & \ \ $^{+1}_{-2}$ \ \ & \ \ $1.00$ \ \ \\ 
 \ \ $346$--$376$ \ \ & \ \ $5.41\cdot 10^{0}$ \ \ & \ \ $0.50$ \ \ & \ \ $0.62$ \ \ & \ \ $^{+4.7}_{-4.3}$ \ \ & \ \ $^{+1.0}_{-0.9}$ \ \ & \ \ $^{+0.8}_{-0.7}$ \ \ & \ \ $^{+2.3}_{-2.2}$ \ \ & \ \ $0.9$ \ \ & \ \ $0.1$ \ \ & \ \ $0.0$ \ \ & & & \ \ $1.01$ \ \ & \ \ $^{+0}_{-2}$ \ \ & \ \ $1.00$ \ \ \\ 
 \ \ $376$--$408$ \ \ & \ \ $3.21\cdot 10^{0}$ \ \ & \ \ $0.28$ \ \ & \ \ $0.47$ \ \ & \ \ $^{+4.5}_{-4.5}$ \ \ & \ \ $^{+1.0}_{-0.9}$ \ \ & \ \ $^{+0.6}_{-0.6}$ \ \ & \ \ $^{+2.0}_{-2.0}$ \ \ & \ \ $0.9$ \ \ & \ \ $0.0$ \ \ & \ \ $0.0$ \ \ & & & \ \ $1.01$ \ \ & \ \ $^{+0}_{-2}$ \ \ & \ \ $1.00$ \ \ \\ 
 \ \ $408$--$442$ \ \ & \ \ $1.93\cdot 10^{0}$ \ \ & \ \ $0.33$ \ \ & \ \ $0.45$ \ \ & \ \ $^{+4.7}_{-4.4}$ \ \ & \ \ $^{+0.8}_{-0.7}$ \ \ & \ \ $^{+0.5}_{-0.4}$ \ \ & \ \ $^{+1.8}_{-1.7}$ \ \ & \ \ $0.9$ \ \ & \ \ $0.0$ \ \ & \ \ $0.0$ \ \ & & & \ \ $1.01$ \ \ & \ \ $^{+0}_{-2}$ \ \ & \ \ $1.00$ \ \ \\ 
 \ \ $442$--$478$ \ \ & \ \ $1.13\cdot 10^{0}$ \ \ & \ \ $0.33$ \ \ & \ \ $0.38$ \ \ & \ \ $^{+4.6}_{-4.5}$ \ \ & \ \ $^{+0.5}_{-0.3}$ \ \ & \ \ $^{+0.3}_{-0.3}$ \ \ & \ \ $^{+1.5}_{-1.5}$ \ \ & \ \ $0.9$ \ \ & \ \ $0.0$ \ \ & \ \ $0.0$ \ \ & & & \ \ $1.01$ \ \ & \ \ $^{+0}_{-2}$ \ \ & \ \ $1.00$ \ \ \\ 
 \ \ $478$--$516$ \ \ & \ \ $6.77\cdot 10^{-1}$ \ \ & \ \ $0.27$ \ \ & \ \ $0.26$ \ \ & \ \ $^{+4.7}_{-4.7}$ \ \ & \ \ $^{+0.2}_{-0.1}$ \ \ & \ \ $^{+0.1}_{-0.1}$ \ \ & \ \ $^{+1.3}_{-1.3}$ \ \ & \ \ $0.8$ \ \ & \ \ $0.0$ \ \ & \ \ $0.0$ \ \ & & & \ \ $1.01$ \ \ & \ \ $^{+0}_{-2}$ \ \ & \ \ $1.00$ \ \ \\ 
 \ \ $516$--$556$ \ \ & \ \ $4.05\cdot 10^{-1}$ \ \ & \ \ $0.34$ \ \ & \ \ $0.26$ \ \ & \ \ $^{+5.1}_{-5.0}$ \ \ & \ \ $^{+0.1}_{-0.0}$ \ \ & \ \ $^{+0.0}_{-0.0}$ \ \ & \ \ $^{+1.1}_{-1.1}$ \ \ & \ \ $0.8$ \ \ & \ \ $0.0$ \ \ & \ \ $0.0$ \ \ & & & \ \ $1.01$ \ \ & \ \ $^{+0}_{-2}$ \ \ & \ \ $1.01$ \ \ \\ 
 \ \ $556$--$598$ \ \ & \ \ $2.41\cdot 10^{-1}$ \ \ & \ \ $0.43$ \ \ & \ \ $0.24$ \ \ & \ \ $^{+5.6}_{-5.5}$ \ \ & \ \ $^{+0.1}_{-0.1}$ \ \ & \ \ $^{+0.0}_{-0.0}$ \ \ & \ \ $^{+1.0}_{-1.0}$ \ \ & \ \ $0.8$ \ \ & \ \ $0.0$ \ \ & \ \ $0.0$ \ \ & & & \ \ $1.01$ \ \ & \ \ $^{+0}_{-1}$ \ \ & \ \ $1.01$ \ \ \\ 
 \ \ $598$--$642$ \ \ & \ \ $1.43\cdot 10^{-1}$ \ \ & \ \ $0.53$ \ \ & \ \ $0.21$ \ \ & \ \ $^{+6.1}_{-6.0}$ \ \ & \ \ $^{+0.1}_{-0.1}$ \ \ & \ \ $^{+0.0}_{-0.0}$ \ \ & \ \ $^{+0.8}_{-0.8}$ \ \ & \ \ $0.9$ \ \ & \ \ $0.0$ \ \ & \ \ $0.0$ \ \ & & & \ \ $1.00$ \ \ & \ \ $^{+0}_{-1}$ \ \ & \ \ $1.01$ \ \ \\ 
 \ \ $642$--$688$ \ \ & \ \ $8.58\cdot 10^{-2}$ \ \ & \ \ $0.69$ \ \ & \ \ $0.22$ \ \ & \ \ $^{+6.7}_{-6.6}$ \ \ & \ \ $^{+0.1}_{-0.1}$ \ \ & \ \ $^{+0.0}_{-0.0}$ \ \ & \ \ $^{+0.7}_{-0.7}$ \ \ & \ \ $0.9$ \ \ & \ \ $0.1$ \ \ & \ \ $0.0$ \ \ & & & \ \ $1.00$ \ \ & \ \ $^{+0}_{-1}$ \ \ & \ \ $1.01$ \ \ \\ 
 \ \ $688$--$736$ \ \ & \ \ $5.13\cdot 10^{-2}$ \ \ & \ \ $0.87$ \ \ & \ \ $0.22$ \ \ & \ \ $^{+7.3}_{-7.2}$ \ \ & \ \ $^{+0.1}_{-0.1}$ \ \ & \ \ $^{+0.0}_{-0.0}$ \ \ & \ \ $^{+0.7}_{-0.7}$ \ \ & \ \ $0.9$ \ \ & \ \ $0.1$ \ \ & \ \ $0.0$ \ \ & & & \ \ $1.00$ \ \ & \ \ $^{+0}_{-1}$ \ \ & \ \ $1.01$ \ \ \\ 
 \ \ $736$--$786$ \ \ & \ \ $3.03\cdot 10^{-2}$ \ \ & \ \ $1.1$ \ \ & \ \ $0.25$ \ \ & \ \ $^{+8.2}_{-8.1}$ \ \ & \ \ $^{+0.1}_{-0.1}$ \ \ & \ \ $^{+0.0}_{-0.0}$ \ \ & \ \ $^{+0.6}_{-0.6}$ \ \ & \ \ $0.9$ \ \ & \ \ $0.1$ \ \ & \ \ $0.0$ \ \ & & & \ \ $1.00$ \ \ & \ \ $^{+0}_{-1}$ \ \ & \ \ $1.02$ \ \ \\ 
 \ \ $786$--$838$ \ \ & \ \ $1.79\cdot 10^{-2}$ \ \ & \ \ $1.4$ \ \ & \ \ $0.24$ \ \ & \ \ $^{+9.0}_{-8.8}$ \ \ & \ \ $^{+0.1}_{-0.1}$ \ \ & \ \ $^{+0.0}_{-0.0}$ \ \ & \ \ $^{+0.6}_{-0.6}$ \ \ & \ \ $0.9$ \ \ & \ \ $0.1$ \ \ & \ \ $0.0$ \ \ & & & \ \ $1.00$ \ \ & \ \ $^{+0}_{-1}$ \ \ & \ \ $1.02$ \ \ \\ 
 \ \ $838$--$894$ \ \ & \ \ $1.08\cdot 10^{-2}$ \ \ & \ \ $1.7$ \ \ & \ \ $0.26$ \ \ & \ \ $^{+9.9}_{-9.6}$ \ \ & \ \ $^{+0.1}_{-0.1}$ \ \ & \ \ $^{+0.0}_{-0.0}$ \ \ & \ \ $^{+0.6}_{-0.6}$ \ \ & \ \ $1.0$ \ \ & \ \ $0.1$ \ \ & \ \ $0.0$ \ \ & & & \ \ $1.00$ \ \ & \ \ $^{+0}_{-1}$ \ \ & \ \ $1.02$ \ \ \\ 
 \ \ $894$--$952$ \ \ & \ \ $6.00\cdot 10^{-3}$ \ \ & \ \ $2.3$ \ \ & \ \ $0.28$ \ \ & \ \ $^{+11}_{-10}$ \ \ & \ \ $^{+0.1}_{-0.1}$ \ \ & \ \ $^{+0.0}_{-0.0}$ \ \ & \ \ $^{+0.7}_{-0.7}$ \ \ & \ \ $1.1$ \ \ & \ \ $0.1$ \ \ & \ \ $0.0$ \ \ & & & \ \ $1.00$ \ \ & \ \ $^{+0}_{-1}$ \ \ & \ \ $1.03$ \ \ \\ 
 \ \ $952$--$1012$ \ \ & \ \ $3.38\cdot 10^{-3}$ \ \ & \ \ $3.1$ \ \ & \ \ $0.35$ \ \ & \ \ $^{+11}_{-11}$ \ \ & \ \ $^{+0.1}_{-0.1}$ \ \ & \ \ $^{+0.0}_{-0.0}$ \ \ & \ \ $^{+0.7}_{-0.7}$ \ \ & \ \ $1.2$ \ \ & \ \ $0.1$ \ \ & \ \ $0.1$ \ \ & & & \ \ $1.00$ \ \ & \ \ $^{+0}_{-1}$ \ \ & \ \ $1.03$ \ \ \\ 
 \ \ $1012$--$1162$ \ \ & \ \ $1.23\cdot 10^{-3}$ \ \ & \ \ $3.7$ \ \ & \ \ $0.26$ \ \ & \ \ $^{+13}_{-12}$ \ \ & \ \ $^{+0.1}_{-0.1}$ \ \ & \ \ $^{+0.0}_{-0.0}$ \ \ & \ \ $^{+0.9}_{-0.9}$ \ \ & \ \ $1.3$ \ \ & \ \ $0.1$ \ \ & \ \ $0.1$ \ \ & & & \ \ $1.00$ \ \ & \ \ $^{+0}_{-1}$ \ \ & \ \ $1.04$ \ \ \\ 
 \ \ $1162$--$1310$ \ \ & \ \ $2.86\cdot 10^{-4}$ \ \ & \ \ $7.7$ \ \ & \ \ $0.33$ \ \ & \ \ $^{+15}_{-14}$ \ \ & \ \ $^{+0.1}_{-0.1}$ \ \ & \ \ $^{+0.0}_{-0.0}$ \ \ & \ \ $^{+1.0}_{-1.0}$ \ \ & \ \ $1.4$ \ \ & \ \ $0.1$ \ \ & \ \ $1.0$ \ \ & & & \ \ $1.00$ \ \ & \ \ $^{+0}_{-1}$ \ \ & \ \ $1.05$ \ \ \\ 
 \ \ $1310$--$1992$ \ \ & \ \ $2.39\cdot 10^{-5}$ \ \ & \ \ $14$ \ \ & \ \ $0.34$ \ \ & \ \ $^{+29}_{-28}$ \ \ & \ \ $^{+0.0}_{-0.1}$ \ \ & \ \ $^{+0.0}_{-0.0}$ \ \ & \ \ $^{+1.1}_{-1.1}$ \ \ & \ \ $2.8$ \ \ & \ \ $0.1$ \ \ & \ \ $4.4$ \ \ & & & \ \ $1.00$ \ \ & \ \ $^{+0}_{-1}$ \ \ & \ \ $1.06$ \ \ \\ 
\hline
\end{tabular}
}
\caption{
Measured double-differential inclusive jet cross-sections for jets with $R=0.6$ in the rapidity bin $0.5\leq|y| < 1.0$. See caption of table~\ref{tab:xsecr06y0} for details.
}
\label{tab:xsecr06y1}
\end{table*}

%% file: sys_R06_Eta3.tex
\renewcommand{\arraystretch}{1.8}
\begin{table*}[!ht]
\tiny
\centering
{\fontfamily{ptm}\selectfont
\begin{tabular}{|@{}c@{}@{}c@{}@{}c@{}@{}r@{}@{}r@{}@{}r@{}@{}r@{}@{}r@{}@{}r@{}@{}r@{}@{}r@{}|@{}r@{}@{}r@{}|@{}r@{}@{}r@{}@{}r@{}|}
\hline
 \ \ \pt range \ \ &
 \ \ $\sigma$ \ \ &
 \ \ $\delta_{\mathrm{stat}}^{\mathrm{data}}$ \ \ &
 \ \ $\delta_{\mathrm{stat}}^{\mathrm{MC}}$ \ \ &
 \ \ $u_{\operatorname{in-situ}}$ \ \ &
 \ \ $u_{\operatorname{pile-up}}$ \ \ &
 \ \ $u_{\operatorname{close-by}}$ \ \ &
\ \ $u_{\mathrm{flavour}}$ \ \ &
 \ \ $u_{\mathrm{JER}}$ \ \ &
 \ \ $u_{\mathrm{JAR}}$ \ \ &
 \ \ $u_{\mathrm{unfold}}$ \ \ &
 \ \ $u_{\mathrm{qual.}}$ \ \ &
 \ \ $u_{\mathrm{lumi}}$ \ \ &
 \ \ NPC \ \ &
 \ \ $u_{\mathrm{NP}}$ \ \ &
 \ \ EWC \ \ \\ \relax
\ \ [GeV] \ \ & \ \ [pb/GeV] \ \ &\ \  \% \ \ & \ \ \% \ \ & \ \  \% \ \ &
 \ \ \% \ \ & \ \ \% \ \ & \ \ \% \ \ & \ \ \% \ \ &
 \ \ \% \ \ & \ \ \% \ \ & \ \ \% \ \ & \ \ \% \ \ & \ \ \ \ \ & \ \ \% \ \ & \ \ \ \ \ \\
\hline
 \ \ $100$--$116$ \ \ & \ \ $4.20\cdot 10^{3}$ \ \ & \ \ $0.72$ \ \ & \ \ $0.81$ \ \ & \ \ $^{+7.6}_{-7.5}$ \ \ & \ \ $^{+1.9}_{-1.9}$ \ \ & \ \ $^{+1.4}_{-1.5}$ \ \ & \ \ $^{+6.8}_{-6.5}$ \ \ & \ \ $7.4$ \ \ & \ \ $0.5$ \ \ & \ \ $0.4$ \ \ & \multirow{26}{*}{\ \ $0.25$ \ \ } & \multirow{26}{*}{\ \ $1.8$ \ \ } & \ \ $1.06$ \ \ & \ \ $^{+1}_{-5}$ \ \ & \ \ $1.00$ \ \ \\ 
 \ \ $116$--$134$ \ \ & \ \ $1.95\cdot 10^{3}$ \ \ & \ \ $0.84$ \ \ & \ \ $0.64$ \ \ & \ \ $^{+7.7}_{-7.4}$ \ \ & \ \ $^{+1.7}_{-1.5}$ \ \ & \ \ $^{+1.3}_{-1.2}$ \ \ & \ \ $^{+6.2}_{-5.9}$ \ \ & \ \ $5.9$ \ \ & \ \ $0.3$ \ \ & \ \ $0.3$ \ \ & & & \ \ $1.05$ \ \ & \ \ $^{+1}_{-5}$ \ \ & \ \ $1.00$ \ \ \\ 
 \ \ $134$--$152$ \ \ & \ \ $9.37\cdot 10^{2}$ \ \ & \ \ $0.86$ \ \ & \ \ $0.46$ \ \ & \ \ $^{+7.5}_{-7.2}$ \ \ & \ \ $^{+1.5}_{-1.3}$ \ \ & \ \ $^{+1.1}_{-1.0}$ \ \ & \ \ $^{+5.6}_{-5.3}$ \ \ & \ \ $4.6$ \ \ & \ \ $0.3$ \ \ & \ \ $0.0$ \ \ & & & \ \ $1.04$ \ \ & \ \ $^{+1}_{-4}$ \ \ & \ \ $1.00$ \ \ \\ 
 \ \ $152$--$172$ \ \ & \ \ $4.64\cdot 10^{2}$ \ \ & \ \ $0.68$ \ \ & \ \ $0.39$ \ \ & \ \ $^{+7.4}_{-7.1}$ \ \ & \ \ $^{+1.3}_{-1.2}$ \ \ & \ \ $^{+1.0}_{-0.9}$ \ \ & \ \ $^{+5.1}_{-4.8}$ \ \ & \ \ $3.8$ \ \ & \ \ $0.2$ \ \ & \ \ $0.0$ \ \ & & & \ \ $1.03$ \ \ & \ \ $^{+1}_{-4}$ \ \ & \ \ $1.00$ \ \ \\ 
 \ \ $172$--$194$ \ \ & \ \ $2.32\cdot 10^{2}$ \ \ & \ \ $0.56$ \ \ & \ \ $0.40$ \ \ & \ \ $^{+7.4}_{-7.0}$ \ \ & \ \ $^{+1.3}_{-1.2}$ \ \ & \ \ $^{+0.9}_{-0.8}$ \ \ & \ \ $^{+4.6}_{-4.4}$ \ \ & \ \ $3.3$ \ \ & \ \ $0.2$ \ \ & \ \ $0.0$ \ \ & & & \ \ $1.03$ \ \ & \ \ $^{+1}_{-3}$ \ \ & \ \ $1.00$ \ \ \\ 
 \ \ $194$--$216$ \ \ & \ \ $1.20\cdot 10^{2}$ \ \ & \ \ $0.71$ \ \ & \ \ $0.44$ \ \ & \ \ $^{+7.3}_{-7.0}$ \ \ & \ \ $^{+1.1}_{-1.1}$ \ \ & \ \ $^{+1.0}_{-0.9}$ \ \ & \ \ $^{+4.2}_{-4.0}$ \ \ & \ \ $2.9$ \ \ & \ \ $0.2$ \ \ & \ \ $0.0$ \ \ & & & \ \ $1.02$ \ \ & \ \ $^{+1}_{-3}$ \ \ & \ \ $1.00$ \ \ \\ 
 \ \ $216$--$240$ \ \ & \ \ $6.36\cdot 10^{1}$ \ \ & \ \ $0.55$ \ \ & \ \ $0.44$ \ \ & \ \ $^{+7.0}_{-7.0}$ \ \ & \ \ $^{+1.0}_{-1.0}$ \ \ & \ \ $^{+1.0}_{-1.1}$ \ \ & \ \ $^{+3.8}_{-3.7}$ \ \ & \ \ $2.6$ \ \ & \ \ $0.1$ \ \ & \ \ $0.0$ \ \ & & & \ \ $1.02$ \ \ & \ \ $^{+1}_{-3}$ \ \ & \ \ $1.00$ \ \ \\ 
 \ \ $240$--$264$ \ \ & \ \ $3.51\cdot 10^{1}$ \ \ & \ \ $0.70$ \ \ & \ \ $0.42$ \ \ & \ \ $^{+6.9}_{-6.7}$ \ \ & \ \ $^{+1.0}_{-0.9}$ \ \ & \ \ $^{+1.1}_{-1.1}$ \ \ & \ \ $^{+3.5}_{-3.4}$ \ \ & \ \ $2.5$ \ \ & \ \ $0.1$ \ \ & \ \ $0.0$ \ \ & & & \ \ $1.02$ \ \ & \ \ $^{+1}_{-3}$ \ \ & \ \ $1.00$ \ \ \\ 
 \ \ $264$--$290$ \ \ & \ \ $1.98\cdot 10^{1}$ \ \ & \ \ $0.93$ \ \ & \ \ $0.40$ \ \ & \ \ $^{+6.7}_{-6.3}$ \ \ & \ \ $^{+1.0}_{-0.9}$ \ \ & \ \ $^{+1.1}_{-1.1}$ \ \ & \ \ $^{+3.2}_{-3.1}$ \ \ & \ \ $2.4$ \ \ & \ \ $0.2$ \ \ & \ \ $0.0$ \ \ & & & \ \ $1.02$ \ \ & \ \ $^{+1}_{-2}$ \ \ & \ \ $1.00$ \ \ \\ 
 \ \ $290$--$318$ \ \ & \ \ $1.12\cdot 10^{1}$ \ \ & \ \ $1.2$ \ \ & \ \ $0.52$ \ \ & \ \ $^{+6.5}_{-6.1}$ \ \ & \ \ $^{+1.0}_{-0.9}$ \ \ & \ \ $^{+1.1}_{-1.1}$ \ \ & \ \ $^{+2.9}_{-2.8}$ \ \ & \ \ $2.4$ \ \ & \ \ $0.1$ \ \ & \ \ $0.0$ \ \ & & & \ \ $1.02$ \ \ & \ \ $^{+1}_{-2}$ \ \ & \ \ $1.00$ \ \ \\ 
 \ \ $318$--$346$ \ \ & \ \ $6.37\cdot 10^{0}$ \ \ & \ \ $1.1$ \ \ & \ \ $0.57$ \ \ & \ \ $^{+6.4}_{-6.1}$ \ \ & \ \ $^{+1.1}_{-1.0}$ \ \ & \ \ $^{+1.0}_{-1.0}$ \ \ & \ \ $^{+2.6}_{-2.5}$ \ \ & \ \ $2.3$ \ \ & \ \ $0.1$ \ \ & \ \ $0.0$ \ \ & & & \ \ $1.01$ \ \ & \ \ $^{+1}_{-2}$ \ \ & \ \ $1.00$ \ \ \\ 
 \ \ $346$--$376$ \ \ & \ \ $3.71\cdot 10^{0}$ \ \ & \ \ $0.59$ \ \ & \ \ $0.65$ \ \ & \ \ $^{+6.3}_{-5.9}$ \ \ & \ \ $^{+1.2}_{-1.1}$ \ \ & \ \ $^{+0.9}_{-0.9}$ \ \ & \ \ $^{+2.3}_{-2.2}$ \ \ & \ \ $2.2$ \ \ & \ \ $0.1$ \ \ & \ \ $0.0$ \ \ & & & \ \ $1.01$ \ \ & \ \ $^{+1}_{-2}$ \ \ & \ \ $1.00$ \ \ \\ 
 \ \ $376$--$408$ \ \ & \ \ $2.15\cdot 10^{0}$ \ \ & \ \ $0.35$ \ \ & \ \ $0.52$ \ \ & \ \ $^{+6.1}_{-5.9}$ \ \ & \ \ $^{+1.2}_{-1.0}$ \ \ & \ \ $^{+0.7}_{-0.7}$ \ \ & \ \ $^{+2.0}_{-1.9}$ \ \ & \ \ $2.2$ \ \ & \ \ $0.1$ \ \ & \ \ $0.0$ \ \ & & & \ \ $1.01$ \ \ & \ \ $^{+0}_{-2}$ \ \ & \ \ $1.00$ \ \ \\ 
 \ \ $408$--$442$ \ \ & \ \ $1.23\cdot 10^{0}$ \ \ & \ \ $0.39$ \ \ & \ \ $0.50$ \ \ & \ \ $^{+5.8}_{-5.8}$ \ \ & \ \ $^{+0.9}_{-0.8}$ \ \ & \ \ $^{+0.5}_{-0.5}$ \ \ & \ \ $^{+1.8}_{-1.7}$ \ \ & \ \ $2.1$ \ \ & \ \ $0.1$ \ \ & \ \ $0.0$ \ \ & & & \ \ $1.01$ \ \ & \ \ $^{+0}_{-2}$ \ \ & \ \ $1.00$ \ \ \\ 
 \ \ $442$--$478$ \ \ & \ \ $7.13\cdot 10^{-1}$ \ \ & \ \ $0.41$ \ \ & \ \ $0.40$ \ \ & \ \ $^{+6.0}_{-5.7}$ \ \ & \ \ $^{+0.5}_{-0.4}$ \ \ & \ \ $^{+0.3}_{-0.3}$ \ \ & \ \ $^{+1.6}_{-1.5}$ \ \ & \ \ $2.2$ \ \ & \ \ $0.1$ \ \ & \ \ $0.0$ \ \ & & & \ \ $1.01$ \ \ & \ \ $^{+0}_{-2}$ \ \ & \ \ $1.00$ \ \ \\ 
 \ \ $478$--$516$ \ \ & \ \ $4.06\cdot 10^{-1}$ \ \ & \ \ $0.35$ \ \ & \ \ $0.32$ \ \ & \ \ $^{+6.2}_{-5.8}$ \ \ & \ \ $^{+0.2}_{-0.1}$ \ \ & \ \ $^{+0.1}_{-0.1}$ \ \ & \ \ $^{+1.5}_{-1.4}$ \ \ & \ \ $2.2$ \ \ & \ \ $0.1$ \ \ & \ \ $0.0$ \ \ & & & \ \ $1.01$ \ \ & \ \ $^{+0}_{-2}$ \ \ & \ \ $1.00$ \ \ \\ 
 \ \ $516$--$556$ \ \ & \ \ $2.29\cdot 10^{-1}$ \ \ & \ \ $0.44$ \ \ & \ \ $0.30$ \ \ & \ \ $^{+6.6}_{-6.2}$ \ \ & \ \ $^{+0.1}_{-0.1}$ \ \ & \ \ $^{+0.0}_{-0.0}$ \ \ & \ \ $^{+1.4}_{-1.3}$ \ \ & \ \ $2.3$ \ \ & \ \ $0.1$ \ \ & \ \ $0.0$ \ \ & & & \ \ $1.01$ \ \ & \ \ $^{+0}_{-2}$ \ \ & \ \ $1.00$ \ \ \\ 
 \ \ $556$--$598$ \ \ & \ \ $1.29\cdot 10^{-1}$ \ \ & \ \ $0.58$ \ \ & \ \ $0.32$ \ \ & \ \ $^{+7.0}_{-6.6}$ \ \ & \ \ $^{+0.1}_{-0.1}$ \ \ & \ \ $^{+0.0}_{-0.0}$ \ \ & \ \ $^{+1.2}_{-1.2}$ \ \ & \ \ $2.4$ \ \ & \ \ $0.1$ \ \ & \ \ $0.0$ \ \ & & & \ \ $1.01$ \ \ & \ \ $^{+0}_{-2}$ \ \ & \ \ $1.00$ \ \ \\ 
 \ \ $598$--$642$ \ \ & \ \ $7.17\cdot 10^{-2}$ \ \ & \ \ $0.75$ \ \ & \ \ $0.28$ \ \ & \ \ $^{+7.6}_{-7.4}$ \ \ & \ \ $^{+0.1}_{-0.1}$ \ \ & \ \ $^{+0.0}_{-0.0}$ \ \ & \ \ $^{+1.1}_{-1.1}$ \ \ & \ \ $2.4$ \ \ & \ \ $0.1$ \ \ & \ \ $0.0$ \ \ & & & \ \ $1.01$ \ \ & \ \ $^{+0}_{-2}$ \ \ & \ \ $1.00$ \ \ \\ 
 \ \ $642$--$688$ \ \ & \ \ $3.90\cdot 10^{-2}$ \ \ & \ \ $0.98$ \ \ & \ \ $0.26$ \ \ & \ \ $^{+8.3}_{-8.2}$ \ \ & \ \ $^{+0.1}_{-0.1}$ \ \ & \ \ $^{+0.0}_{-0.0}$ \ \ & \ \ $^{+1.0}_{-1.0}$ \ \ & \ \ $2.5$ \ \ & \ \ $0.1$ \ \ & \ \ $0.0$ \ \ & & & \ \ $1.01$ \ \ & \ \ $^{+0}_{-2}$ \ \ & \ \ $1.00$ \ \ \\ 
 \ \ $688$--$736$ \ \ & \ \ $2.13\cdot 10^{-2}$ \ \ & \ \ $1.3$ \ \ & \ \ $0.28$ \ \ & \ \ $^{+9.3}_{-9.1}$ \ \ & \ \ $^{+0.1}_{-0.1}$ \ \ & \ \ $^{+0.0}_{-0.0}$ \ \ & \ \ $^{+1.0}_{-1.0}$ \ \ & \ \ $2.6$ \ \ & \ \ $0.1$ \ \ & \ \ $0.0$ \ \ & & & \ \ $1.01$ \ \ & \ \ $^{+0}_{-2}$ \ \ & \ \ $1.00$ \ \ \\ 
 \ \ $736$--$786$ \ \ & \ \ $1.17\cdot 10^{-2}$ \ \ & \ \ $1.7$ \ \ & \ \ $0.30$ \ \ & \ \ $^{+10}_{-10}$ \ \ & \ \ $^{+0.1}_{-0.1}$ \ \ & \ \ $^{+0.0}_{-0.0}$ \ \ & \ \ $^{+0.9}_{-1.0}$ \ \ & \ \ $2.8$ \ \ & \ \ $0.1$ \ \ & \ \ $0.1$ \ \ & & & \ \ $1.01$ \ \ & \ \ $^{+0}_{-2}$ \ \ & \ \ $1.00$ \ \ \\ 
 \ \ $786$--$838$ \ \ & \ \ $5.96\cdot 10^{-3}$ \ \ & \ \ $2.3$ \ \ & \ \ $0.34$ \ \ & \ \ $^{+11}_{-11}$ \ \ & \ \ $^{+0.1}_{-0.1}$ \ \ & \ \ $^{+0.0}_{-0.0}$ \ \ & \ \ $^{+0.9}_{-1.0}$ \ \ & \ \ $3.0$ \ \ & \ \ $0.1$ \ \ & \ \ $0.1$ \ \ & & & \ \ $1.01$ \ \ & \ \ $^{+0}_{-2}$ \ \ & \ \ $1.00$ \ \ \\ 
 \ \ $838$--$894$ \ \ & \ \ $3.05\cdot 10^{-3}$ \ \ & \ \ $3.2$ \ \ & \ \ $0.45$ \ \ & \ \ $^{+12}_{-12}$ \ \ & \ \ $^{+0.1}_{-0.1}$ \ \ & \ \ $^{+0.0}_{-0.0}$ \ \ & \ \ $^{+1.0}_{-1.0}$ \ \ & \ \ $3.1$ \ \ & \ \ $0.1$ \ \ & \ \ $0.9$ \ \ & & & \ \ $1.01$ \ \ & \ \ $^{+0}_{-2}$ \ \ & \ \ $1.01$ \ \ \\ 
 \ \ $894$--$1012$ \ \ & \ \ $1.06\cdot 10^{-3}$ \ \ & \ \ $4.3$ \ \ & \ \ $0.33$ \ \ & \ \ $^{+14}_{-13}$ \ \ & \ \ $^{+0.1}_{-0.1}$ \ \ & \ \ $^{+0.0}_{-0.0}$ \ \ & \ \ $^{+1.0}_{-1.0}$ \ \ & \ \ $3.4$ \ \ & \ \ $0.1$ \ \ & \ \ $2.4$ \ \ & & & \ \ $1.01$ \ \ & \ \ $^{+0}_{-2}$ \ \ & \ \ $1.01$ \ \ \\ 
 \ \ $1012$--$1992$ \ \ & \ \ $3.59\cdot 10^{-5}$ \ \ & \ \ $8.8$ \ \ & \ \ $0.50$ \ \ & \ \ $^{+19}_{-18}$ \ \ & \ \ $^{+0.1}_{-0.2}$ \ \ & \ \ $^{+0.0}_{-0.0}$ \ \ & \ \ $^{+1.5}_{-1.7}$ \ \ & \ \ $4.4$ \ \ & \ \ $0.1$ \ \ & \ \ $2.0$ \ \ & & & \ \ $1.01$ \ \ & \ \ $^{+0}_{-2}$ \ \ & \ \ $1.01$ \ \ \\ 
\hline
\end{tabular}
}
\caption{
Measured double-differential inclusive jet cross-sections for jets with $R=0.6$ in the rapidity bin $1.0\leq|y| < 1.5$. See caption of table~\ref{tab:xsecr06y0} for details.
}
\label{tab:xsecr06y2}
\end{table*}

%% file: sys_R06_Eta4.tex
\renewcommand{\arraystretch}{1.8}
\begin{table*}[!ht]
\tiny
\centering
{\fontfamily{ptm}\selectfont
\begin{tabular}{|@{}c@{}@{}c@{}@{}c@{}@{}r@{}@{}r@{}@{}r@{}@{}r@{}@{}r@{}@{}r@{}@{}r@{}@{}r@{}|@{}r@{}@{}r@{}|@{}r@{}@{}r@{}@{}r@{}|}
\hline
 \ \ \pt range \ \ &
 \ \ $\sigma$ \ \ &
 \ \ $\delta_{\mathrm{stat}}^{\mathrm{data}}$ \ \ &
 \ \ $\delta_{\mathrm{stat}}^{\mathrm{MC}}$ \ \ &
 \ \ $u_{\operatorname{in-situ}}$ \ \ &
 \ \ $u_{\operatorname{pile-up}}$ \ \ &
 \ \ $u_{\operatorname{close-by}}$ \ \ &
\ \ $u_{\mathrm{flavour}}$ \ \ &
 \ \ $u_{\mathrm{JER}}$ \ \ &
 \ \ $u_{\mathrm{JAR}}$ \ \ &
 \ \ $u_{\mathrm{unfold}}$ \ \ &
 \ \ $u_{\mathrm{qual.}}$ \ \ &
 \ \ $u_{\mathrm{lumi}}$ \ \ &
 \ \ NPC \ \ &
 \ \ $u_{\mathrm{NP}}$ \ \ &
 \ \ EWC \ \ \\ \relax
\ \ [GeV] \ \ & \ \ [pb/GeV] \ \ &\ \  \% \ \ & \ \ \% \ \ & \ \  \% \ \ &
 \ \ \% \ \ & \ \ \% \ \ & \ \ \% \ \ & \ \ \% \ \ &
 \ \ \% \ \ & \ \ \% \ \ & \ \ \% \ \ & \ \ \% \ \ & \ \ \ \ \ & \ \ \% \ \ & \ \ \ \ \ \\
\hline
 \ \ $100$--$116$ \ \ & \ \ $3.25\cdot 10^{3}$ \ \ & \ \ $0.63$ \ \ & \ \ $0.99$ \ \ & \ \ $^{+11}_{-10}$ \ \ & \ \ $^{+2.0}_{-2.0}$ \ \ & \ \ $^{+1.6}_{-1.4}$ \ \ & \ \ $^{+6.9}_{-6.6}$ \ \ & \ \ $9.1$ \ \ & \ \ $1.0$ \ \ & \ \ $0.4$ \ \ & \multirow{23}{*}{\ \ $0.25$ \ \ } & \multirow{23}{*}{\ \ $1.8$ \ \ } & \ \ $1.05$ \ \ & \ \ $^{+1}_{-5}$ \ \ & \ \ $1.00$ \ \ \\ 
 \ \ $116$--$134$ \ \ & \ \ $1.47\cdot 10^{3}$ \ \ & \ \ $0.87$ \ \ & \ \ $0.66$ \ \ & \ \ $^{+11}_{-10}$ \ \ & \ \ $^{+1.8}_{-1.5}$ \ \ & \ \ $^{+1.5}_{-1.2}$ \ \ & \ \ $^{+6.1}_{-5.6}$ \ \ & \ \ $7.1$ \ \ & \ \ $0.7$ \ \ & \ \ $0.2$ \ \ & & & \ \ $1.04$ \ \ & \ \ $^{+2}_{-4}$ \ \ & \ \ $1.00$ \ \ \\ 
 \ \ $134$--$152$ \ \ & \ \ $6.91\cdot 10^{2}$ \ \ & \ \ $1.1$ \ \ & \ \ $0.54$ \ \ & \ \ $^{+12}_{-10}$ \ \ & \ \ $^{+1.8}_{-1.3}$ \ \ & \ \ $^{+1.2}_{-1.0}$ \ \ & \ \ $^{+5.6}_{-4.9}$ \ \ & \ \ $5.9$ \ \ & \ \ $0.6$ \ \ & \ \ $0.0$ \ \ & & & \ \ $1.03$ \ \ & \ \ $^{+2}_{-4}$ \ \ & \ \ $1.00$ \ \ \\ 
 \ \ $152$--$172$ \ \ & \ \ $3.42\cdot 10^{2}$ \ \ & \ \ $0.81$ \ \ & \ \ $0.43$ \ \ & \ \ $^{+12}_{-11}$ \ \ & \ \ $^{+1.6}_{-1.3}$ \ \ & \ \ $^{+1.0}_{-1.0}$ \ \ & \ \ $^{+5.0}_{-4.6}$ \ \ & \ \ $5.2$ \ \ & \ \ $0.5$ \ \ & \ \ $0.0$ \ \ & & & \ \ $1.03$ \ \ & \ \ $^{+2}_{-3}$ \ \ & \ \ $1.00$ \ \ \\ 
 \ \ $172$--$194$ \ \ & \ \ $1.66\cdot 10^{2}$ \ \ & \ \ $0.67$ \ \ & \ \ $0.45$ \ \ & \ \ $^{+12}_{-11}$ \ \ & \ \ $^{+1.4}_{-1.3}$ \ \ & \ \ $^{+1.0}_{-1.0}$ \ \ & \ \ $^{+4.4}_{-4.1}$ \ \ & \ \ $4.5$ \ \ & \ \ $0.4$ \ \ & \ \ $0.0$ \ \ & & & \ \ $1.03$ \ \ & \ \ $^{+2}_{-3}$ \ \ & \ \ $1.00$ \ \ \\ 
 \ \ $194$--$216$ \ \ & \ \ $8.16\cdot 10^{1}$ \ \ & \ \ $0.92$ \ \ & \ \ $0.47$ \ \ & \ \ $^{+11}_{-11}$ \ \ & \ \ $^{+1.3}_{-1.2}$ \ \ & \ \ $^{+1.0}_{-1.0}$ \ \ & \ \ $^{+3.8}_{-3.6}$ \ \ & \ \ $3.9$ \ \ & \ \ $0.3$ \ \ & \ \ $0.0$ \ \ & & & \ \ $1.02$ \ \ & \ \ $^{+1}_{-3}$ \ \ & \ \ $1.00$ \ \ \\ 
 \ \ $216$--$240$ \ \ & \ \ $4.28\cdot 10^{1}$ \ \ & \ \ $0.69$ \ \ & \ \ $0.43$ \ \ & \ \ $^{+11}_{-10}$ \ \ & \ \ $^{+1.1}_{-1.1}$ \ \ & \ \ $^{+1.1}_{-1.1}$ \ \ & \ \ $^{+3.4}_{-3.3}$ \ \ & \ \ $3.8$ \ \ & \ \ $0.3$ \ \ & \ \ $0.0$ \ \ & & & \ \ $1.02$ \ \ & \ \ $^{+1}_{-2}$ \ \ & \ \ $1.00$ \ \ \\ 
 \ \ $240$--$264$ \ \ & \ \ $2.28\cdot 10^{1}$ \ \ & \ \ $0.90$ \ \ & \ \ $0.42$ \ \ & \ \ $^{+11}_{-10}$ \ \ & \ \ $^{+1.1}_{-1.0}$ \ \ & \ \ $^{+1.2}_{-1.1}$ \ \ & \ \ $^{+3.2}_{-3.1}$ \ \ & \ \ $3.7$ \ \ & \ \ $0.3$ \ \ & \ \ $0.0$ \ \ & & & \ \ $1.02$ \ \ & \ \ $^{+1}_{-2}$ \ \ & \ \ $1.00$ \ \ \\ 
 \ \ $264$--$290$ \ \ & \ \ $1.23\cdot 10^{1}$ \ \ & \ \ $1.2$ \ \ & \ \ $0.50$ \ \ & \ \ $^{+11}_{-9.6}$ \ \ & \ \ $^{+1.1}_{-1.0}$ \ \ & \ \ $^{+1.3}_{-1.1}$ \ \ & \ \ $^{+3.1}_{-2.9}$ \ \ & \ \ $3.7$ \ \ & \ \ $0.3$ \ \ & \ \ $0.0$ \ \ & & & \ \ $1.02$ \ \ & \ \ $^{+1}_{-2}$ \ \ & \ \ $1.00$ \ \ \\ 
 \ \ $290$--$318$ \ \ & \ \ $6.38\cdot 10^{0}$ \ \ & \ \ $1.6$ \ \ & \ \ $0.52$ \ \ & \ \ $^{+10}_{-9.3}$ \ \ & \ \ $^{+1.2}_{-1.1}$ \ \ & \ \ $^{+1.2}_{-1.1}$ \ \ & \ \ $^{+2.9}_{-2.7}$ \ \ & \ \ $3.6$ \ \ & \ \ $0.3$ \ \ & \ \ $0.1$ \ \ & & & \ \ $1.02$ \ \ & \ \ $^{+1}_{-2}$ \ \ & \ \ $1.00$ \ \ \\ 
 \ \ $318$--$346$ \ \ & \ \ $3.48\cdot 10^{0}$ \ \ & \ \ $1.6$ \ \ & \ \ $0.66$ \ \ & \ \ $^{+9.9}_{-9.3}$ \ \ & \ \ $^{+1.2}_{-1.2}$ \ \ & \ \ $^{+1.1}_{-1.0}$ \ \ & \ \ $^{+2.6}_{-2.6}$ \ \ & \ \ $3.7$ \ \ & \ \ $0.3$ \ \ & \ \ $0.0$ \ \ & & & \ \ $1.02$ \ \ & \ \ $^{+1}_{-2}$ \ \ & \ \ $1.00$ \ \ \\ 
 \ \ $346$--$376$ \ \ & \ \ $1.93\cdot 10^{0}$ \ \ & \ \ $0.84$ \ \ & \ \ $0.78$ \ \ & \ \ $^{+9.7}_{-9.5}$ \ \ & \ \ $^{+1.3}_{-1.2}$ \ \ & \ \ $^{+0.9}_{-0.9}$ \ \ & \ \ $^{+2.3}_{-2.4}$ \ \ & \ \ $3.9$ \ \ & \ \ $0.3$ \ \ & \ \ $0.0$ \ \ & & & \ \ $1.02$ \ \ & \ \ $^{+0}_{-2}$ \ \ & \ \ $1.00$ \ \ \\ 
 \ \ $376$--$408$ \ \ & \ \ $1.02\cdot 10^{0}$ \ \ & \ \ $0.50$ \ \ & \ \ $0.80$ \ \ & \ \ $^{+9.7}_{-9.1}$ \ \ & \ \ $^{+1.3}_{-1.2}$ \ \ & \ \ $^{+0.7}_{-0.8}$ \ \ & \ \ $^{+2.2}_{-2.2}$ \ \ & \ \ $4.0$ \ \ & \ \ $0.3$ \ \ & \ \ $0.0$ \ \ & & & \ \ $1.02$ \ \ & \ \ $^{+0}_{-2}$ \ \ & \ \ $0.99$ \ \ \\ 
 \ \ $408$--$442$ \ \ & \ \ $5.31\cdot 10^{-1}$ \ \ & \ \ $0.63$ \ \ & \ \ $0.67$ \ \ & \ \ $^{+9.7}_{-8.8}$ \ \ & \ \ $^{+1.1}_{-1.0}$ \ \ & \ \ $^{+0.5}_{-0.5}$ \ \ & \ \ $^{+2.1}_{-2.0}$ \ \ & \ \ $4.0$ \ \ & \ \ $0.3$ \ \ & \ \ $0.0$ \ \ & & & \ \ $1.02$ \ \ & \ \ $^{+0}_{-2}$ \ \ & \ \ $0.99$ \ \ \\ 
 \ \ $442$--$478$ \ \ & \ \ $2.74\cdot 10^{-1}$ \ \ & \ \ $0.69$ \ \ & \ \ $0.59$ \ \ & \ \ $^{+10}_{-8.9}$ \ \ & \ \ $^{+0.7}_{-0.6}$ \ \ & \ \ $^{+0.3}_{-0.3}$ \ \ & \ \ $^{+2.0}_{-1.9}$ \ \ & \ \ $4.2$ \ \ & \ \ $0.3$ \ \ & \ \ $0.0$ \ \ & & & \ \ $1.02$ \ \ & \ \ $^{+0}_{-2}$ \ \ & \ \ $0.99$ \ \ \\ 
 \ \ $478$--$516$ \ \ & \ \ $1.37\cdot 10^{-1}$ \ \ & \ \ $0.60$ \ \ & \ \ $0.52$ \ \ & \ \ $^{+10}_{-9.3}$ \ \ & \ \ $^{+0.3}_{-0.2}$ \ \ & \ \ $^{+0.1}_{-0.1}$ \ \ & \ \ $^{+1.9}_{-1.8}$ \ \ & \ \ $4.5$ \ \ & \ \ $0.3$ \ \ & \ \ $0.0$ \ \ & & & \ \ $1.02$ \ \ & \ \ $^{+0}_{-2}$ \ \ & \ \ $0.99$ \ \ \\ 
 \ \ $516$--$556$ \ \ & \ \ $6.73\cdot 10^{-2}$ \ \ & \ \ $0.81$ \ \ & \ \ $0.58$ \ \ & \ \ $^{+11}_{-9.7}$ \ \ & \ \ $^{+0.1}_{-0.1}$ \ \ & \ \ $^{+0.0}_{-0.0}$ \ \ & \ \ $^{+1.8}_{-1.7}$ \ \ & \ \ $5.0$ \ \ & \ \ $0.3$ \ \ & \ \ $0.1$ \ \ & & & \ \ $1.02$ \ \ & \ \ $^{+0}_{-2}$ \ \ & \ \ $0.99$ \ \ \\ 
 \ \ $556$--$598$ \ \ & \ \ $3.17\cdot 10^{-2}$ \ \ & \ \ $1.2$ \ \ & \ \ $0.57$ \ \ & \ \ $^{+11}_{-11}$ \ \ & \ \ $^{+0.2}_{-0.1}$ \ \ & \ \ $^{+0.0}_{-0.0}$ \ \ & \ \ $^{+1.7}_{-1.7}$ \ \ & \ \ $5.5$ \ \ & \ \ $0.3$ \ \ & \ \ $0.1$ \ \ & & & \ \ $1.02$ \ \ & \ \ $^{+0}_{-2}$ \ \ & \ \ $0.99$ \ \ \\ 
 \ \ $598$--$642$ \ \ & \ \ $1.44\cdot 10^{-2}$ \ \ & \ \ $1.6$ \ \ & \ \ $0.52$ \ \ & \ \ $^{+12}_{-12}$ \ \ & \ \ $^{+0.2}_{-0.2}$ \ \ & \ \ $^{+0.0}_{-0.0}$ \ \ & \ \ $^{+1.6}_{-1.6}$ \ \ & \ \ $6.1$ \ \ & \ \ $0.3$ \ \ & \ \ $0.1$ \ \ & & & \ \ $1.02$ \ \ & \ \ $^{+0}_{-2}$ \ \ & \ \ $0.99$ \ \ \\ 
 \ \ $642$--$688$ \ \ & \ \ $6.47\cdot 10^{-3}$ \ \ & \ \ $2.5$ \ \ & \ \ $0.64$ \ \ & \ \ $^{+13}_{-13}$ \ \ & \ \ $^{+0.1}_{-0.2}$ \ \ & \ \ $^{+0.0}_{-0.0}$ \ \ & \ \ $^{+1.5}_{-1.6}$ \ \ & \ \ $6.8$ \ \ & \ \ $0.3$ \ \ & \ \ $0.1$ \ \ & & & \ \ $1.02$ \ \ & \ \ $^{+0}_{-2}$ \ \ & \ \ $0.99$ \ \ \\ 
 \ \ $688$--$736$ \ \ & \ \ $2.66\cdot 10^{-3}$ \ \ & \ \ $3.6$ \ \ & \ \ $0.61$ \ \ & \ \ $^{+14}_{-14}$ \ \ & \ \ $^{+0.1}_{-0.2}$ \ \ & \ \ $^{+0.0}_{-0.0}$ \ \ & \ \ $^{+1.4}_{-1.5}$ \ \ & \ \ $7.4$ \ \ & \ \ $0.3$ \ \ & \ \ $1.3$ \ \ & & & \ \ $1.02$ \ \ & \ \ $^{+0}_{-2}$ \ \ & \ \ $0.99$ \ \ \\ 
 \ \ $736$--$894$ \ \ & \ \ $5.36\cdot 10^{-4}$ \ \ & \ \ $5.3$ \ \ & \ \ $0.55$ \ \ & \ \ $^{+16}_{-16}$ \ \ & \ \ $^{+0.0}_{-0.1}$ \ \ & \ \ $^{+0.0}_{-0.0}$ \ \ & \ \ $^{+1.2}_{-1.2}$ \ \ & \ \ $8.5$ \ \ & \ \ $0.3$ \ \ & \ \ $2.7$ \ \ & & & \ \ $1.02$ \ \ & \ \ $^{+0}_{-2}$ \ \ & \ \ $0.99$ \ \ \\ 
 \ \ $894$--$1992$ \ \ & \ \ $2.70\cdot 10^{-6}$ \ \ & \ \ $27$ \ \ & \ \ $2.9$ \ \ & \ \ $^{+30}_{-27}$ \ \ & \ \ $^{+0.0}_{-0.5}$ \ \ & \ \ $^{+0.0}_{-0.0}$ \ \ & \ \ $^{+1.3}_{-1.4}$ \ \ & \ \ $15$ \ \ & \ \ $0.6$ \ \ & \ \ $4.3$ \ \ & & & \ \ $1.02$ \ \ & \ \ $^{+0}_{-2}$ \ \ & \ \ $0.99$ \ \ \\ 
\hline
\end{tabular}
}
\caption{
Measured double-differential inclusive jet cross-sections for jets with $R=0.6$ in the rapidity bin $1.5\leq|y| < 2.0$. See caption of table~\ref{tab:xsecr06y0} for details.
}
\label{tab:xsecr06y3}
\end{table*}

%% file: sys_R06_Eta5.tex
\renewcommand{\arraystretch}{1.8}
\begin{table*}[!ht]
\tiny
\centering
{\fontfamily{ptm}\selectfont
\begin{tabular}{|@{}c@{}@{}c@{}@{}c@{}@{}r@{}@{}r@{}@{}r@{}@{}r@{}@{}r@{}@{}r@{}@{}r@{}@{}r@{}|@{}r@{}@{}r@{}|@{}r@{}@{}r@{}@{}r@{}|}
\hline
 \ \ \pt range \ \ &
 \ \ $\sigma$ \ \ &
 \ \ $\delta_{\mathrm{stat}}^{\mathrm{data}}$ \ \ &
 \ \ $\delta_{\mathrm{stat}}^{\mathrm{MC}}$ \ \ &
 \ \ $u_{\operatorname{in-situ}}$ \ \ &
 \ \ $u_{\operatorname{pile-up}}$ \ \ &
 \ \ $u_{\operatorname{close-by}}$ \ \ &
\ \ $u_{\mathrm{flavour}}$ \ \ &
 \ \ $u_{\mathrm{JER}}$ \ \ &
 \ \ $u_{\mathrm{JAR}}$ \ \ &
 \ \ $u_{\mathrm{unfold}}$ \ \ &
 \ \ $u_{\mathrm{qual.}}$ \ \ &
 \ \ $u_{\mathrm{lumi}}$ \ \ &
 \ \ NPC \ \ &
 \ \ $u_{\mathrm{NP}}$ \ \ &
 \ \ EWC \ \ \\ \relax
\ \ [GeV] \ \ & \ \ [pb/GeV] \ \ &\ \  \% \ \ & \ \ \% \ \ & \ \  \% \ \ &
 \ \ \% \ \ & \ \ \% \ \ & \ \ \% \ \ & \ \ \% \ \ &
 \ \ \% \ \ & \ \ \% \ \ & \ \ \% \ \ & \ \ \% \ \ & \ \ \ \ \ & \ \ \% \ \ & \ \ \ \ \ \\
\hline
 \ \ $100$--$116$ \ \ & \ \ $2.26\cdot 10^{3}$ \ \ & \ \ $0.86$ \ \ & \ \ $1.0$ \ \ & \ \ $^{+15}_{-14}$ \ \ & \ \ $^{+2.0}_{-2.3}$ \ \ & \ \ $^{+1.6}_{-1.3}$ \ \ & \ \ $^{+4.8}_{-4.5}$ \ \ & \ \ $12$ \ \ & \ \ $1.5$ \ \ & \ \ $0.4$ \ \ & \multirow{19}{*}{\ \ $0.25$ \ \ } & \multirow{19}{*}{\ \ $1.8$ \ \ } & \ \ $1.05$ \ \ & \ \ $^{+2}_{-4}$ \ \ & \ \ $1.00$ \ \ \\ 
 \ \ $116$--$134$ \ \ & \ \ $9.78\cdot 10^{2}$ \ \ & \ \ $1.1$ \ \ & \ \ $0.97$ \ \ & \ \ $^{+15}_{-14}$ \ \ & \ \ $^{+2.0}_{-1.9}$ \ \ & \ \ $^{+1.2}_{-1.1}$ \ \ & \ \ $^{+4.2}_{-3.9}$ \ \ & \ \ $9.7$ \ \ & \ \ $1.1$ \ \ & \ \ $0.1$ \ \ & & & \ \ $1.04$ \ \ & \ \ $^{+2}_{-3}$ \ \ & \ \ $1.00$ \ \ \\ 
 \ \ $134$--$152$ \ \ & \ \ $4.45\cdot 10^{2}$ \ \ & \ \ $1.4$ \ \ & \ \ $0.82$ \ \ & \ \ $^{+15}_{-14}$ \ \ & \ \ $^{+1.8}_{-1.6}$ \ \ & \ \ $^{+1.0}_{-1.0}$ \ \ & \ \ $^{+3.8}_{-3.5}$ \ \ & \ \ $7.6$ \ \ & \ \ $0.9$ \ \ & \ \ $0.0$ \ \ & & & \ \ $1.03$ \ \ & \ \ $^{+2}_{-3}$ \ \ & \ \ $1.00$ \ \ \\ 
 \ \ $152$--$172$ \ \ & \ \ $1.98\cdot 10^{2}$ \ \ & \ \ $1.1$ \ \ & \ \ $1.1$ \ \ & \ \ $^{+15}_{-15}$ \ \ & \ \ $^{+1.6}_{-1.5}$ \ \ & \ \ $^{+1.0}_{-1.0}$ \ \ & \ \ $^{+3.4}_{-3.3}$ \ \ & \ \ $6.4$ \ \ & \ \ $0.8$ \ \ & \ \ $0.0$ \ \ & & & \ \ $1.03$ \ \ & \ \ $^{+2}_{-3}$ \ \ & \ \ $1.00$ \ \ \\ 
 \ \ $172$--$194$ \ \ & \ \ $9.36\cdot 10^{1}$ \ \ & \ \ $0.92$ \ \ & \ \ $0.51$ \ \ & \ \ $^{+16}_{-15}$ \ \ & \ \ $^{+1.6}_{-1.5}$ \ \ & \ \ $^{+1.0}_{-0.9}$ \ \ & \ \ $^{+3.2}_{-3.1}$ \ \ & \ \ $5.9$ \ \ & \ \ $0.8$ \ \ & \ \ $0.0$ \ \ & & & \ \ $1.03$ \ \ & \ \ $^{+2}_{-2}$ \ \ & \ \ $1.00$ \ \ \\ 
 \ \ $194$--$216$ \ \ & \ \ $4.22\cdot 10^{1}$ \ \ & \ \ $1.3$ \ \ & \ \ $0.55$ \ \ & \ \ $^{+17}_{-15}$ \ \ & \ \ $^{+1.6}_{-1.4}$ \ \ & \ \ $^{+1.0}_{-1.0}$ \ \ & \ \ $^{+3.0}_{-2.9}$ \ \ & \ \ $5.4$ \ \ & \ \ $0.7$ \ \ & \ \ $0.1$ \ \ & & & \ \ $1.02$ \ \ & \ \ $^{+2}_{-2}$ \ \ & \ \ $1.00$ \ \ \\ 
 \ \ $216$--$240$ \ \ & \ \ $2.00\cdot 10^{1}$ \ \ & \ \ $1.0$ \ \ & \ \ $0.77$ \ \ & \ \ $^{+17}_{-16}$ \ \ & \ \ $^{+1.5}_{-1.4}$ \ \ & \ \ $^{+1.1}_{-1.1}$ \ \ & \ \ $^{+2.8}_{-2.8}$ \ \ & \ \ $5.2$ \ \ & \ \ $0.7$ \ \ & \ \ $0.0$ \ \ & & & \ \ $1.02$ \ \ & \ \ $^{+1}_{-2}$ \ \ & \ \ $1.00$ \ \ \\ 
 \ \ $240$--$264$ \ \ & \ \ $9.67\cdot 10^{0}$ \ \ & \ \ $1.5$ \ \ & \ \ $0.59$ \ \ & \ \ $^{+18}_{-16}$ \ \ & \ \ $^{+1.3}_{-1.3}$ \ \ & \ \ $^{+1.2}_{-1.2}$ \ \ & \ \ $^{+2.7}_{-2.6}$ \ \ & \ \ $5.2$ \ \ & \ \ $0.7$ \ \ & \ \ $0.0$ \ \ & & & \ \ $1.02$ \ \ & \ \ $^{+1}_{-2}$ \ \ & \ \ $1.00$ \ \ \\ 
 \ \ $264$--$290$ \ \ & \ \ $4.69\cdot 10^{0}$ \ \ & \ \ $2.1$ \ \ & \ \ $0.71$ \ \ & \ \ $^{+18}_{-15}$ \ \ & \ \ $^{+1.3}_{-1.2}$ \ \ & \ \ $^{+1.3}_{-1.2}$ \ \ & \ \ $^{+2.5}_{-2.5}$ \ \ & \ \ $5.5$ \ \ & \ \ $0.7$ \ \ & \ \ $0.0$ \ \ & & & \ \ $1.02$ \ \ & \ \ $^{+1}_{-2}$ \ \ & \ \ $0.99$ \ \ \\ 
 \ \ $290$--$318$ \ \ & \ \ $2.14\cdot 10^{0}$ \ \ & \ \ $3.0$ \ \ & \ \ $0.76$ \ \ & \ \ $^{+17}_{-15}$ \ \ & \ \ $^{+1.4}_{-1.3}$ \ \ & \ \ $^{+1.2}_{-1.2}$ \ \ & \ \ $^{+2.5}_{-2.4}$ \ \ & \ \ $5.8$ \ \ & \ \ $0.8$ \ \ & \ \ $0.1$ \ \ & & & \ \ $1.02$ \ \ & \ \ $^{+1}_{-2}$ \ \ & \ \ $0.99$ \ \ \\ 
 \ \ $318$--$346$ \ \ & \ \ $9.48\cdot 10^{-1}$ \ \ & \ \ $3.4$ \ \ & \ \ $0.86$ \ \ & \ \ $^{+17}_{-15}$ \ \ & \ \ $^{+1.6}_{-1.4}$ \ \ & \ \ $^{+1.1}_{-1.0}$ \ \ & \ \ $^{+2.3}_{-2.3}$ \ \ & \ \ $6.2$ \ \ & \ \ $0.7$ \ \ & \ \ $0.0$ \ \ & & & \ \ $1.02$ \ \ & \ \ $^{+1}_{-2}$ \ \ & \ \ $0.99$ \ \ \\ 
 \ \ $346$--$376$ \ \ & \ \ $4.32\cdot 10^{-1}$ \ \ & \ \ $1.9$ \ \ & \ \ $1.2$ \ \ & \ \ $^{+17}_{-15}$ \ \ & \ \ $^{+1.8}_{-1.6}$ \ \ & \ \ $^{+0.9}_{-0.9}$ \ \ & \ \ $^{+2.2}_{-2.2}$ \ \ & \ \ $6.6$ \ \ & \ \ $0.6$ \ \ & \ \ $0.1$ \ \ & & & \ \ $1.02$ \ \ & \ \ $^{+1}_{-2}$ \ \ & \ \ $0.99$ \ \ \\ 
 \ \ $376$--$408$ \ \ & \ \ $1.83\cdot 10^{-1}$ \ \ & \ \ $1.2$ \ \ & \ \ $1.1$ \ \ & \ \ $^{+17}_{-16}$ \ \ & \ \ $^{+1.8}_{-1.6}$ \ \ & \ \ $^{+0.7}_{-0.7}$ \ \ & \ \ $^{+2.1}_{-2.3}$ \ \ & \ \ $7.2$ \ \ & \ \ $0.5$ \ \ & \ \ $0.0$ \ \ & & & \ \ $1.02$ \ \ & \ \ $^{+1}_{-2}$ \ \ & \ \ $0.99$ \ \ \\ 
 \ \ $408$--$442$ \ \ & \ \ $7.48\cdot 10^{-2}$ \ \ & \ \ $1.8$ \ \ & \ \ $1.6$ \ \ & \ \ $^{+18}_{-17}$ \ \ & \ \ $^{+1.6}_{-1.3}$ \ \ & \ \ $^{+0.5}_{-0.5}$ \ \ & \ \ $^{+2.1}_{-2.3}$ \ \ & \ \ $7.8$ \ \ & \ \ $0.4$ \ \ & \ \ $0.0$ \ \ & & & \ \ $1.02$ \ \ & \ \ $^{+0}_{-2}$ \ \ & \ \ $0.99$ \ \ \\ 
 \ \ $442$--$478$ \ \ & \ \ $2.88\cdot 10^{-2}$ \ \ & \ \ $2.4$ \ \ & \ \ $1.4$ \ \ & \ \ $^{+20}_{-18}$ \ \ & \ \ $^{+1.1}_{-0.8}$ \ \ & \ \ $^{+0.3}_{-0.3}$ \ \ & \ \ $^{+2.1}_{-2.2}$ \ \ & \ \ $8.5$ \ \ & \ \ $0.4$ \ \ & \ \ $0.1$ \ \ & & & \ \ $1.02$ \ \ & \ \ $^{+0}_{-2}$ \ \ & \ \ $0.99$ \ \ \\ 
 \ \ $478$--$516$ \ \ & \ \ $1.01\cdot 10^{-2}$ \ \ & \ \ $2.3$ \ \ & \ \ $1.3$ \ \ & \ \ $^{+22}_{-20}$ \ \ & \ \ $^{+0.5}_{-0.4}$ \ \ & \ \ $^{+0.2}_{-0.1}$ \ \ & \ \ $^{+2.1}_{-2.1}$ \ \ & \ \ $9.4$ \ \ & \ \ $0.4$ \ \ & \ \ $0.1$ \ \ & & & \ \ $1.02$ \ \ & \ \ $^{+0}_{-2}$ \ \ & \ \ $0.99$ \ \ \\ 
 \ \ $516$--$556$ \ \ & \ \ $3.29\cdot 10^{-3}$ \ \ & \ \ $4.0$ \ \ & \ \ $2.2$ \ \ & \ \ $^{+24}_{-22}$ \ \ & \ \ $^{+0.1}_{-0.1}$ \ \ & \ \ $^{+0.1}_{-0.0}$ \ \ & \ \ $^{+1.9}_{-2.0}$ \ \ & \ \ $11$ \ \ & \ \ $0.4$ \ \ & \ \ $0.8$ \ \ & & & \ \ $1.02$ \ \ & \ \ $^{+0}_{-2}$ \ \ & \ \ $0.99$ \ \ \\ 
 \ \ $556$--$642$ \ \ & \ \ $6.57\cdot 10^{-4}$ \ \ & \ \ $6.6$ \ \ & \ \ $2.4$ \ \ & \ \ $^{+27}_{-22}$ \ \ & \ \ $^{+0.1}_{-0.2}$ \ \ & \ \ $^{+0.0}_{-0.0}$ \ \ & \ \ $^{+1.4}_{-1.8}$ \ \ & \ \ $13$ \ \ & \ \ $0.4$ \ \ & \ \ $1.2$ \ \ & & & \ \ $1.02$ \ \ & \ \ $^{+0}_{-3}$ \ \ & \ \ $0.99$ \ \ \\ 
 \ \ $642$--$894$ \ \ & \ \ $1.20\cdot 10^{-5}$ \ \ & \ \ $29$ \ \ & \ \ $5.5$ \ \ & \ \ $^{+48}_{-38}$ \ \ & \ \ $^{+0.2}_{-0.3}$ \ \ & \ \ $^{+0.0}_{-0.0}$ \ \ & \ \ $^{+0.2}_{-0.7}$ \ \ & \ \ $31$ \ \ & \ \ $0.4$ \ \ & \ \ $1.7$ \ \ & & & \ \ $1.02$ \ \ & \ \ $^{+0}_{-3}$ \ \ & \ \ $0.98$ \ \ \\ 
\hline
\end{tabular}
}
\caption{
Measured double-differential inclusive jet cross-sections for jets with $R=0.6$ in the rapidity bin $2.0\leq|y| < 2.5$. See caption of table~\ref{tab:xsecr06y0} for details.
}
\label{tab:xsecr06y4}
\end{table*}

%% file: sys_R06_Eta6.tex
\renewcommand{\arraystretch}{1.8}
\begin{table*}[!ht]
\tiny
\centering
{\fontfamily{ptm}\selectfont
\begin{tabular}{|@{}c@{}@{}c@{}@{}c@{}@{}r@{}@{}r@{}@{}r@{}@{}r@{}@{}r@{}@{}r@{}@{}r@{}@{}r@{}|@{}r@{}@{}r@{}|@{}r@{}@{}r@{}@{}r@{}|}
\hline
 \ \ \pt range \ \ &
 \ \ $\sigma$ \ \ &
 \ \ $\delta_{\mathrm{stat}}^{\mathrm{data}}$ \ \ &
 \ \ $\delta_{\mathrm{stat}}^{\mathrm{MC}}$ \ \ &
 \ \ $u_{\operatorname{in-situ}}$ \ \ &
 \ \ $u_{\operatorname{pile-up}}$ \ \ &
 \ \ $u_{\operatorname{close-by}}$ \ \ &
\ \ $u_{\mathrm{flavour}}$ \ \ &
 \ \ $u_{\mathrm{JER}}$ \ \ &
 \ \ $u_{\mathrm{JAR}}$ \ \ &
 \ \ $u_{\mathrm{unfold}}$ \ \ &
 \ \ $u_{\mathrm{qual.}}$ \ \ &
 \ \ $u_{\mathrm{lumi}}$ \ \ &
 \ \ NPC \ \ &
 \ \ $u_{\mathrm{NP}}$ \ \ &
 \ \ EWC \ \ \\ \relax
\ \ [GeV] \ \ & \ \ [pb/GeV] \ \ &\ \  \% \ \ & \ \ \% \ \ & \ \  \% \ \ &
 \ \ \% \ \ & \ \ \% \ \ & \ \ \% \ \ & \ \ \% \ \ &
 \ \ \% \ \ & \ \ \% \ \ & \ \ \% \ \ & \ \ \% \ \ & \ \ \ \ \ & \ \ \% \ \ & \ \ \ \ \ \\
\hline
 \ \ $100$--$116$ \ \ & \ \ $1.36\cdot 10^{3}$ \ \ & \ \ $1.0$ \ \ & \ \ $1.1$ \ \ & \ \ $^{+21}_{-18}$ \ \ & \ \ $^{+2.6}_{-1.9}$ \ \ & \ \ $^{+1.4}_{-1.4}$ \ \ & \ \ $^{+3.5}_{-3.0}$ \ \ & \ \ $13$ \ \ & \ \ $2.6$ \ \ & \ \ $0.1$ \ \ & \multirow{12}{*}{\ \ $0.25$ \ \ } & \multirow{12}{*}{\ \ $1.8$ \ \ } & \ \ $1.04$ \ \ & \ \ $^{+2}_{-3}$ \ \ & \ \ $1.00$ \ \ \\ 
 \ \ $116$--$134$ \ \ & \ \ $5.20\cdot 10^{2}$ \ \ & \ \ $1.5$ \ \ & \ \ $1.1$ \ \ & \ \ $^{+22}_{-19}$ \ \ & \ \ $^{+2.1}_{-2.1}$ \ \ & \ \ $^{+1.3}_{-1.1}$ \ \ & \ \ $^{+2.8}_{-2.9}$ \ \ & \ \ $10$ \ \ & \ \ $2.3$ \ \ & \ \ $0.1$ \ \ & & & \ \ $1.03$ \ \ & \ \ $^{+2}_{-2}$ \ \ & \ \ $1.00$ \ \ \\ 
 \ \ $134$--$152$ \ \ & \ \ $2.04\cdot 10^{2}$ \ \ & \ \ $2.1$ \ \ & \ \ $0.83$ \ \ & \ \ $^{+22}_{-20}$ \ \ & \ \ $^{+2.1}_{-2.4}$ \ \ & \ \ $^{+1.1}_{-1.0}$ \ \ & \ \ $^{+2.6}_{-2.9}$ \ \ & \ \ $9.0$ \ \ & \ \ $2.1$ \ \ & \ \ $0.0$ \ \ & & & \ \ $1.03$ \ \ & \ \ $^{+2}_{-2}$ \ \ & \ \ $1.00$ \ \ \\ 
 \ \ $152$--$172$ \ \ & \ \ $8.20\cdot 10^{1}$ \ \ & \ \ $1.7$ \ \ & \ \ $0.83$ \ \ & \ \ $^{+23}_{-21}$ \ \ & \ \ $^{+2.1}_{-2.6}$ \ \ & \ \ $^{+1.0}_{-1.0}$ \ \ & \ \ $^{+2.5}_{-2.9}$ \ \ & \ \ $8.4$ \ \ & \ \ $1.9$ \ \ & \ \ $0.0$ \ \ & & & \ \ $1.03$ \ \ & \ \ $^{+2}_{-2}$ \ \ & \ \ $1.00$ \ \ \\ 
 \ \ $172$--$194$ \ \ & \ \ $3.10\cdot 10^{1}$ \ \ & \ \ $1.6$ \ \ & \ \ $0.92$ \ \ & \ \ $^{+25}_{-22}$ \ \ & \ \ $^{+2.1}_{-2.5}$ \ \ & \ \ $^{+1.0}_{-1.1}$ \ \ & \ \ $^{+2.4}_{-2.8}$ \ \ & \ \ $8.4$ \ \ & \ \ $2.0$ \ \ & \ \ $0.0$ \ \ & & & \ \ $1.03$ \ \ & \ \ $^{+2}_{-2}$ \ \ & \ \ $1.00$ \ \ \\ 
 \ \ $194$--$216$ \ \ & \ \ $1.14\cdot 10^{1}$ \ \ & \ \ $2.5$ \ \ & \ \ $1.1$ \ \ & \ \ $^{+28}_{-23}$ \ \ & \ \ $^{+2.2}_{-2.2}$ \ \ & \ \ $^{+1.1}_{-1.1}$ \ \ & \ \ $^{+2.6}_{-2.6}$ \ \ & \ \ $8.9$ \ \ & \ \ $2.3$ \ \ & \ \ $0.1$ \ \ & & & \ \ $1.03$ \ \ & \ \ $^{+1}_{-2}$ \ \ & \ \ $1.00$ \ \ \\ 
 \ \ $216$--$240$ \ \ & \ \ $4.30\cdot 10^{0}$ \ \ & \ \ $2.2$ \ \ & \ \ $1.6$ \ \ & \ \ $^{+31}_{-25}$ \ \ & \ \ $^{+2.3}_{-2.2}$ \ \ & \ \ $^{+1.2}_{-1.1}$ \ \ & \ \ $^{+2.8}_{-2.6}$ \ \ & \ \ $9.7$ \ \ & \ \ $2.7$ \ \ & \ \ $0.0$ \ \ & & & \ \ $1.03$ \ \ & \ \ $^{+1}_{-2}$ \ \ & \ \ $0.99$ \ \ \\ 
 \ \ $240$--$264$ \ \ & \ \ $1.50\cdot 10^{0}$ \ \ & \ \ $3.5$ \ \ & \ \ $1.6$ \ \ & \ \ $^{+34}_{-27}$ \ \ & \ \ $^{+2.5}_{-2.2}$ \ \ & \ \ $^{+1.4}_{-1.3}$ \ \ & \ \ $^{+2.9}_{-2.7}$ \ \ & \ \ $10$ \ \ & \ \ $2.9$ \ \ & \ \ $0.0$ \ \ & & & \ \ $1.03$ \ \ & \ \ $^{+1}_{-2}$ \ \ & \ \ $0.99$ \ \ \\ 
 \ \ $264$--$290$ \ \ & \ \ $4.86\cdot 10^{-1}$ \ \ & \ \ $5.7$ \ \ & \ \ $2.3$ \ \ & \ \ $^{+36}_{-28}$ \ \ & \ \ $^{+2.5}_{-2.3}$ \ \ & \ \ $^{+1.5}_{-1.4}$ \ \ & \ \ $^{+2.7}_{-2.5}$ \ \ & \ \ $12$ \ \ & \ \ $2.9$ \ \ & \ \ $0.1$ \ \ & & & \ \ $1.03$ \ \ & \ \ $^{+1}_{-3}$ \ \ & \ \ $0.99$ \ \ \\ 
 \ \ $290$--$318$ \ \ & \ \ $1.77\cdot 10^{-1}$ \ \ & \ \ $9.2$ \ \ & \ \ $2.1$ \ \ & \ \ $^{+38}_{-29}$ \ \ & \ \ $^{+2.4}_{-2.4}$ \ \ & \ \ $^{+1.6}_{-1.4}$ \ \ & \ \ $^{+2.6}_{-2.4}$ \ \ & \ \ $13$ \ \ & \ \ $3.0$ \ \ & \ \ $0.0$ \ \ & & & \ \ $1.03$ \ \ & \ \ $^{+0}_{-3}$ \ \ & \ \ $0.99$ \ \ \\ 
 \ \ $318$--$376$ \ \ & \ \ $2.89\cdot 10^{-2}$ \ \ & \ \ $14$ \ \ & \ \ $2.3$ \ \ & \ \ $^{+42}_{-30}$ \ \ & \ \ $^{+2.6}_{-2.9}$ \ \ & \ \ $^{+1.5}_{-1.2}$ \ \ & \ \ $^{+2.9}_{-2.6}$ \ \ & \ \ $16$ \ \ & \ \ $3.6$ \ \ & \ \ $0.1$ \ \ & & & \ \ $1.03$ \ \ & \ \ $^{+0}_{-3}$ \ \ & \ \ $0.99$ \ \ \\ 
 \ \ $376$--$478$ \ \ & \ \ $8.92\cdot 10^{-4}$ \ \ & \ \ $9.9$ \ \ & \ \ $6.9$ \ \ & \ \ $^{+55}_{-38}$ \ \ & \ \ $^{+6.3}_{-5.6}$ \ \ & \ \ $^{+1.4}_{-0.3}$ \ \ & \ \ $^{+3.0}_{-3.4}$ \ \ & \ \ $29$ \ \ & \ \ $11$ \ \ & \ \ $0.3$ \ \ & & & \ \ $1.03$ \ \ & \ \ $^{+0}_{-3}$ \ \ & \ \ $0.99$ \ \ \\ 
\hline
\end{tabular}
}
\caption{
Measured double-differential inclusive jet cross-sections for jets with $R=0.6$ in the rapidity bin $2.5\leq|y| < 3.0$. See caption of table~\ref{tab:xsecr06y0} for details.
}
\label{tab:xsecr06y5}
\end{table*}

%% file: atlas_authlist.tex
\begin{flushleft}
{\Large The ATLAS Collaboration}

\bigskip

G.~Aad$^{\rm 84}$,
B.~Abbott$^{\rm 112}$,
J.~Abdallah$^{\rm 152}$,
S.~Abdel~Khalek$^{\rm 116}$,
O.~Abdinov$^{\rm 11}$,
R.~Aben$^{\rm 106}$,
B.~Abi$^{\rm 113}$,
M.~Abolins$^{\rm 89}$,
O.S.~AbouZeid$^{\rm 159}$,
H.~Abramowicz$^{\rm 154}$
% ...
\bigskip
\\
% ...
$^{11}$ Institute of Physics, Azerbaijan Academy of Sciences, Baku, Azerbaijan\\
% ...
$^{84}$ CPPM, Aix-Marseille Universit{\'e} and CNRS/IN2P3, Marseille, France\\
% ...
$^{89}$ Department of Physics and Astronomy, Michigan State University, East Lansing MI, United States of America\\
% ...
$^{106}$ Nikhef National Institute for Subatomic Physics and University of Amsterdam, Amsterdam, Netherlands\\
% ...
$^{112}$ Homer L. Dodge Department of Physics and Astronomy, University of Oklahoma, Norman OK, United States of America\\
$^{113}$ Department of Physics, Oklahoma State University, Stillwater OK, United States of America\\
% ...
$^{116}$ LAL, Universit{\'e} Paris-Sud and CNRS/IN2P3, Orsay, France\\
% ...
$^{152}$ Institute of Physics, Academia Sinica, Taipei, Taiwan\\
% ...
$^{154}$ Raymond and Beverly Sackler School of Physics and Astronomy, Tel Aviv University, Tel Aviv, Israel\\
% ...
$^{159}$ Department of Physics, University of Toronto, Toronto ON, Canada\\
% ...
\end{flushleft}
